\documentclass [manuscript]{aastex}
\usepackage{amsmath}
\usepackage{graphicx}

\newcommand{\lya}{Lyman-$\alpha$}

\shorttitle{THE INFLUENCE OF THE EUV SED ON THE ATMOSPHERE OF EXOPLANET}
\shortauthors{Guo \& Ben-Jaffel}

\begin{document}

\title{The influence of the Extreme Ultraviolet spectral energy distribution on the structure and composition of the upper atmosphere of exoplanets}
\author{J. H. Guo\altaffilmark{1} }
\affil{Yunnan Observatories,
Chinese Academy of Sciences, P.O. Box 110, Kunming 650011, China; guojh@ynao.ac.cn}
\author{Lotfi Ben-Jaffel\altaffilmark{2}}
\affil{Sorbonne UniversitŽ\'es, UPMC Univ Paris 6 et CNRS, UMR 7095, Institut  Astrophysique de Paris,
F-75014 Paris, France; bjaffel@iap.fr}



\altaffiltext{1}{Key Laboratory for the Structure and Evolution of
Celestial Objects, CAS, Kunming 650011, China}
\altaffiltext{2}{Visiting Research Scientist, University of Arizona, Lunar \& Planetary Laboratory, Tucson, USA}
\begin{abstract}
By varying the profiles of stellar extreme ultraviolet (EUV) spectral energy distribution (SED), we tested the influences of stellar EUV SEDs on the physical and chemical properties of the escaping atmosphere. We apply our model to study four exoplanets, HD\,189733b, HD\,209458b, GJ \,436b, and Kepler-11b. We found that the total mass loss rates of an exoplanet, which are determined mainly by the integrated fluxes, are moderately affected by the profiles of the EUV SED, but the composition and species distributions in the atmosphere can be dramatically modified by the different profiles of the EUV SED. For exoplanets with a high hydrodynamic escape parameter ($\lambda$), the amount of atomic hydrogen produced by photoionization at different altitudes can vary by one to two orders of magnitude with the variation of stellar EUV SEDs. The effect of photoionization of H is prominent when the EUV SED is dominated by the low-energy spectral region (400-900${\AA}$), which pushes the transition of H/H$^{+}$ to low altitudes. On the contrary, the transitions of H/H$^{+}$ move to higher altitudes when most photons concentrate in the high-energy spectral region (50-400${\AA}$). For exoplanets with a low $\lambda$, the lower temperatures of the atmosphere make many chemical reactions so important that photoionization alone can no longer determine the composition of the escaping atmosphere.
For HD 189733b, it is possible to explain the time variability of \lya\ between 2010 and 2011 by a change in the EUV SED of the host K star, yet invoking only thermal H\,I in the atmosphere.


\end{abstract}


\section{INTRODUCTION AND MOTIVATION}
Stellar irradiation can be the most important source of energy in planetary atmospheres. For exoplanets orbiting near to their parent stars, the stellar irradiation largely affects the physical and chemical properties of their atmospheres, which ultimately determines their structure and composition. If an exoplanet orbits at a very close orbital distance ($<$ 0.15AU) and receives strong XUV (1-912$\AA$) irradiation from its host star, the atmosphere could be unstable \citep[]{kos07}. This means that atmospheric species can escape the bounds of the planet \citep[]{lammer03}.

Planetary transits provide good opportunities to probe the atmospheric composition and structure of exoplanets, which may constrain their escape processes. All ultraviolet observations confirmed a strong absorption of the Lyman $\alpha$ line during the transit of HD 209458b, HD 189733b, and GJ 436b \citep[]{vidal03,ben07,lecavelier10,bou13a,kulow14}.  Heavier elements such as OI and CII were also detected by HST during the transit of HD 209458b \citep[]{vidal04,ben10}.
Recently, \citet[]{ball15} reported that Si III ions have not been detected in the thermosphere of HD 209458b. Apparently, Silicon may exist in the upper atmosphere but with a lower ionization state. For HD189733b, \citet[]{ben13} also detected neutral oxygen and possibly ionized carbon in the upper atmosphere. For HD 209458b, most theoretical models attributed the observed transit absorption directly to a mass loss rate in the order of magnitude of 10$^{10}$g s$^{-1}$ \citep[]{yelle04,tian05,gar07,penz08,murray09,guo11,guo13,bou13b,kos13}. However, while the role of the stellar wind could be important, its impact on the Lyman $\alpha$ transit absorption is not clear. For instance, the stellar wind's pressure can confine the atmospheric escape of the planet and change the geometrical distribution of particles \citep[]{stone09}. The charge exchange between the stellar wind and planetary wind can also provide hot atomic hydrogen that can produce significant absorption in Lyman $\alpha$ if the process is efficient \citep[]{holm08, trem13}. Finally, recent modeling of atmospheric escape included the effect of a planetary magnetic field, which, if strong enough, may confine the atmosphere \citep[]{adams11, trammell14, khoda15}.

Among the above studies, some models use the solar EUV spectra. Generally, the EUV band is defined in the range of 100-912$\AA$, yet here we restrict its range to the 50-900$\AA$ band as the radiative input (see Section 2.1 for more details).
In fact, the profiles of the spectral energy distribution in the EUV band (EUV SED) depend on the type and activity level of the star. Since stellar EUV fluxes are very difficult to measure, \citet[]{sanz11} developed a technique to reconstruct the EUV fluxes based on an emission measure analysis of observed stellar X-ray spectra. Subsequently, \citet[]{linsky14} also reconstructed the EUV SED by relating the ratio of the fluxes of Lyman $\alpha$ with the EUV, yet some discrepancies appear with the analysis of \citet[]{sanz11}. For reference, \citet[]{linsky13} reported that the fluxes of Lyman $\alpha$ are related to the stellar temperatures and rotational periods. More generally, all the stellar intrinsic parameters should influence the stellar radiative input, a fundamental yet difficult problem that is becoming crucial for explanatory studies because of the tremendous impact of the EUV SED on their atmospheres.

For the Sun, the EUV SEDs are dominated by the higher-energy spectral regions in its young age, but the higher-energy regions decrease with the Sun's evolution \citep[]{ribas05}.
\citet[]{claire12} developed a method of parameterization to depict the evolution of solar flux, which can provide quantitative estimates of the wavelength dependence of the flux for G-type stars. In addition, stellar flares can result in the emission of higher-energy photons \citep[]{sanz02,osten04}. This means that the EUV SED of a star not only varies with the long evolutionary time, but is also related to short-timescale activity.

On the theoretical side, some models \citep[]{penz08,murray09,guo11} use an integrated flux to denote the stellar irradiation. The gray approximation to the heating and photoionization should be inspected further. In addition, it is important to understand the influences of different profiles of the EUV SED on the mass loss rates and the chemical properties because they determine the remnant planetary mass and the composition of the atmosphere with the planetary evolution. The composition of the upper atmosphere plays an important role in detecting escape of the atmosphere. For hydrogen-dominated atmospheres, the amount of neutral hydrogen atoms (or degree of ionization) determines the level of the \lya\ transit absorption that can be detected, which in turn helps constrain the strength of the escape processes that operate in the upper atmosphere. Therefore, an interesting question presents itself:
Are the mass loss rate and the atmospheric composition of exoplanets variable with the variation of the EUV SED's profile.

In the following, we focus on the effects of the EUV SED on the mass loss rates and the chemical composition of escaping atmosphere. In Section 2, we present the model and describe how to construct the EUV SED profiles. In Section 3, we discuss our atmospheric model results for a set of exoplanets, including two hot Jupiters, HD 189733b and HD 209458b, the hot Neptune-like GJ 436b, and the super-Earth Kepler-11b. In Section 4.1, we focus our discussion on the sensitivity of atmospheric properties to the hydrodynamic escape parameter. The effect of second ionization is discussed in Section 4.2. In Section 5, we discuss with particularity the example of time variability of the HD\,189733b \lya\ transit absorption observed in 2010 and 2011 in terms of a potential variability in the EUV SED index.

\section{Constructing the EUV SED and Atmospheric Model}

\subsection{The EUV SED}
The EUV irradiation is the main energy in heating and ionizing the upper atmospheres of exoplanets orbiting near their parent stars. The stellar EUV SED is wavelength-dependent, making exoplanets, at their orbits, receive different energies at different wavelengths. If the energies at different wavelengths are integrated in total EUV bands, we can obtain the integrated flux. The amount of integrated flux in the EUV band determines how much energy can be deposited depending on the atmosphere's composition. In an ionized atmosphere, the photons with different energy can penetrate to different depths because the absorption cross-section, $\sigma$, is also wavelength-dependent. Generally, the absorption cross-section of higher-energy photons is smaller than that of lower-energy photons for H, H$_{2}$, and He. Thus, the higher-energy photons in the hydrogen-dominated atmosphere can penetrate to deeper altitudes.
It is important to quantitatively understand the impact of the EUV SED on different planets. In that frame, our study covers a wide set of star-planet systems. We test a sample of four typical and particular planets: HD 189733b and HD 209458b (Hot-Jupiter), GJ 436b (Hot-Neptune), and Kepler-11b (Super-earth).
These four cases cover different masses, sizes, and orbital distances of exoplanets. The EUV flux received at their orbits are also very diverse. Transit absorptions have been clearly detected for HD 189733b, HD 209458b, and GJ 436b, revealing their main composition. For the HD 189733 system, the \lya\ transit absorption varied over a one-year period, which may offer an opportunity to constrain the EUV radiation of its host K-star for different periods (Section 4). Although there have been no spectroscopic investigations for Kepler -11b, the relatively far orbital distance (see Table 2) of the planet means that it receives fainter EUV irradiation.  Moreover, the atmospheric temperature of Kepler-11b is lower than that of HD 209458b \citep[]{lammer13}.
Compared to hot planets, the factors mentioned above can affect the response of the atmosphere to the EUV SED and result in different atmospheric scenarios. Therefore, the study we are implementing with this sample can give us a better understanding of how the atmospheric composition and mass loss of the planets with different properties are affected by the EUV SED of a host star. For reference, we note that we exclude from our study exoplanets orbiting farther from their host stars because their atmospheres should be so stable that the hydrodynamic escape would not occur.

\begin{figure}
  \begin{center}
\centering
\includegraphics[width=5.5in,height=4.in]{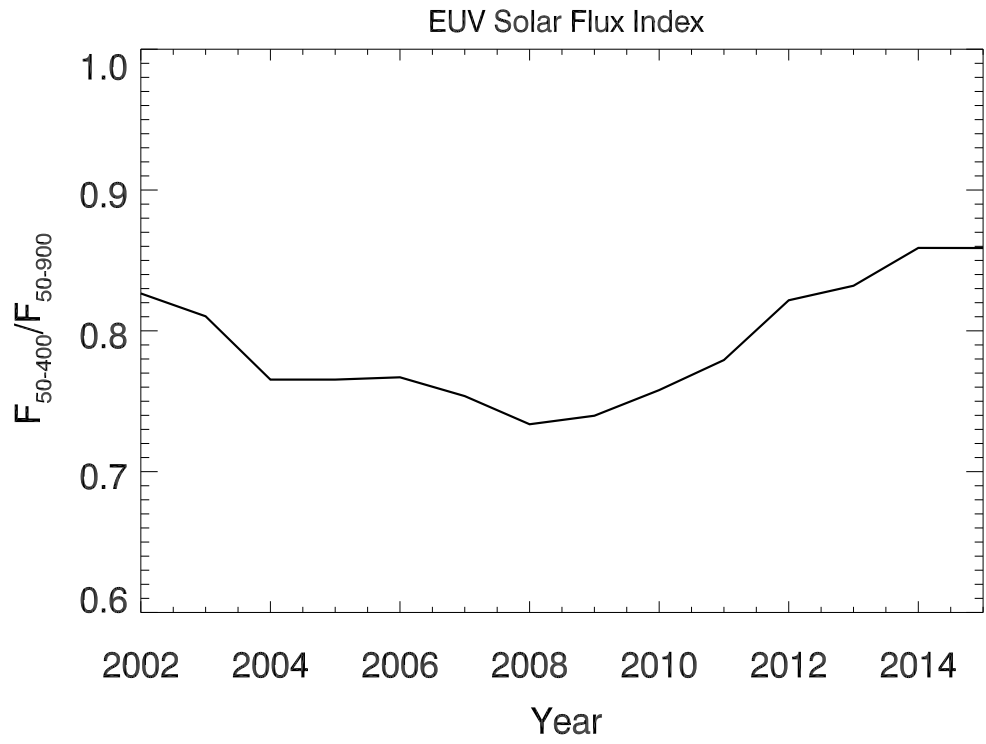}
\caption{The variations of solar EUV spectral index $\beta$ with time.}
  \end{center}
\end{figure}

In order to probe the influence of the EUV SED on the upper atmosphere, we artificially construct the spectral energy distributions, while the integrated EUV fluxes are set as constants. It is indeed possible that the integrated flux remains unchanged, but the profile of the EUV SED may change over time or from one target to another.
Using the EUV spectrum of the Sun is a reasonable approximation for HD 209458 and Kepler-11 because the two stars are solar-like. As an M-type star, GJ 436 has a spectrum that should be different from that of a G-star.
Unfortunately, we cannot find any reliable EUV spectrum in the literature, which forces us to use the solar spectrum as an approximation for that target.

\begin{figure}
\begin{minipage}[t]{0.5\linewidth}
\centering
\includegraphics[width=3.6in,height=2.6in]{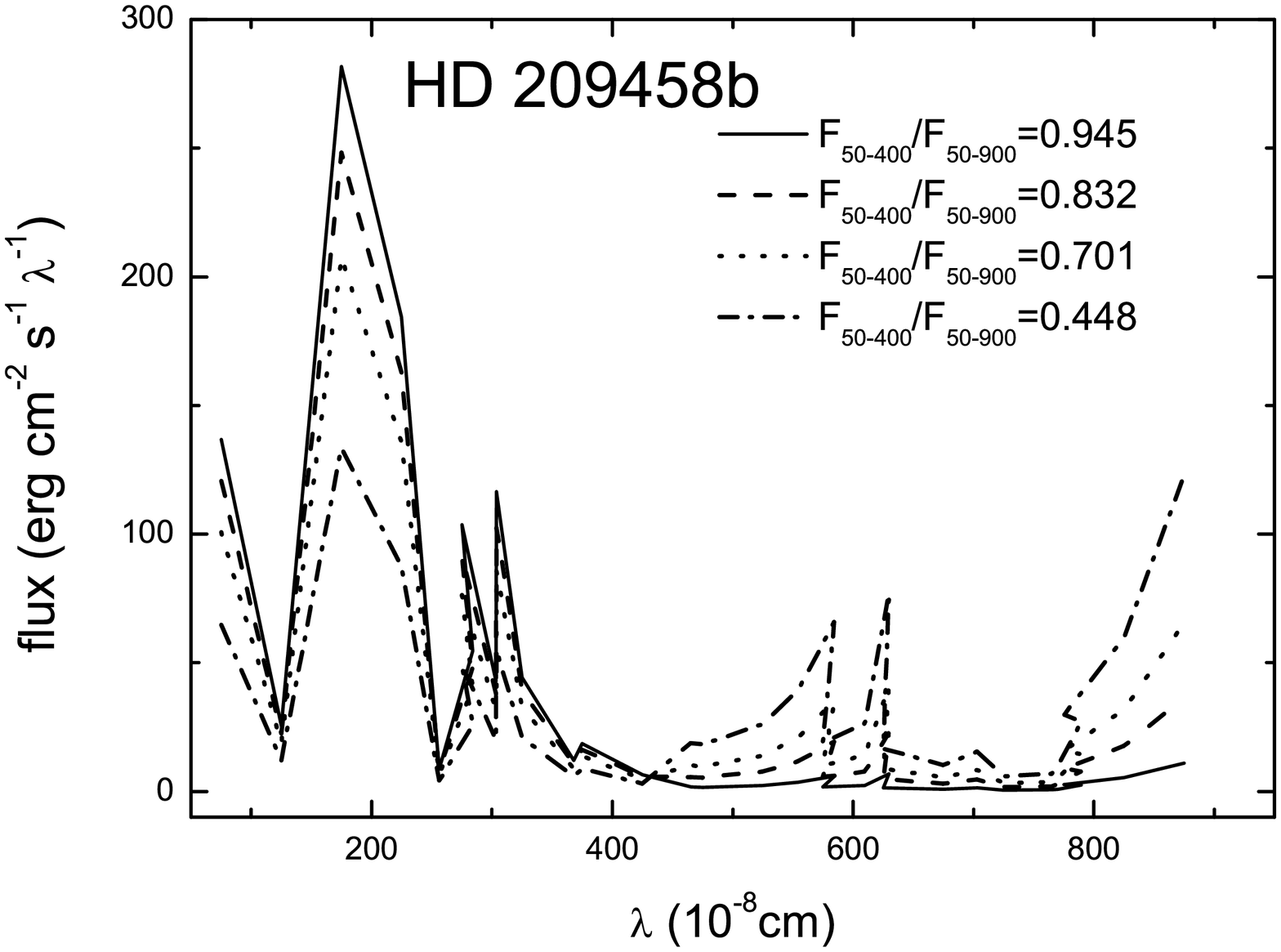}
\end{minipage}
\begin{minipage}[t]{0.5\linewidth}
\centering
\includegraphics[width=3.6in,height=2.6in]{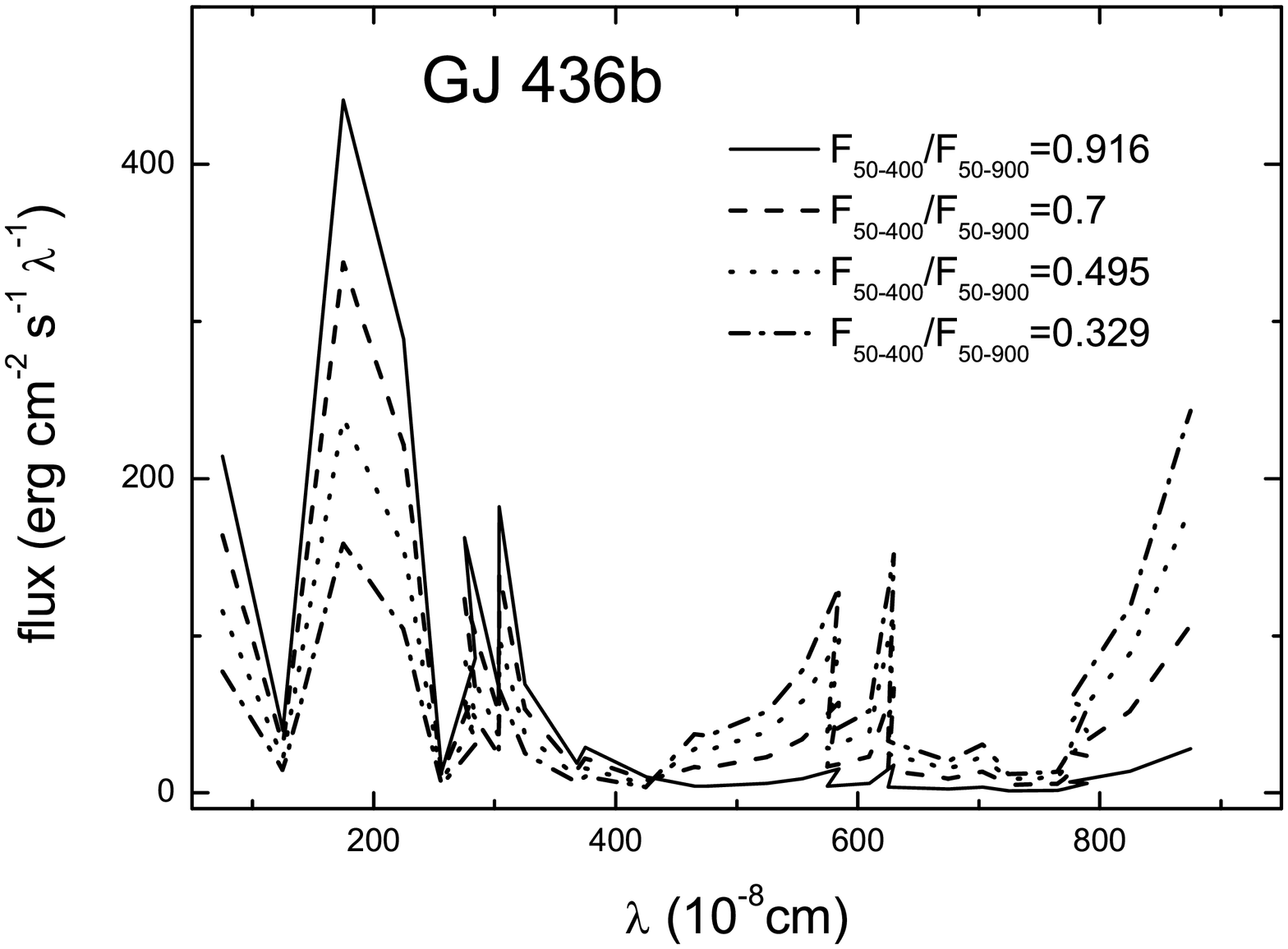}
\end{minipage}
\begin{minipage}[t]{0.5\linewidth}
\centering
\includegraphics[width=3.6in,height=2.6in]{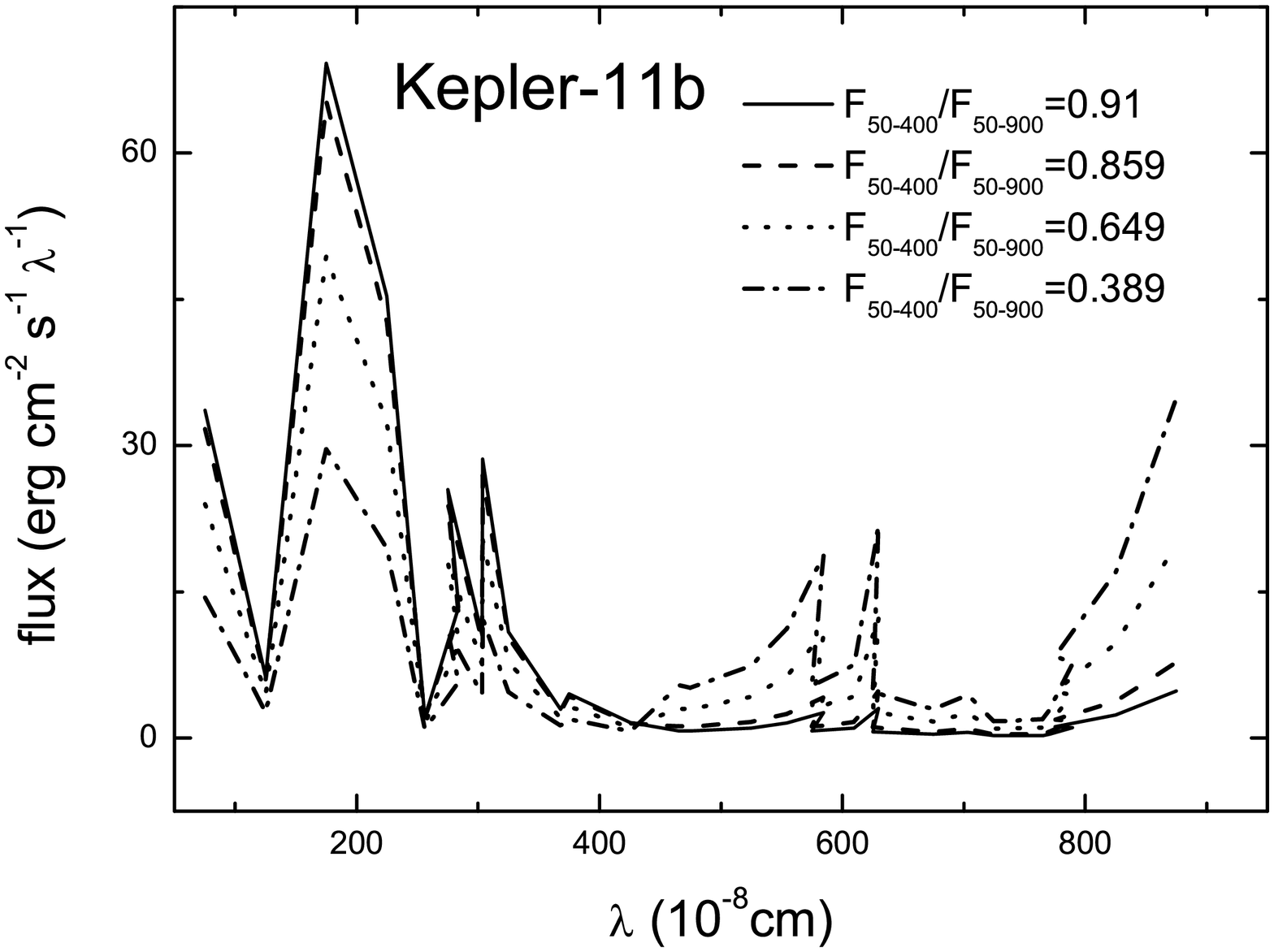}
\end{minipage}
\begin{minipage}[t]{0.5\linewidth}
\centering
\includegraphics[width=3.6in,height=2.6in]{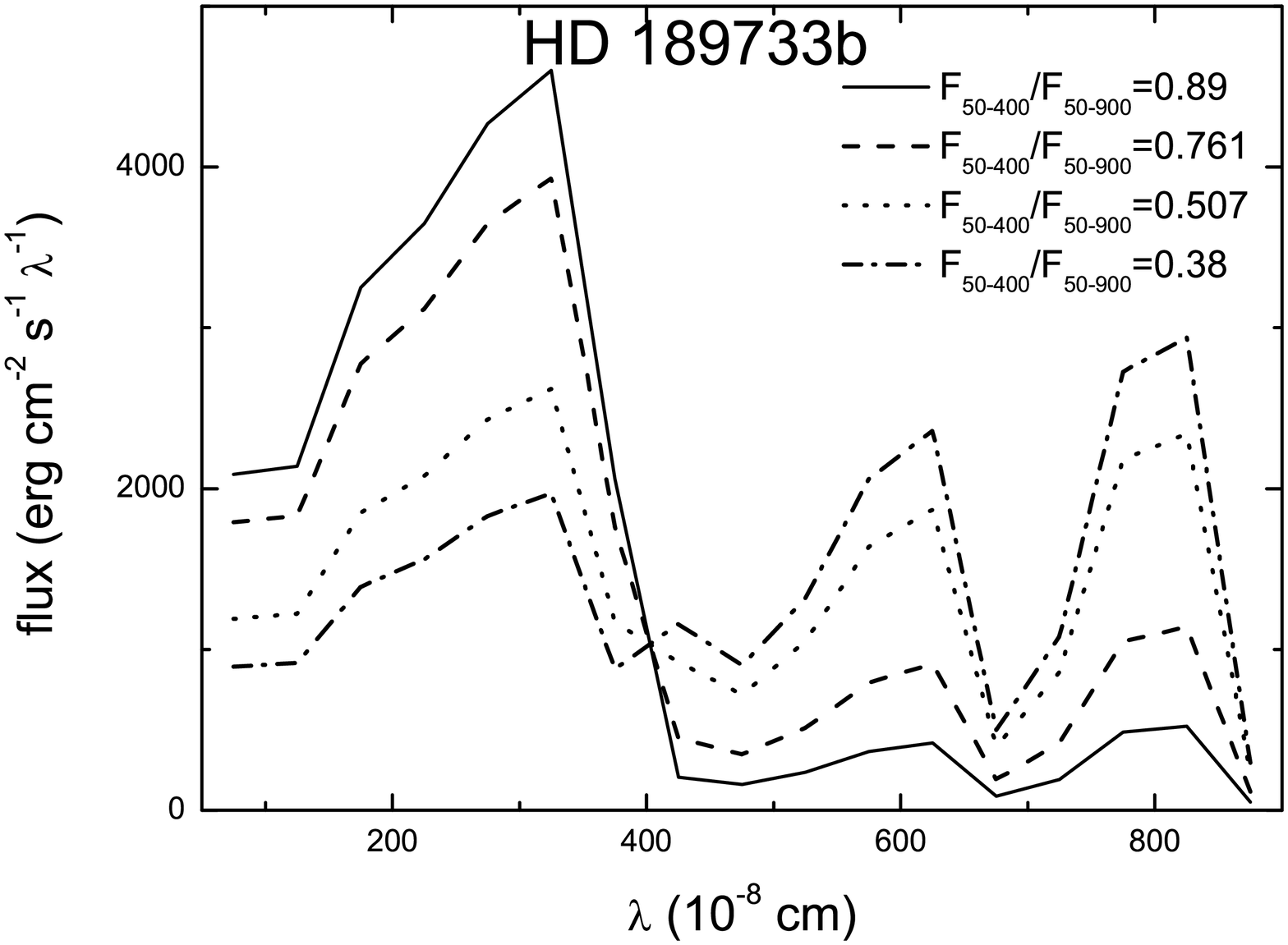}
\end{minipage}
\caption{The spectral energy distributions received by our planets sample. For HD 209458, GJ 436, and
Kepler-11, the corresponding EUV SEDs are constructed from the solar EUV spectrum.
Their spectra are composed with 31 bins. The flux is photon energy integrated over a spectral bin of $\bigtriangleup \lambda$.
The EUV SEDs of HD 189733 is based on a similar K-star $\epsilon$\, Eridani spectrum  \citep{ribas05}, which is composed with 17 bins. }
\end{figure}

To construct the radiative input, we used the solar spectral model of \citet[]{richards94} as our tool to produce the needed EUV spectra, which provides fluxes in the 37 wavelength bins. The model covers the range of 50-1050 $\AA$, which explains our selection of the 50-900$\AA$ band as the radiative input. Using the \citet[]{richards94} model is convenient in constructing the shape of the EUV SED and should be adequate here because the aim of this paper is to discuss theoretically the approximate effect of the EUV SED on the mass loss rate and the composition of exoplanet. In the case of HD 189733, we use a more realistic EUV SED of the K-star for the comparison with the HST/STIS observations. More generally, for more accurate calculations and comparisons to observations, the models of \citet[]{sanz11,claire12,linsky14} could be more appropriate, yet the effort of obtaining a consistent EUV stellar spectrum is rather difficult.

The solar spectra of \citet[]{richards94} are determined by the solar proxy p$_{10.7}$.
By modifying the solar proxy P$_{10.7}$, the solar spectra at different levels of solar activity can be obtained. We define the range of 50-400${\AA}$ as the high-energy spectral region, and the rest, (400-900${\AA}$), as the low-energy part.
The ratio of $\beta$=F$_{50-400}$/F$_{50-900}$ is defined as the spectral index that characterizes the EUV SED of a selected star (F$_{50-400}$ and F$_{50-900}$ are the integrated fluxes in the range of 50-400${\AA}$ and 50-900${\AA}$, respectively). Thus, the value of $\beta$ uniquely reflects the energy distribution of the EUV SED. For reference, we show the solar spectral index $\beta$ variation with time that we derive from the solar irradiance time series compiled in http://lasp.colorado.edu/lizird/ for the 2002-2015 period (Figure 1). The spectral index  $\beta$ for the selected wavelength window varies from 0.74 to 0.86. This means that the EUV SED of the Sun is indeed variable with time and dominated by the high-energy band $< 400\, \AA$. However, no concrete evidence shows that the EUV SEDs of other G-stars are similar to that of the Sun because the accurate measures for other stars are still scanty, particularly at a few stellar radii from the corona.
In order to obtain the EUV SED needed for each of our targets, we set different values of the solar proxy for the high- and low-energy spectral regions, respectively. For the high-energy regions, the value of P$_{10.7}$ is set as 230. It is set as 80 for the low-energy part. Further, the spectra obtained are scaled by a different factor. This means that the two spectral regions are demarcated with different values of P$_{10.7}$--namely, with different levels of solar activity.
Figure 2 shows the spectral energy distributions constructed by that method. Clearly, the spectral shapes are modified by the different values of $\beta$.
In the case of a high spectral index, the high-energy region dominates the spectra.
With the decrease of $\beta$, the low-energy region contributes a significant portion.

In contrast to the solar-like star HD 209458, HD 189733 is a young ($\sim 1.2$\,Gy) K0-K1-type star that shows high variability on short time scales \citep{sanz11,ben13,pillitteri15}. In addition, FUV observations and nearly simultaneous X-rays and FUV observations of the star have shown flaring activity \citep{lecavelier12,pillitteri15}. The EUV spectrum of the star is not available, yet we can get a good insight from a similar star for which a spectrum was estimated. For instance, Epsilon Eridanis is also a young ($\sim 1.1$\,Gy) K2V star.  For HD 189733, XMM-Newton X-ray observations and stellar coronal models have been used to derive luminosities log L$_{X}$ = 28.18 and log L$_{EUV}$ = 28.48 in the X-ray (0.5-10 nm) and EUV (10-92 nm) \citep{sanz11,ben13}. These values are comparable to luminosities (log L$_{X}$ $\sim 28.2$ and log L$_{EUV}$ $\sim  28.4$) reported for the star Epsilon Eridanis. In addition, the effective temperature of HD 189733 ($\sim $ 4875K) is comparable to that of $\epsilon$ Eri ($\sim$ 4900K). It is important to stress here that the $\epsilon$ Eri EUV spectrum is different from the Sun EUV spectrum  \citep{chadney15}.
Therefore, while a scaled solar-type spectrum is a good approximation for solar-like stars, using the EUV spectrum of $\epsilon$ Eri is a better approximation for the study of the EUV spectrum's impact on the atmosphere of HD 189733b proposed here. As shown in Figure 2, the spectrum of a K-type star is different from the solar spectrum, which justifies our decision to use the $\epsilon$ Eri spectrum as a reference for our comparison to HST/STIS data (Section 4).

\subsection{The Atmospheric Model}
To discuss the issue, we complement our previous 1D model \citep[]{guo13} with the photochemistry of hydrogen and helium. The current model includes the neutral species H$_{2}$, He, and H, as well as the ionized species
H$^{+}$, H$^{+}_{2}$ , H$^{+}_{3}$ , He$^{+}$, and HeH$^{+}$. Electrons are also included. The relevant chemical processes are listed in Table 1. We include the photoionization and photoabsorption heating of H$_{2}$, H and He and cooling of H$_{3}^{+}$ (we fit the uppermost curve in Figure 2 of \citet[]{Neale96}.).
\begin{table*}
 \centering
 \begin{minipage}{200mm}
  \caption{Chemical reactions}
  \begin{tabular}{llllllll}
  \hline
Reaction &        & & & rate\footnote{Photolysis rates are in s$^{-1}$. Two body rates are in cm$^{3}$ s$^{-1}$. Three body rates are in cm$^{6}$ s$^{-1}$. }             &      &Reference\\
 \hline
 R1 & H$_{2}$ + h$\nu$ &$\longrightarrow$ &H$_{2}^{+}$ + e&$\sum_{\nu}\frac{F_{\nu}e^{-\tau_{\nu}}}{h\nu}\sigma_{\nu,H_{2}^{+}}$ & & This paper\\
 R2 & H$_{2}$ + h$\nu$ &$\longrightarrow$ &H$^{+}$ + H +e&$\sum_{\nu}\frac{F_{\nu}e^{-\tau_{\nu}}}{h\nu}\sigma_{\nu,H^{+}}$ & & This paper \\
 R3 & H + h$\nu$ &$\longrightarrow$ &H$^{+}$ + e&.. & & \citet[]{ricotti02}  \\
 R4 & He + h$\nu$ &$\longrightarrow$ &He$^{+}$ + e&.. & & \citet[]{ricotti02}  \\
 R5 & H$_{2}$ + M &$\longrightarrow$ &H + H + M&1.5$\times$10$^{-9}$e$^{-48000/T}$ & &\citet[]{bau92}\\
 R6 & H + H + M&$\longrightarrow$ &H$_{2}$ + M&8.0 $\times$10$^{-33}$(300/T )$^{0.6}$ & &\citet[]{ham70} \\
 R7 & H$_{2}^{+}$ + H$_{2}$ &$\longrightarrow$ &H$_{3}^{+}$ + H&2.0$\times$10$^{-9}$ & & \citet[]{thread74} \\
 R8 & H$_{3}^{+}$ + H&$\longrightarrow$ &H$_{2}^{+}$ + H$_{2}$&2.0$\times$10$^{-9}$  && Estimated by \citet[]{yelle04}\\
 R9 & H$_{2}^{+}$ + H &$\longrightarrow$ &H$_{2}$ + H$^{+}$&6.4$\times$10$^{-10}$ & & \citet[]{kap79}  \\
 R10 & H$^{+}$ + H$_{2}$ &$\longrightarrow$ &H$_{2}^{+}$ + H&1.0$\times$10$^{-9}$e$^{-21900/T}$  & & Estimated by \citet[]{yelle04} \\
 R11 & He$^{+}$ + H$_{2}$ &$\longrightarrow$ &He + H$^{+}$ + H&4.2$\times$10$^{-13}$ & & \citet[]{schz89}\\
 R12 & He$^{+}$ + H$_{2}$ &$\longrightarrow$ &H$^{+}$ + H + He&8.8$\times$10$^{-14}$ & & \citet[]{schz89}\\
 R13 & HeH$^{+}$+ H$_{2}$ &$\longrightarrow$ &H$_{3}^{+}$ + He&1.5$\times$10$^{-9}$ & & \citet[]{boh80}\\
 R14 & HeH$^{+}$+ H &$\longrightarrow$ &H$_{2}^{+}$ + He&9.1$\times$10$^{-10}$ & &\citet[]{kap79} \\
 R15 & H$^{+}$ + e &$\longrightarrow$ &H + h$\nu$&4.0$\times$10$^{-12}$(300/T)$^{0.64}$ & & \citet[]{sto95} \\
 R16 & He$^{+}$ + e &$\longrightarrow$ &He + h$\nu$&4.6$\times$10$^{-12}$(300/T)$^{0.64}$ & &\citet[]{sto95} \\
 R17 & H$_{2}^{+}$ + e &$\longrightarrow$ &H + H&2.3$\times$10$^{-8}$(300/T)$^{0.4}$ & & \citet[]{aue77} \\
 R18 & H$_{3}^{+}$ + e &$\longrightarrow$ &H$_{2}$ + H&2.9$\times$10$^{-8}$(300/T)$^{0.65}$  & & \citet[]{sun94} \\
 R19 & H$_{3}^{+}$ + e &$\longrightarrow$ &H + H + H&8.6$\times$10$^{-8}$(300/T)$^{0.65}$ & & \citet[]{datz95}\\
 R20 & HeH$^{+}$ + e &$\longrightarrow$ &He + H&1.0$\times$10$^{-8}$(300/T)$^{0.6}$ & & \citet[]{yousif89} \\
\hline
\end{tabular}
\end{minipage}
\end{table*}

The total photoabsorption heating of the atmosphere can be written as

\begin{equation}
Heat=\sum_{s}n_{s}\sum_{\lambda}\eta_{\lambda,s} F_{\lambda} e^{-\tau_{\lambda}}\sigma_{\lambda,s},
\end{equation}
where $\lambda$ is the mid-wavelength of every wavelength bin \citep[]{richards94}. The
$n_{s}$ is the number density of H$_{2}$, H and He. $F_{\lambda}$ is the photon energy integrated over a spectral bin of $\bigtriangleup \lambda$,
$\tau_{\lambda}$ is the optical depth at mid-wavelength $\lambda$. $\sigma_{\lambda,s}$ is the photoabsorption cross-section of the sth  species \citep[]{sch00}.

In our previous models \citep[]{guo11,guo13}, we defined the net heating efficiency (which for species s is defined as the fraction of the absorbed stellar energy that heats the atmosphere) as $\eta_{\nu,s}=(h\nu I_{ion,s})/h\nu$ ($I_{ion,s }$ is the ionization potential of the sth species), which is the maximum heating efficiency. In fact, photoionization/photodissociation of H, He, and H$_{2}$ can produce energetic photoelectrons that share their energies with other particles via ionization, excitation, and dissociation, and they can heat gas before their energies are thermalized. Thus, the net heating efficiency should be smaller than the above definition.
Generally, most energy's fraction of a photoelectron with energy \textit{E} can be deposited to heat the gas when the electron abundance (mixing ratio) in the gas is higher. On the contrary, it is favorable for the second ionization when the electron abundance fraction in the gas is lower (Figure 13 of \citet []{ricotti02}).
Thus, the secondary ionization of atoms produced by energetic photoelectrons can be important in the bottom of an atmosphere where the gas is almost neutral. A full calculation for the process is beyond the scope of this paper.
Here we used the analytic fittings of \citet[]{ricotti02} to calculate the net heating efficiency and secondary ionization of H and He. For the net heating efficiency of H$_{2}$, the fitting results of \citet[]{dalgarno99} are adopted. The net heating efficiency in our previous models only depends on the wavelength. In this paper, however, the net heating efficiency is a function both of wavelength and of altitude because the net heating efficiency obtained by \citet[]{ricotti02} and \citet[]{dalgarno99} is related to the electron abundance fraction in the gas and the wavelength of incident photons. The second ionization that affects the ionization degree of hydrogen will be discussed in Section 4.2.  The model is suitable to calculate the escape of molecules, atoms, and ions of hydrogen and helium. In this paper, it is used to calculate the escape of hydrogen-dominated atmospheres.

\section{Sensitivity Study of Atmospheric Composition of EUV SED: Results}

We obtain the parameters of HD 189733b, HD 209458b, and GJ 436b from http://exoplanet.eu, and those of Kepler-11b from \citet[]{liss13}. All parameters are listed in Table 2.
We use the quiet solar integrated flux as the integrated flux of HD 209458b (F$_{EUV}$=1086 erg cm $^{-2}$ s$^{-1}$). For GJ 436b and Kepler-11b, the integrated fluxes are set as 1760 erg cm $^{-2}$ s$^{-1}$ \citep[]{kulow14} and 278 erg cm $^{-2}$ s$^{-1}$ \citep[]{lammer13}. According to \citet[]{lecavelier10}, the EUV integrated flux of HD 189733b should be to 10-40 times the solar value. \citet[]{lecavelier12} estimates the total luminosity of the XUV band as 7.1$\times$10$^{28}$ erg s$^{-1}$. Here, we use 24778 erg cm $^{-2}$ s$^{-1}$ as the integrated EUV flux of HD 189733b.  For HD 189733b, HD 209458b, GJ 436b, and Kepler-11b, the temperatures at the lower boundary are set as 1200k \citep[]{knu07}, 1200K \citep[]{gar07}, 717k \citep[]{dem07}, and 900k \citep[]{lammer13}.

\begin{table*}
 \centering
 \begin{minipage}{140mm}
  \caption{The Properties of HD 209458b, GJ 436b, Kepler-11b, and HD 189733b.}
  \begin{tabular}{cccccccc}
  \hline
   Planet & Mass       &  Radius   & Separation & Integrated Flux       & T$_{0}\footnote{The temperature at the lower boundary.}$ & $\lambda$\footnote{Hydrodynamic escape parameter $\lambda=\frac{GM_{p}\mu}{R_{p}\kappa T_{0}}$ at the lower boundary. }\\
          & (M$_{J}$)  & (R$_{J}$) & (AU)       & (erg cm$^{-2}$s$^{-1}$) &  (K)    \\
 \hline
 HD 209458b & 0.69 & 1.38 & 0.047 & 1086 & 1200&179 \\
 GJ 436b & 0.07 & 0.38 & 0.02887 & 1760 & 717&115   \\
 Kepler-11b & 0.0135 & 0.1759 & 0.091 & 278 & 900&36  \\
 HD 189733b & 1.138 & 1.138 & 0.03 & 24778 & 1200&358 \\
\hline
\end{tabular}
\end{minipage}
\end{table*}

\subsection{Chemical Composition}
\subsubsection{HD\, 209458b}
\begin{figure}
\begin{minipage}[t]{0.5\linewidth}
\centering
\includegraphics[width=3.6in,height=2.6in]{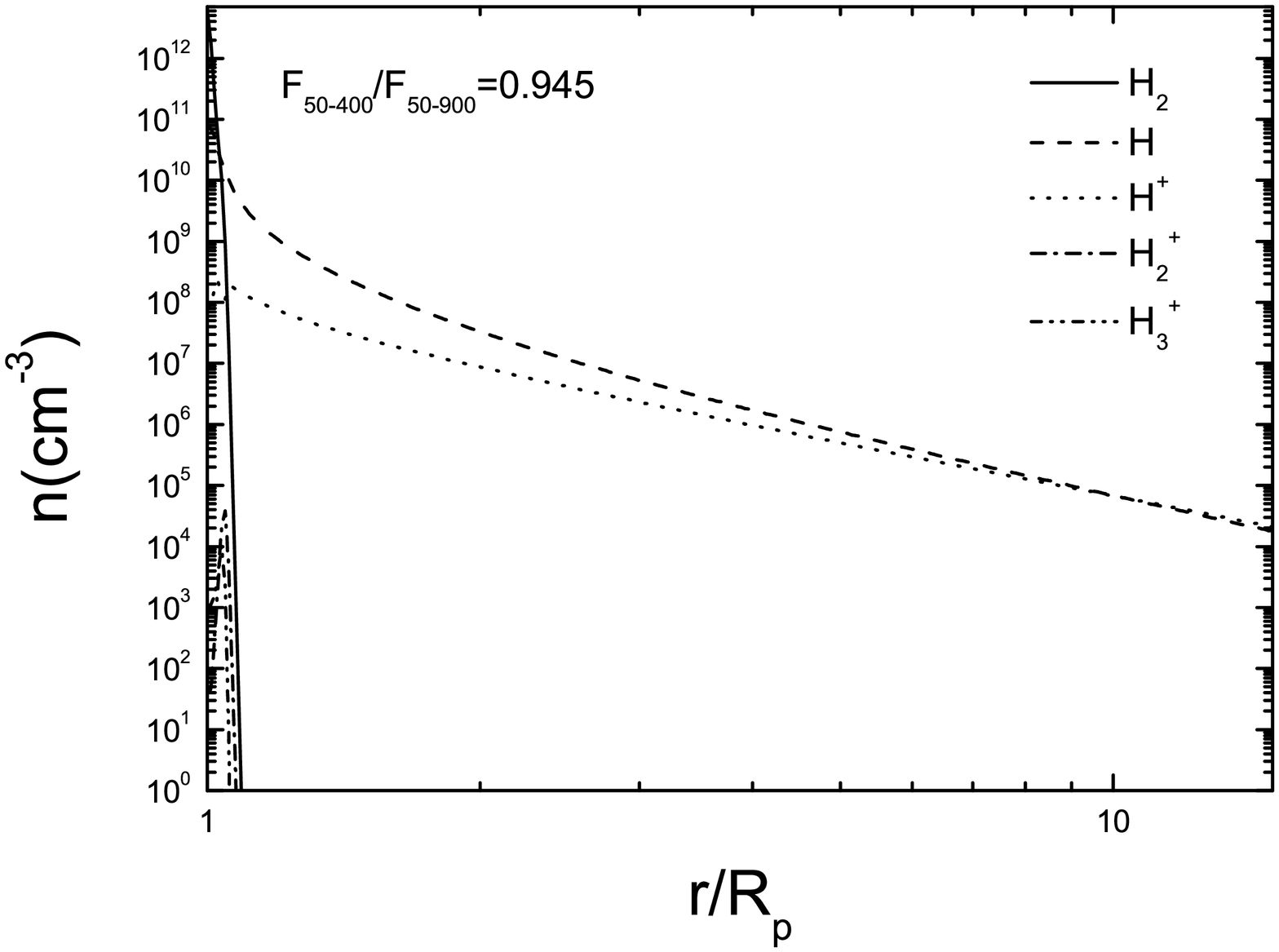}
\end{minipage}
\begin{minipage}[t]{0.5\linewidth}
\centering
\includegraphics[width=3.6in,height=2.6in]{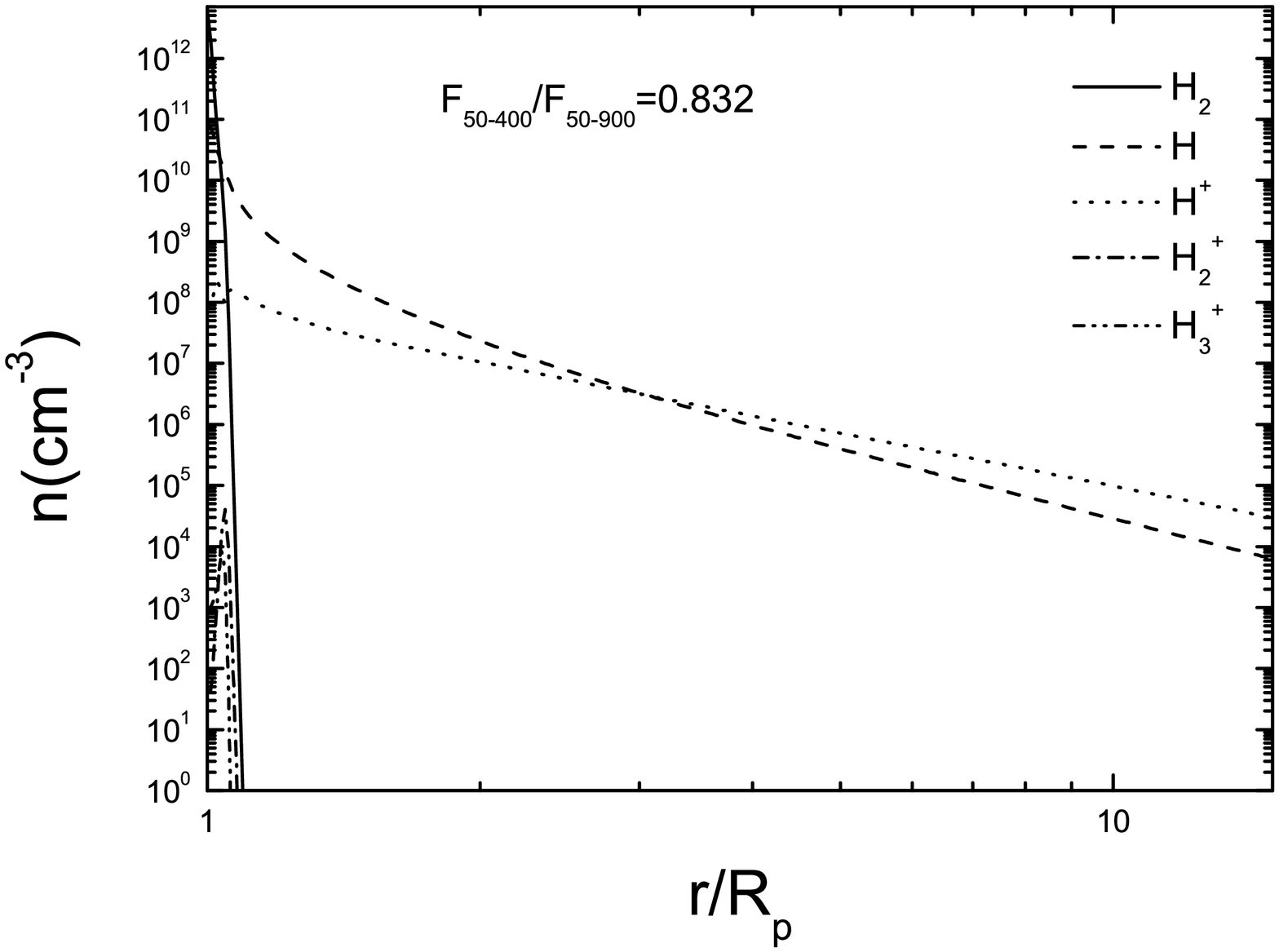}
\end{minipage}
\begin{minipage}[t]{0.5\linewidth}
\centering
\includegraphics[width=3.6in,height=2.6in]{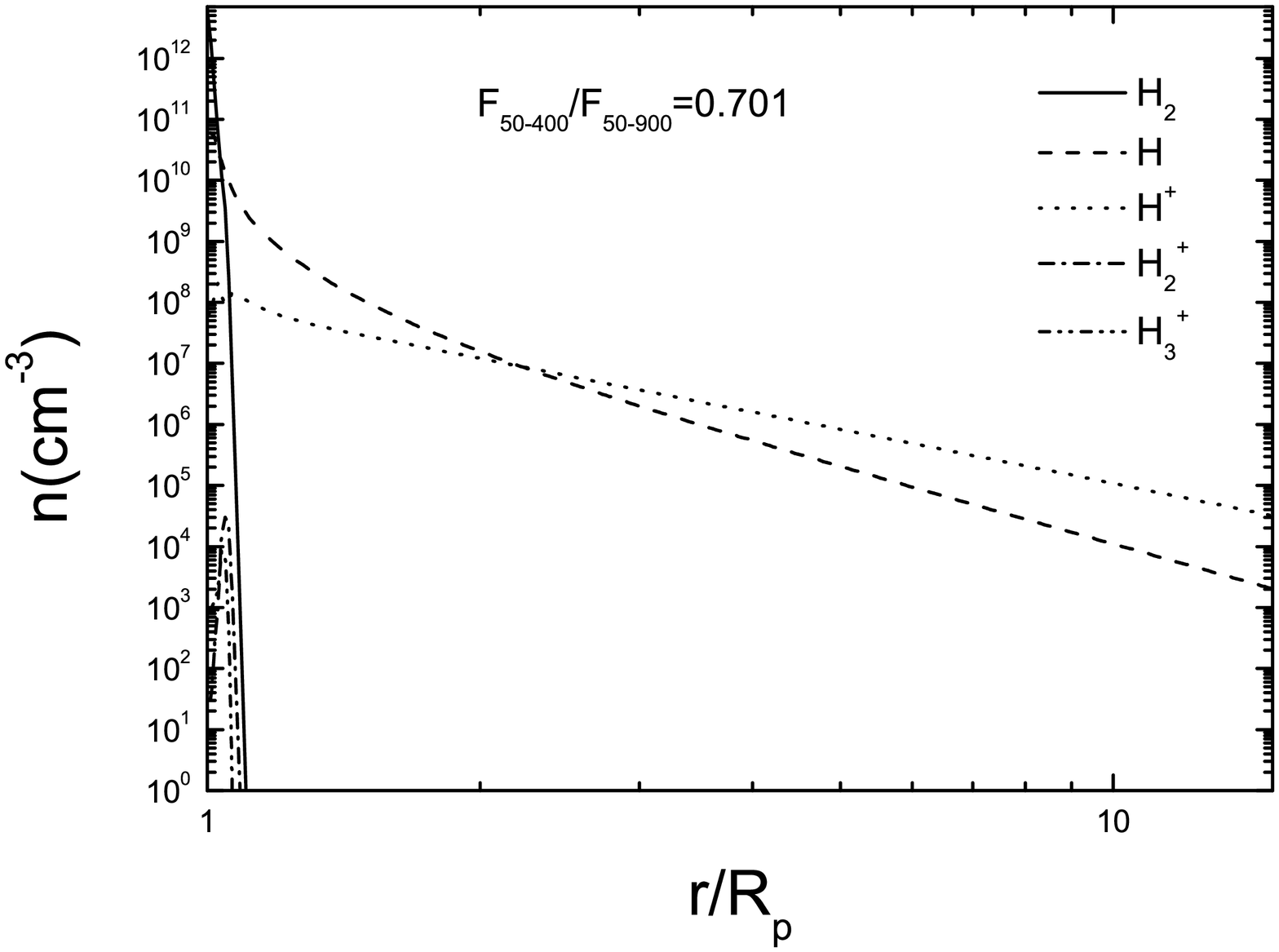}
\end{minipage}
\begin{minipage}[t]{0.5\linewidth}
\centering
\includegraphics[width=3.6in,height=2.6in]{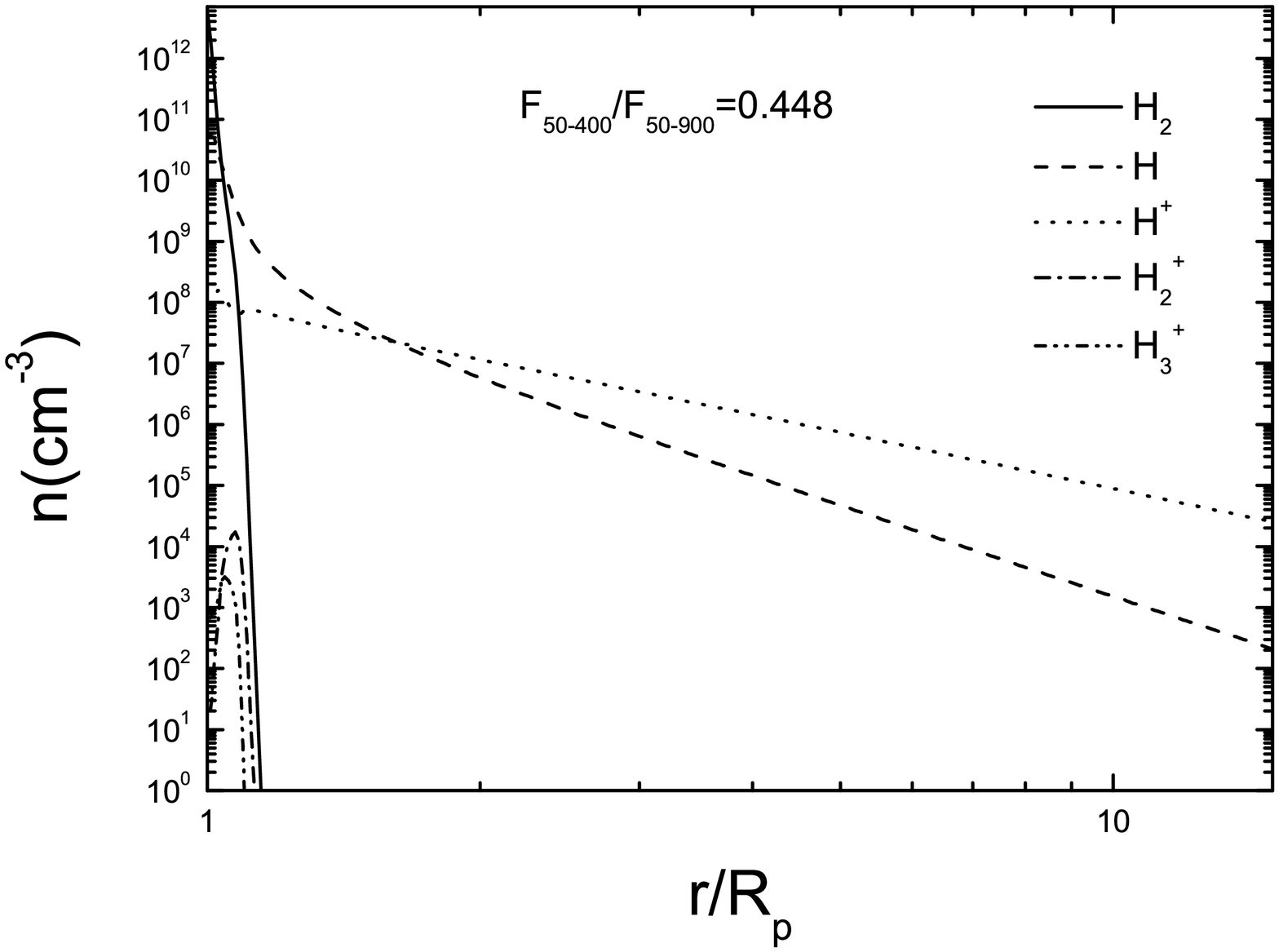}
\end{minipage}
\caption{The composition of HD 209458b with different spectral index. }
\end{figure}

\begin{figure}
\begin{minipage}[t]{0.5\linewidth}
\centering
\includegraphics[width=3.6in,height=2.6in]{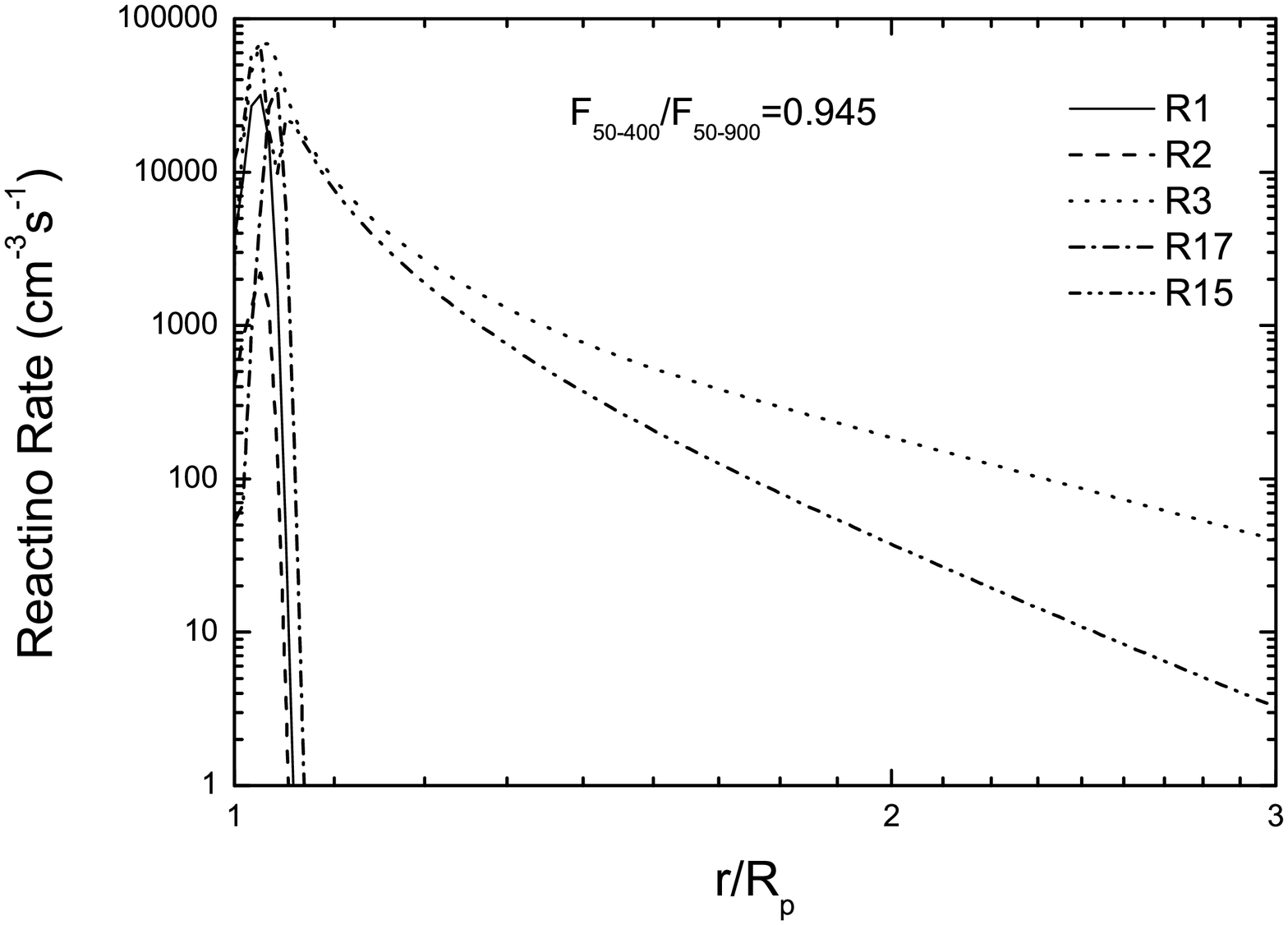}
\end{minipage}
\begin{minipage}[t]{0.5\linewidth}
\centering
\includegraphics[width=3.6in,height=2.6in]{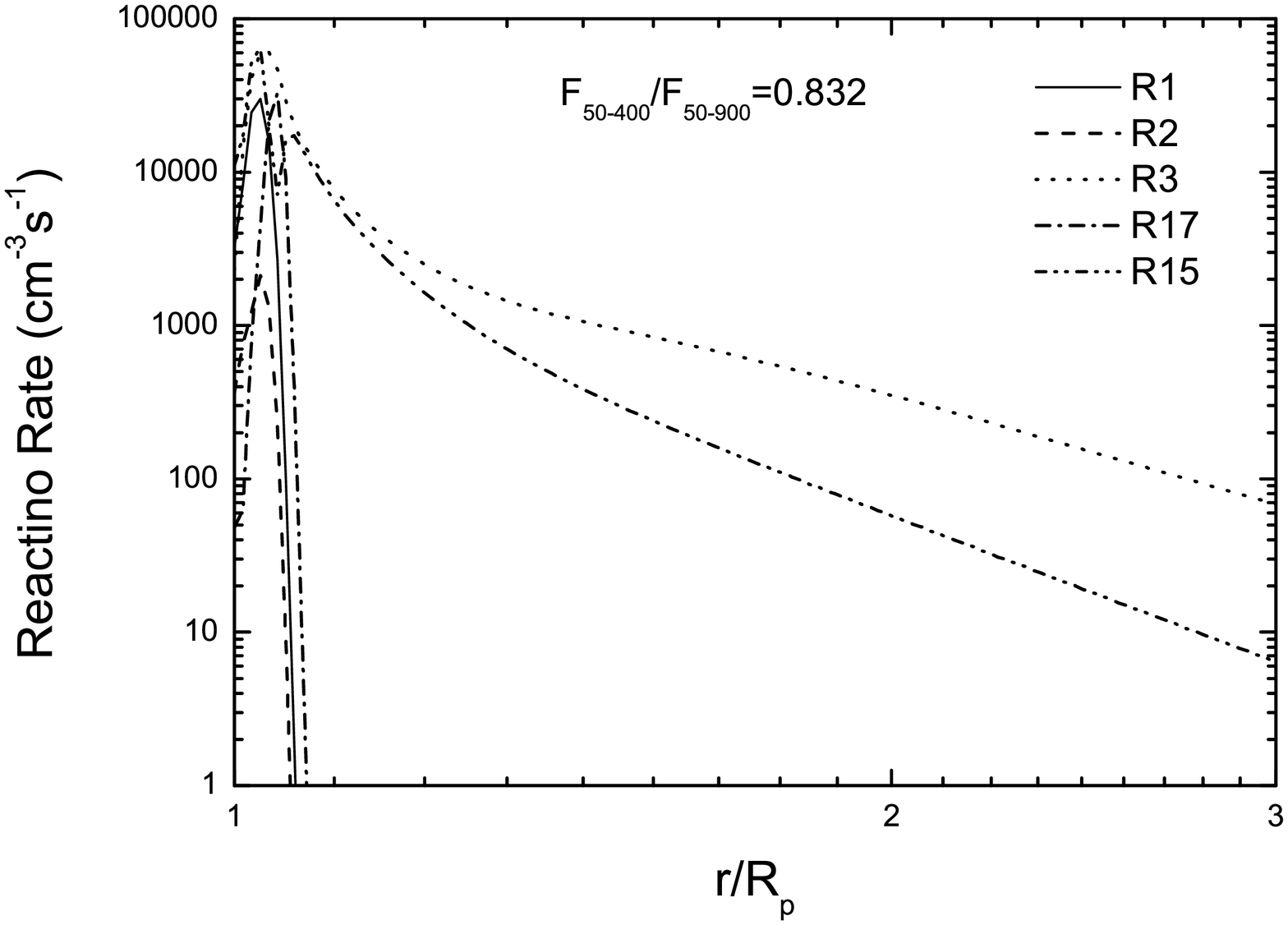}
\end{minipage}
\begin{minipage}[t]{0.5\linewidth}
\centering
\includegraphics[width=3.6in,height=2.6in]{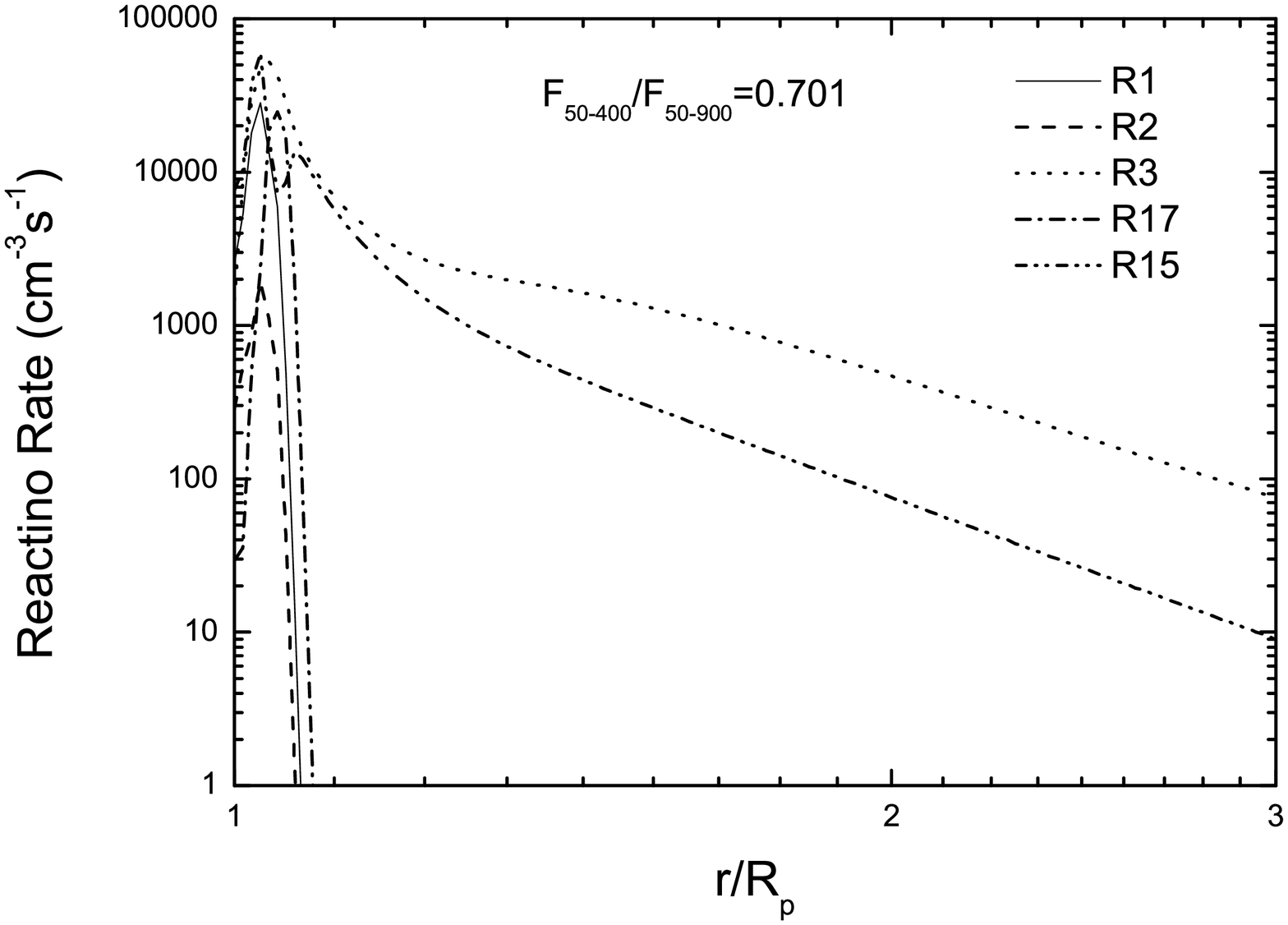}
\end{minipage}
\begin{minipage}[t]{0.5\linewidth}
\centering
\includegraphics[width=3.6in,height=2.6in]{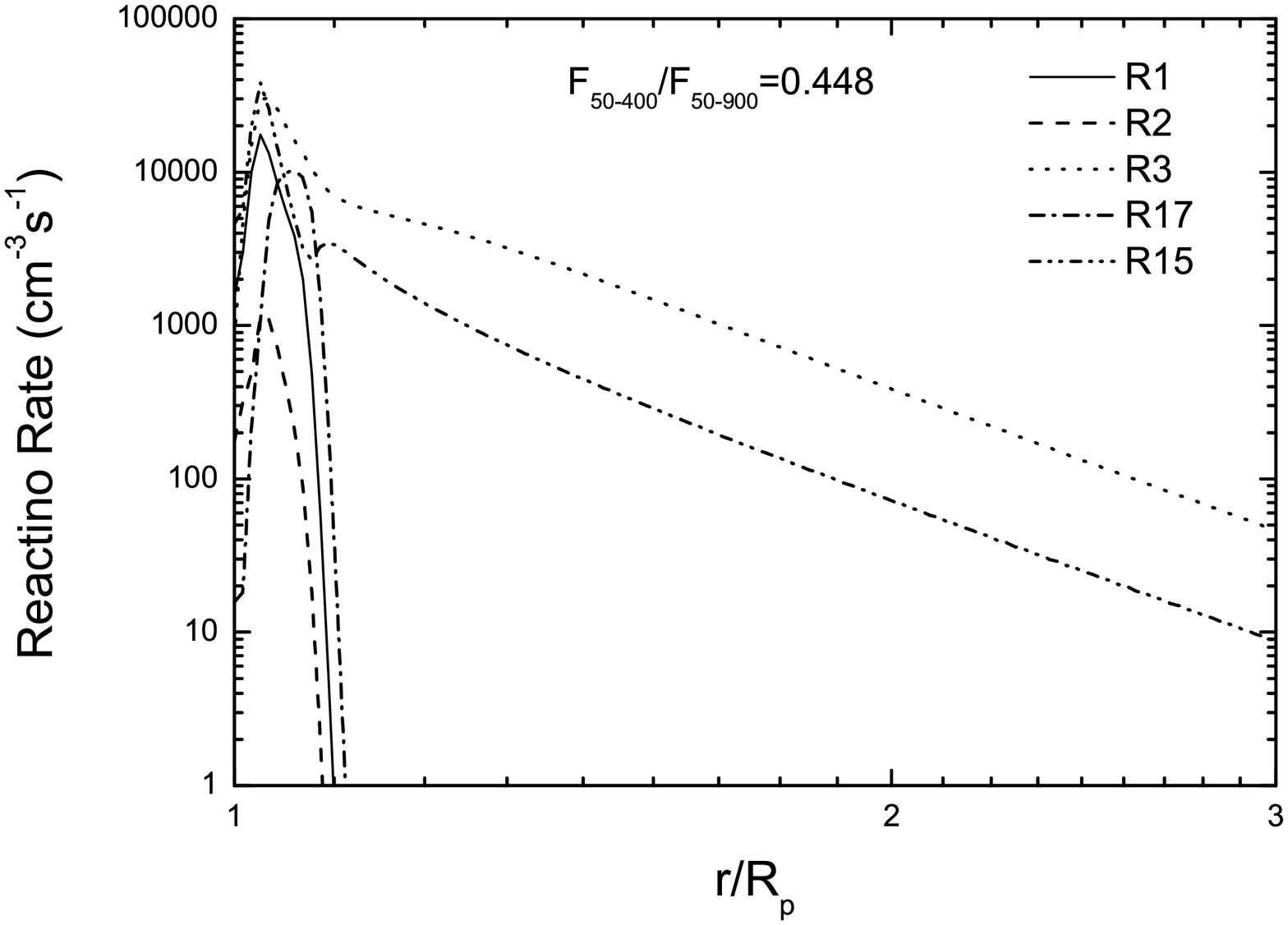}
\end{minipage}
\caption{The primary chemical reaction rates in the atmosphere of HD 209458b. Note that
not all chemical reactions are shown.}
\end{figure}

Figure 3 shows the chemical composition of HD 209458b. The dominant species at the bottom of the atmosphere is H$_{2}$ that is converted to H at the location of 1.1-1.2 R$_{p}$ due to strong EUV irradiation (see Figure 4; also see Yelle 2004). In the regions of r $>$ 1.2 R$_{p}$, H and H$^{+}$ are the dominant species. The number densities of H$_{2}$, H$_{2}^{+}$, and H$_{3}^{+}$ are lower than 10$^{5}$cm$^{-3}$, and the corresponding chemical reaction rates involving those species are also low (some chemical reactions are not shown in the paper.).
The H/H$^{+}$ transition (where the number density of H$^{+}$ exceeds the number density of H) occurs at the altitude near 10R$_{p}$ in the case of $\beta$=0.945. With decreasing values of $\beta$, the H/H$^{+}$ transition moves to lower altitudes. For the case of $\beta$=0.448, the H/H$^{+}$ transition appears at 1.6R$_{p}$. Obviously, the location of H/H$^{+}$ is very sensitive to the profiles of the EUV SED.

We show the primary chemical reactions of HD 209458b in Figure 4. In the bottom of the atmosphere, the photodissociation of H$_{2}$ is the important chemical reaction that decomposes H$_{2}$ into H$_{2}^{+}$ (R1) , H, and H$^{+}$ (R2). However, the photoionization of H and recombination of H$^{+}$ dominate the chemical processes in the regions of r $>$ 1.2R$_{p}$, and the reaction rate of photoionization of H is almost 10 times the recombination of H$^{+}$ in the regions of 2 $<$ r/R$_{p}<$3. This fact clearly shows that the atmospheric photochemistry of HD 209458b is controlled by the processes of the photoionization of the EUV photons.

\subsubsection{GJ\, 436b}
\begin{figure}
\begin{minipage}[t]{0.5\linewidth}
\centering
\includegraphics[width=3.6in,height=2.6in]{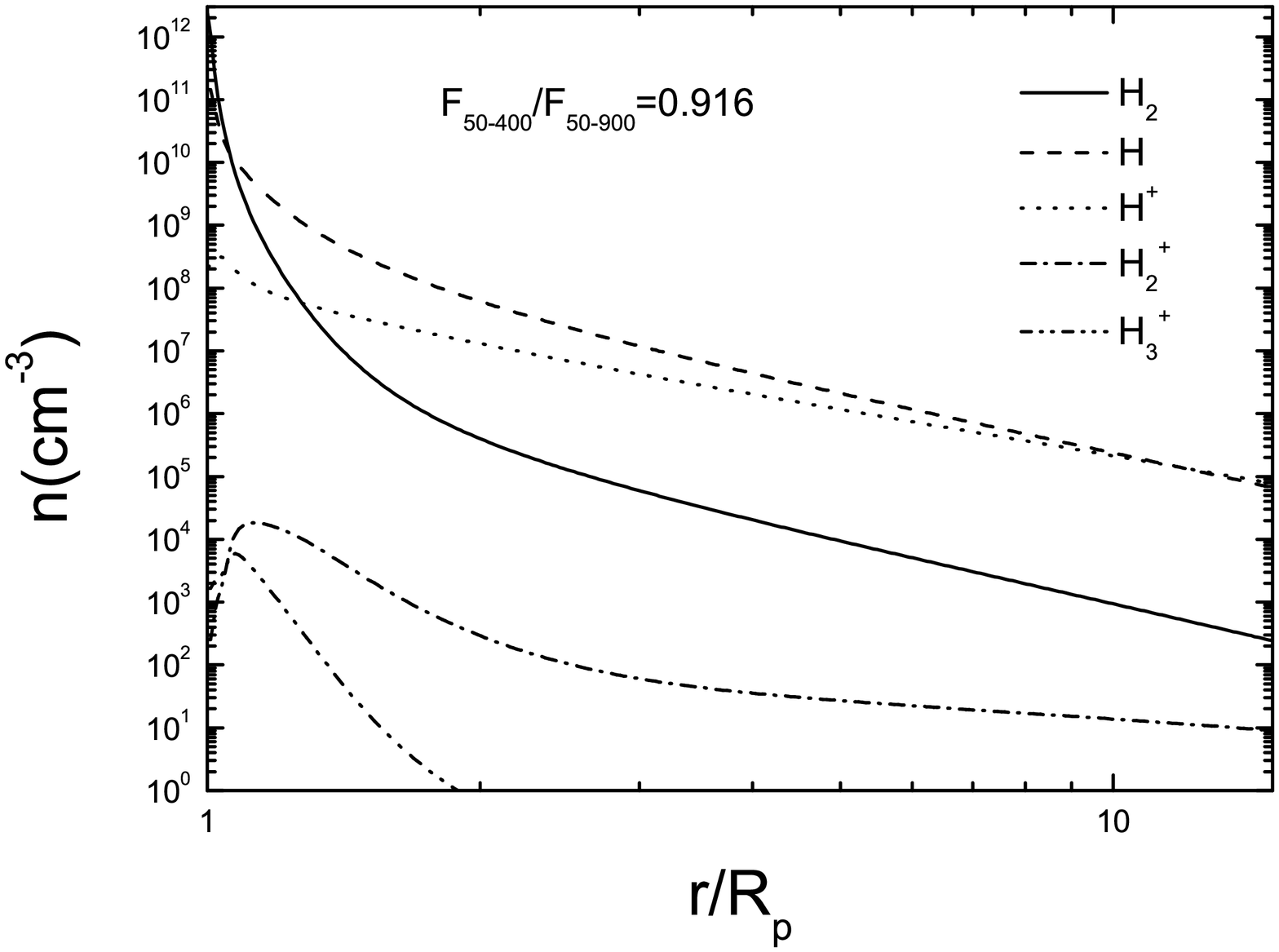}
\end{minipage}
\begin{minipage}[t]{0.5\linewidth}
\centering
\includegraphics[width=3.6in,height=2.6in]{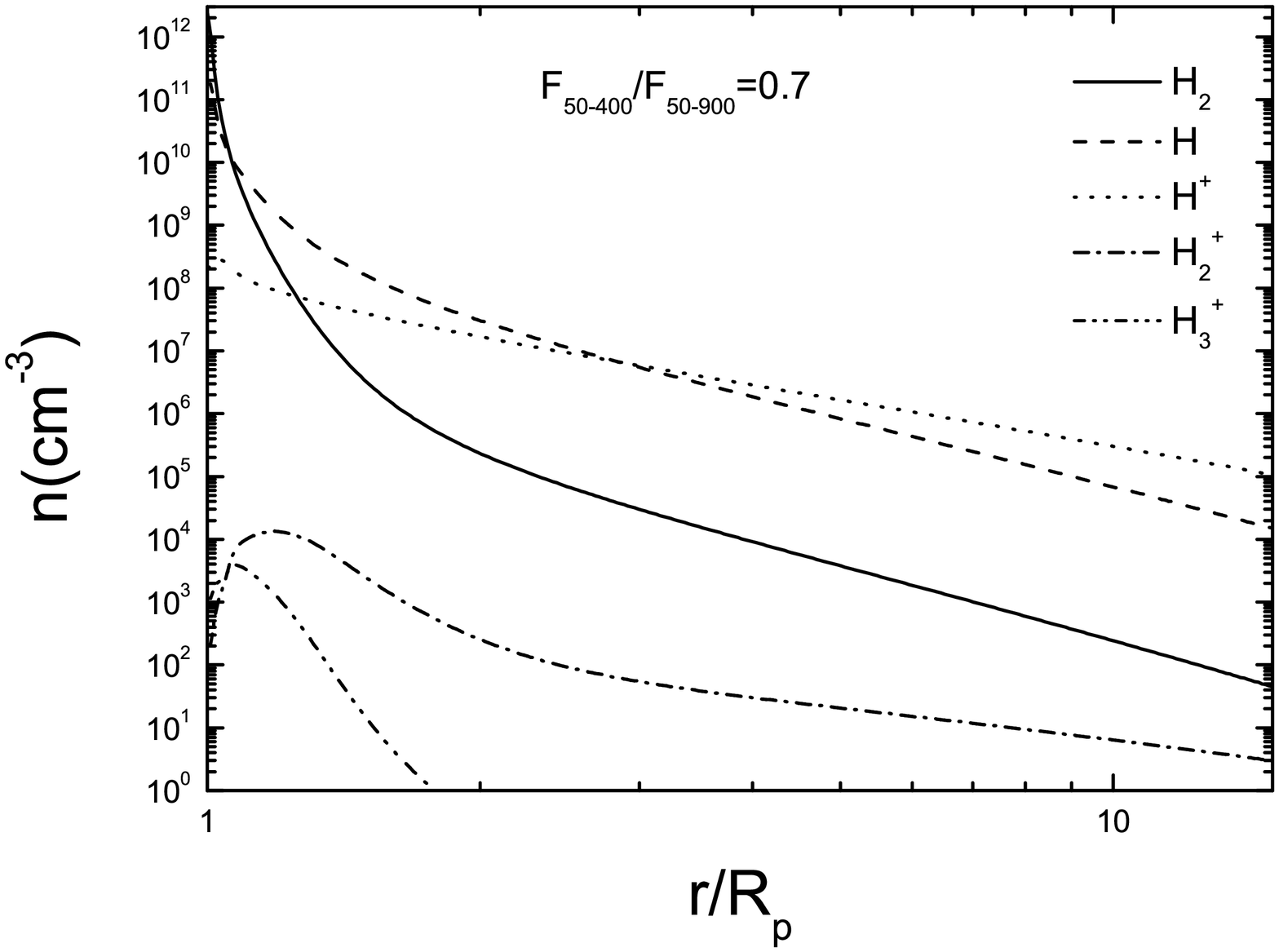}
\end{minipage}
\begin{minipage}[t]{0.5\linewidth}
\centering
\includegraphics[width=3.6in,height=2.6in]{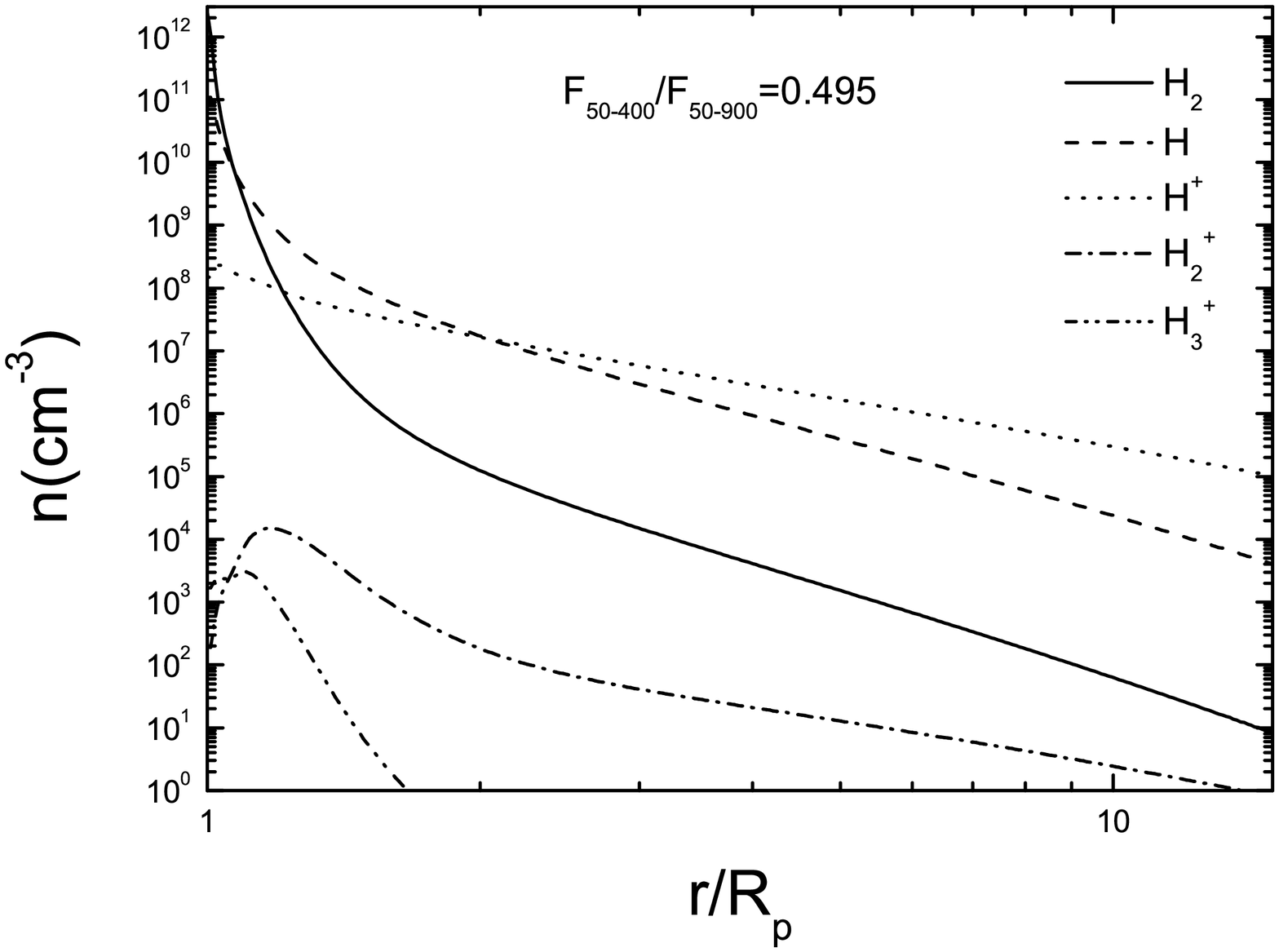}
\end{minipage}
\begin{minipage}[t]{0.5\linewidth}
\centering
\includegraphics[width=3.6in,height=2.6in]{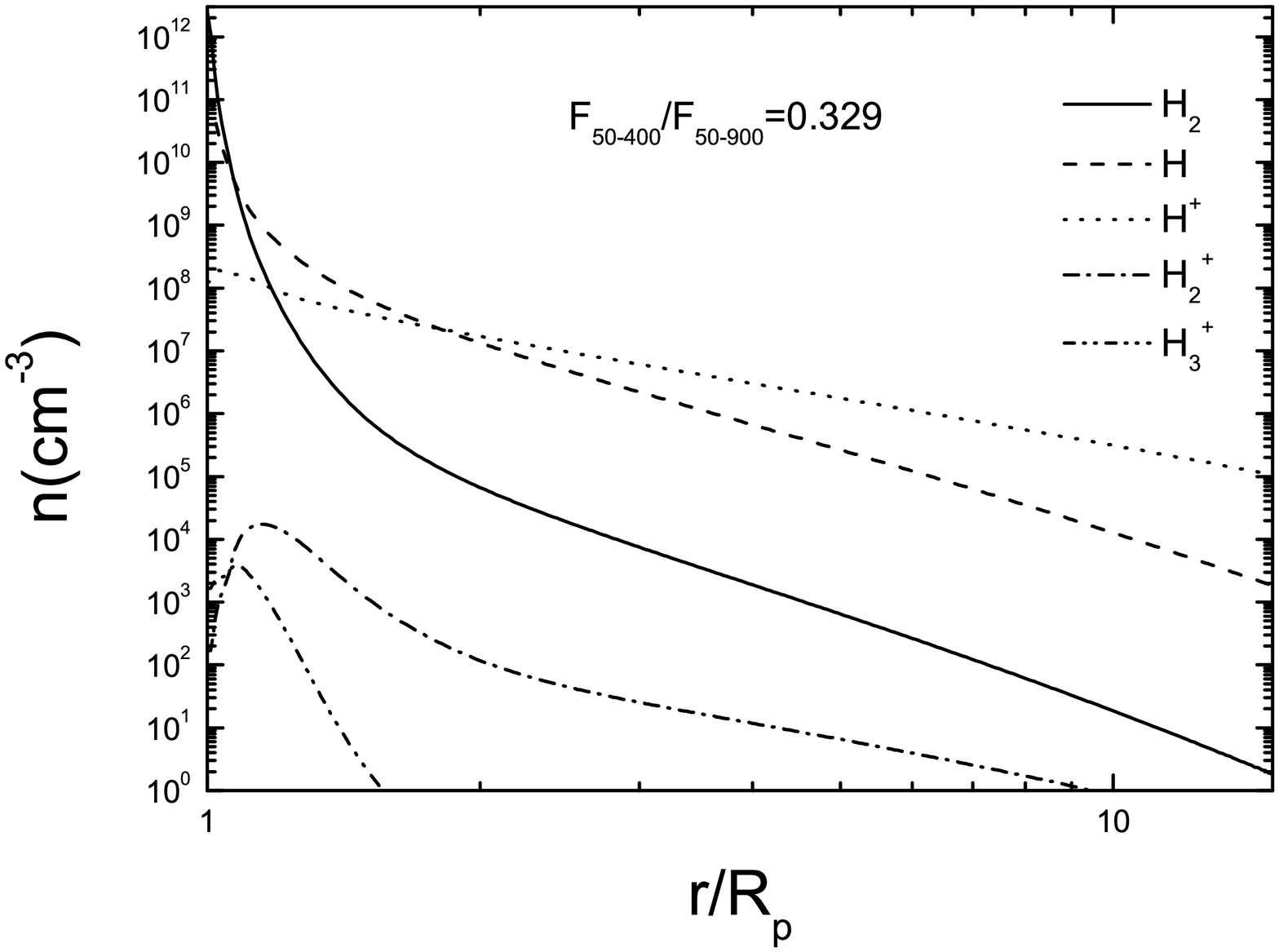}
\end{minipage}
\caption{The composition of GJ\, 436b with different spectral index. }
\end{figure}

\begin{figure}
\begin{minipage}[t]{0.5\linewidth}
\centering
\includegraphics[width=3.6in,height=2.6in]{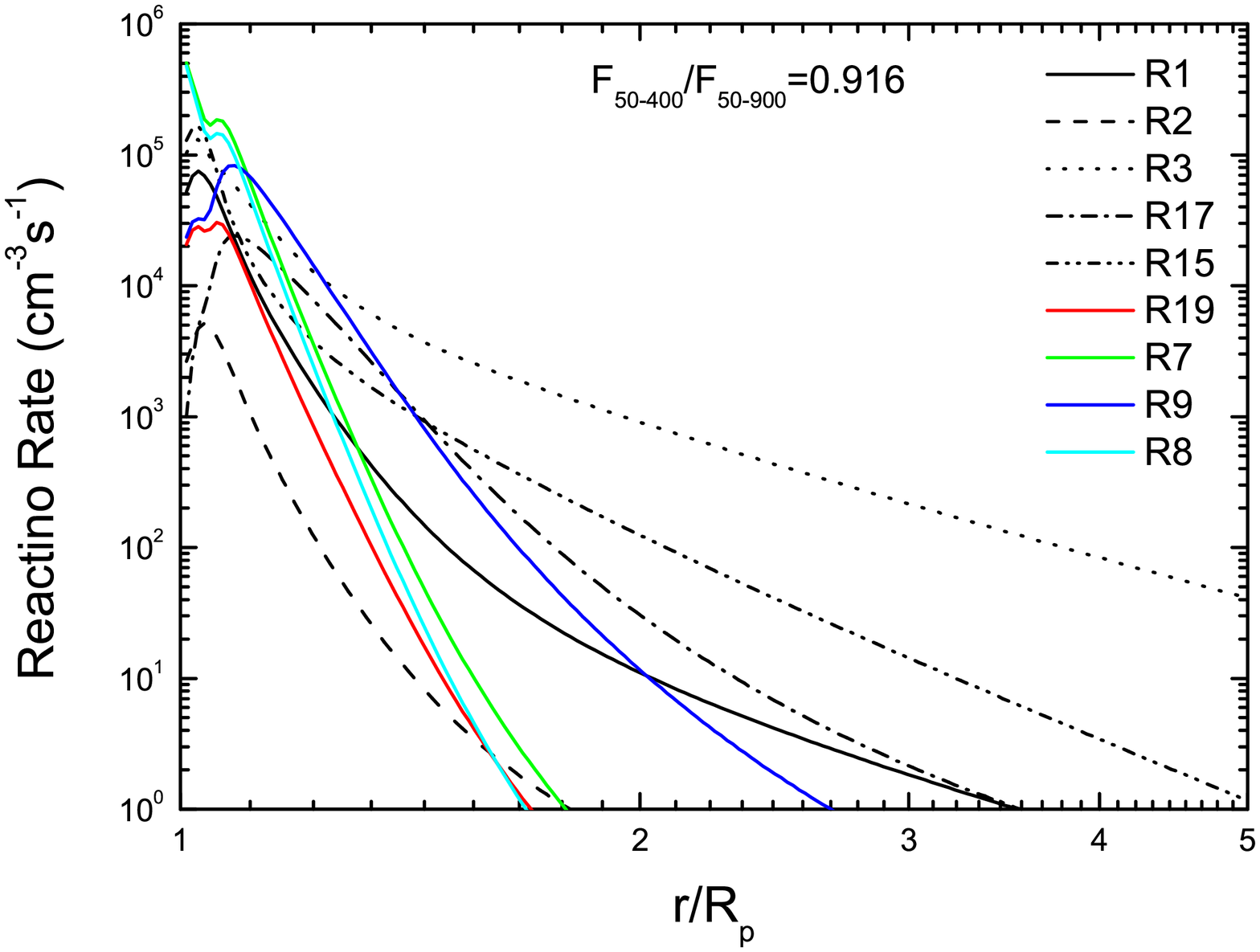}
\end{minipage}
\begin{minipage}[t]{0.5\linewidth}
\centering
\includegraphics[width=3.6in,height=2.6in]{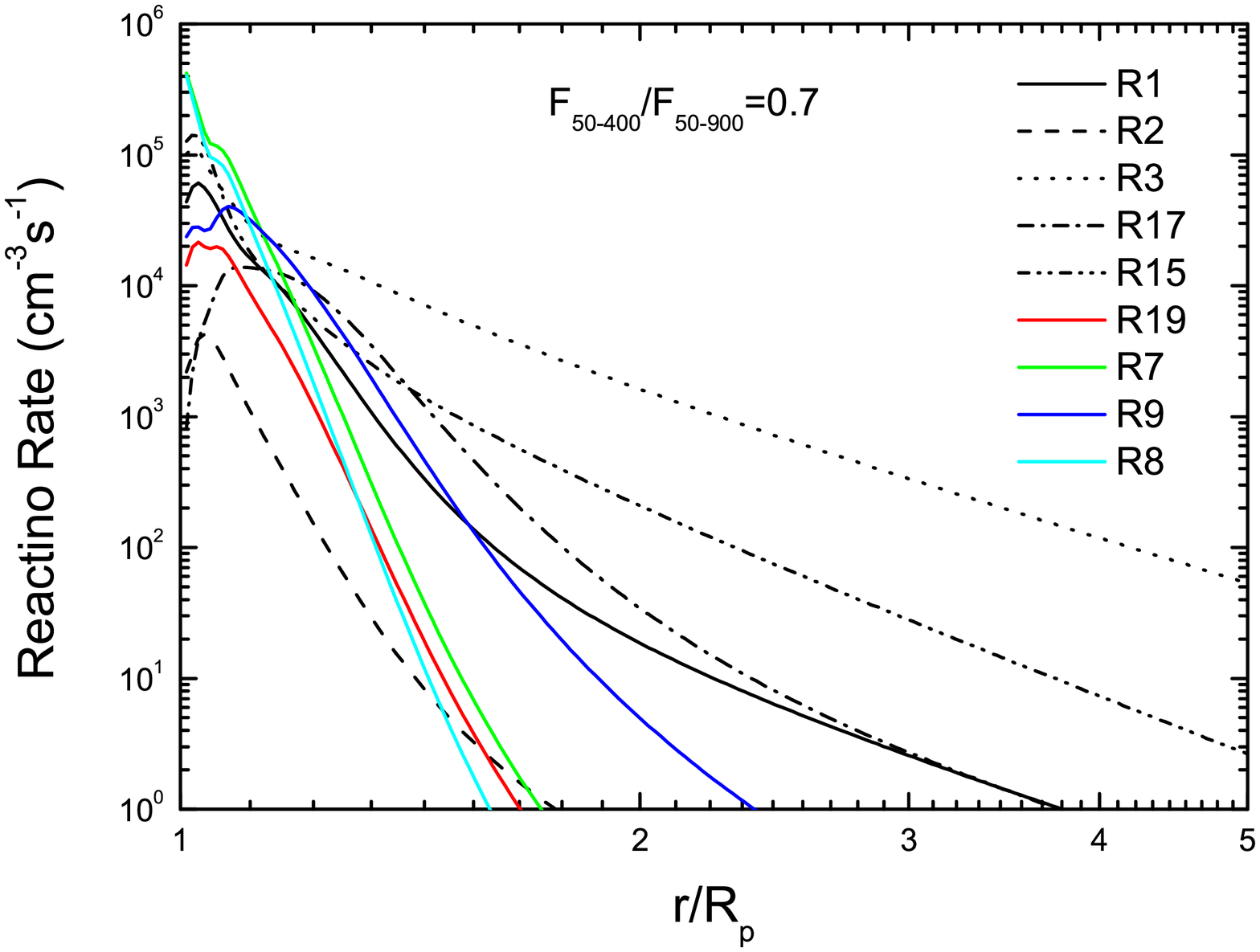}
\end{minipage}
\begin{minipage}[t]{0.5\linewidth}
\centering
\includegraphics[width=3.6in,height=2.6in]{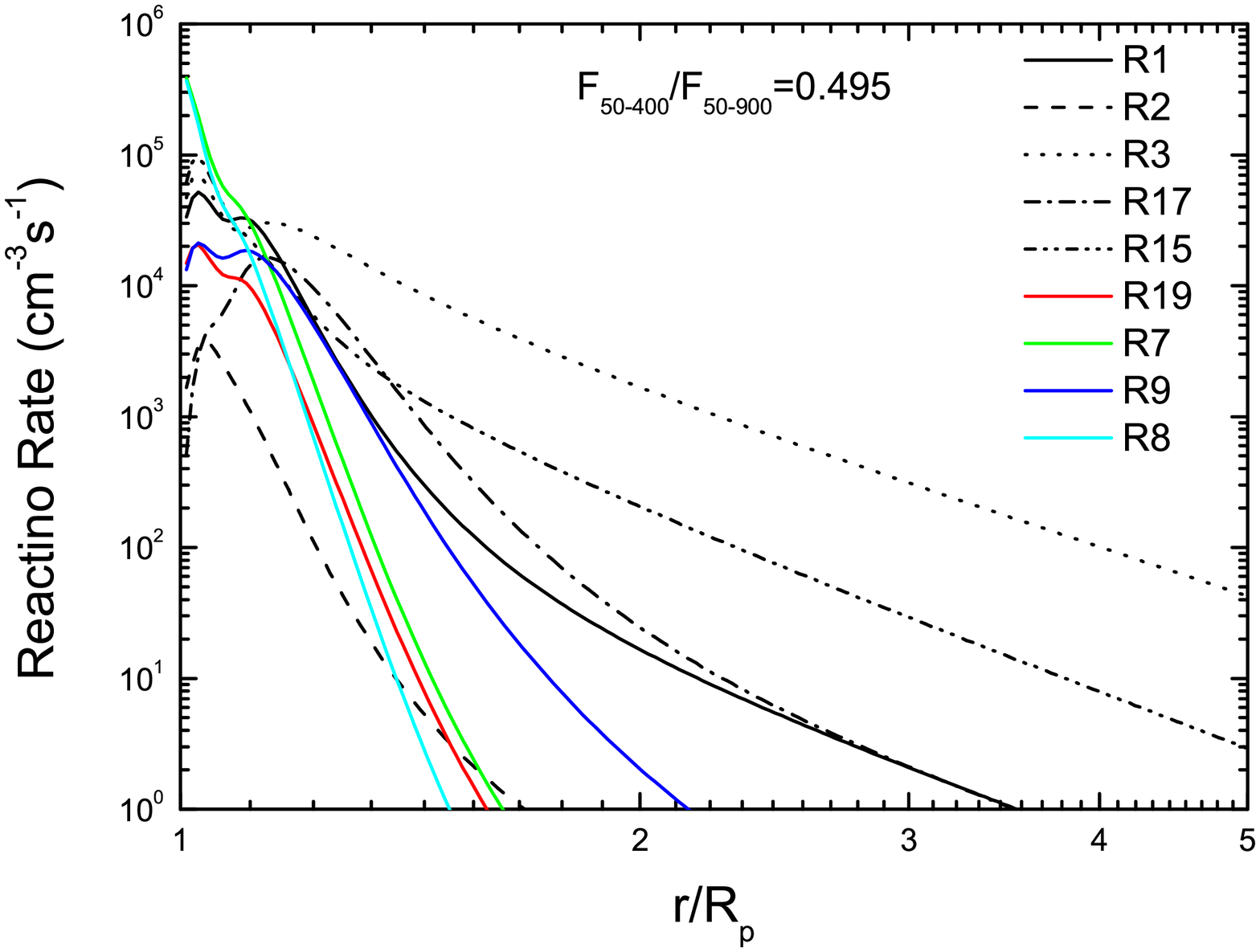}
\end{minipage}
\begin{minipage}[t]{0.5\linewidth}
\centering
\includegraphics[width=3.6in,height=2.6in]{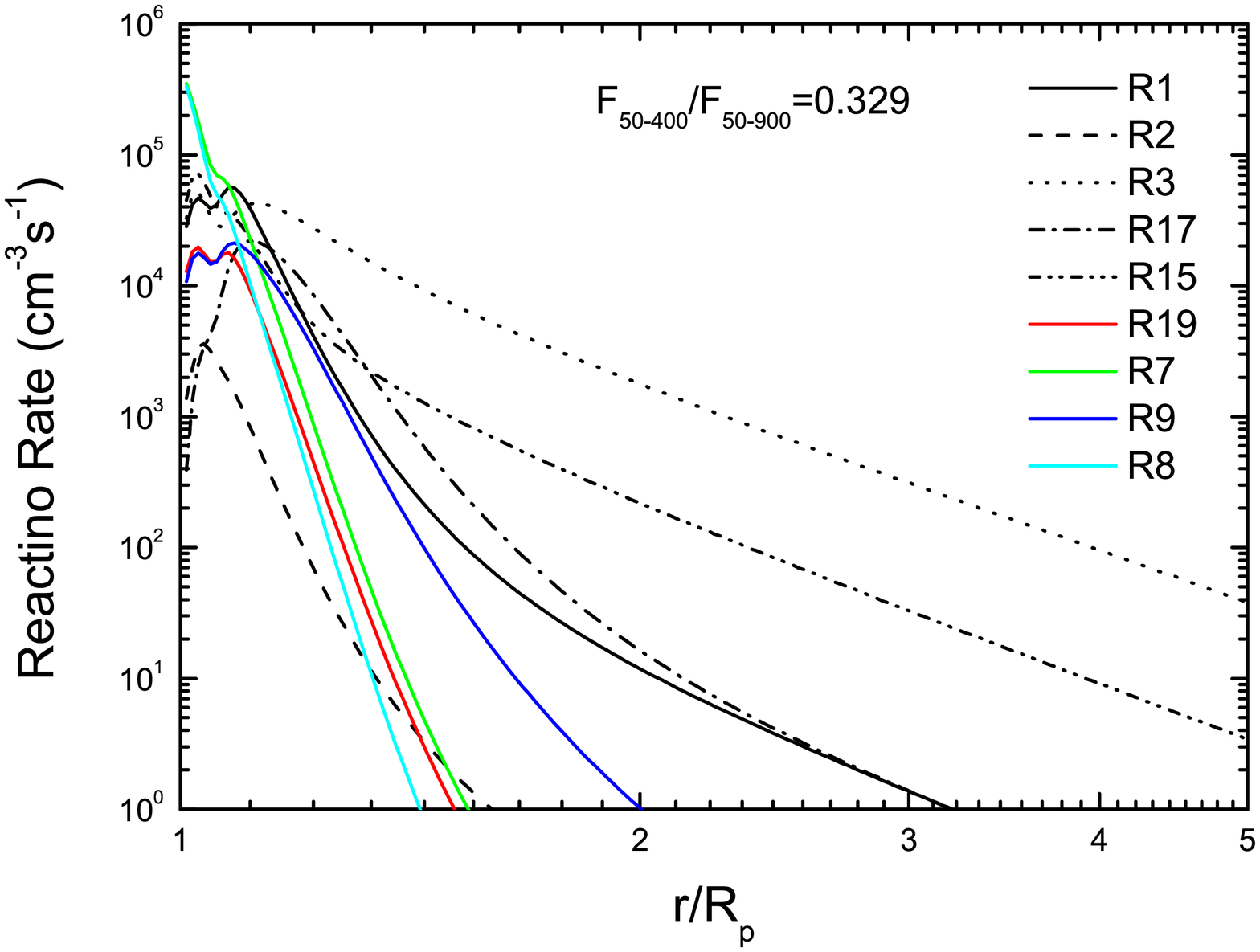}
\end{minipage}
\caption{The primary chemical reaction rates in the atmosphere of GJ\, 436b. Note that not all chemical reactions are shown.}
\end{figure}

For GJ\, 436b, the chemical composition of the atmosphere is somewhat different from that of HD 209458b. As shown in Figure 5, the atmosphere is composed mainly of H and H$^{+}$. The transitions of H/H$^{+}$ move to a lower altitude with the decrease of $\beta$. However, the number density of H$_{2}$ is not negligible. For the case of $\beta$=0.916, the number density of H$_{2}$ at 2R$_{p}$ is about 10$^{6}$ cm$^{-3}$. In contrast, the number density of H$_{2}$ at 2R$_{p}$ decreases to 10$^{5}$ cm$^{-3}$ when the value of $\beta$ decreases to 0.389. We also note that the number densities of H$_{2}^{+}$ and H$_{3}^{+}$ attain 10$^{3}$-10$^{4}$ cm$^{-3}$ near 1.2R$_{p}$ and decrease with the increase of altitude.
Beyond 2R$_{p}$, the number density of H$_{3}^{+}$ is very low. However, the number density of H$_{2}^{+}$ in the upper atmosphere is significant (about 10$^{1}$ -10$^{2}$ cm$^{-3}$). The vertical profiles of H$_{2}^{+}$ and H$_{3}^{+}$ are also affected by the EUV SED. It is clear from Figure 5 that the number densities of H$_{2}^{+}$ and H$_{3}^{+}$ decline faster in the case of $\beta$=0.329 than that of $\beta$=0.916.

The primary chemical reactions of GJ\, 436b are shown in Figure 6. As in the case of HD 209458b, the photodissociation of H$_{2}$ and the photoionization of H (R3) decompose H$_{2}$ and H into H$_{2}^{+}$, H$^{+}$ in the lower altitude (r$<$1.3R$_{p}$). At the same time, their inverse reactions (R15 and R17) recombine H$_{2}^{+}$, H$^{+}$ and electrons into H$_{2}$ and H. Those chemical reaction rates are comparable in the bottom of the atmosphere.
In addition, the chemical reactions involving H$_{2}^{+}$ and H$_{3}^{+}$ (for example, R7 and R8) also play important roles in the photochemical processes. With increasing altitudes, the photoionization rate of H (R3) becomes higher than other chemical reaction rates and dominates the photochemical processes above 1.3R$_{p}$ for the value of  $\beta$ is 0.916. In the case of $\beta$=0.329, the photoionization of H becomes even more dominant at r=1.1R$_{p}$.
The findings shown in Figures 5 and 6 affirm that the composition of the upper atmosphere of GJ 436b is sensitive to the EUV SED.

\subsubsection{Kepler-11b}
\begin{figure}
\begin{minipage}[t]{0.5\linewidth}
\centering
\includegraphics[width=3.6in,height=2.6in]{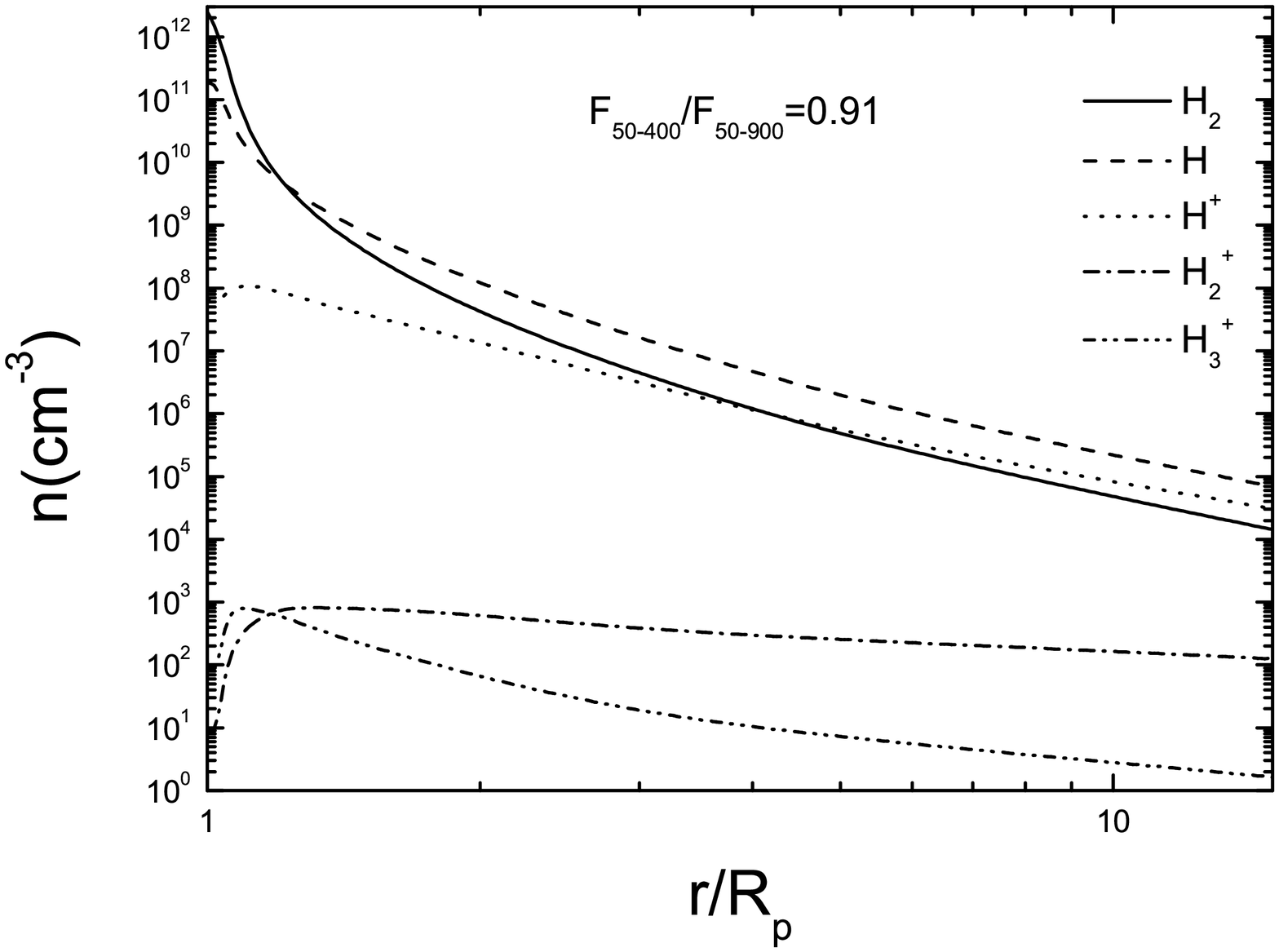}
\end{minipage}
\begin{minipage}[t]{0.5\linewidth}
\centering
\includegraphics[width=3.6in,height=2.6in]{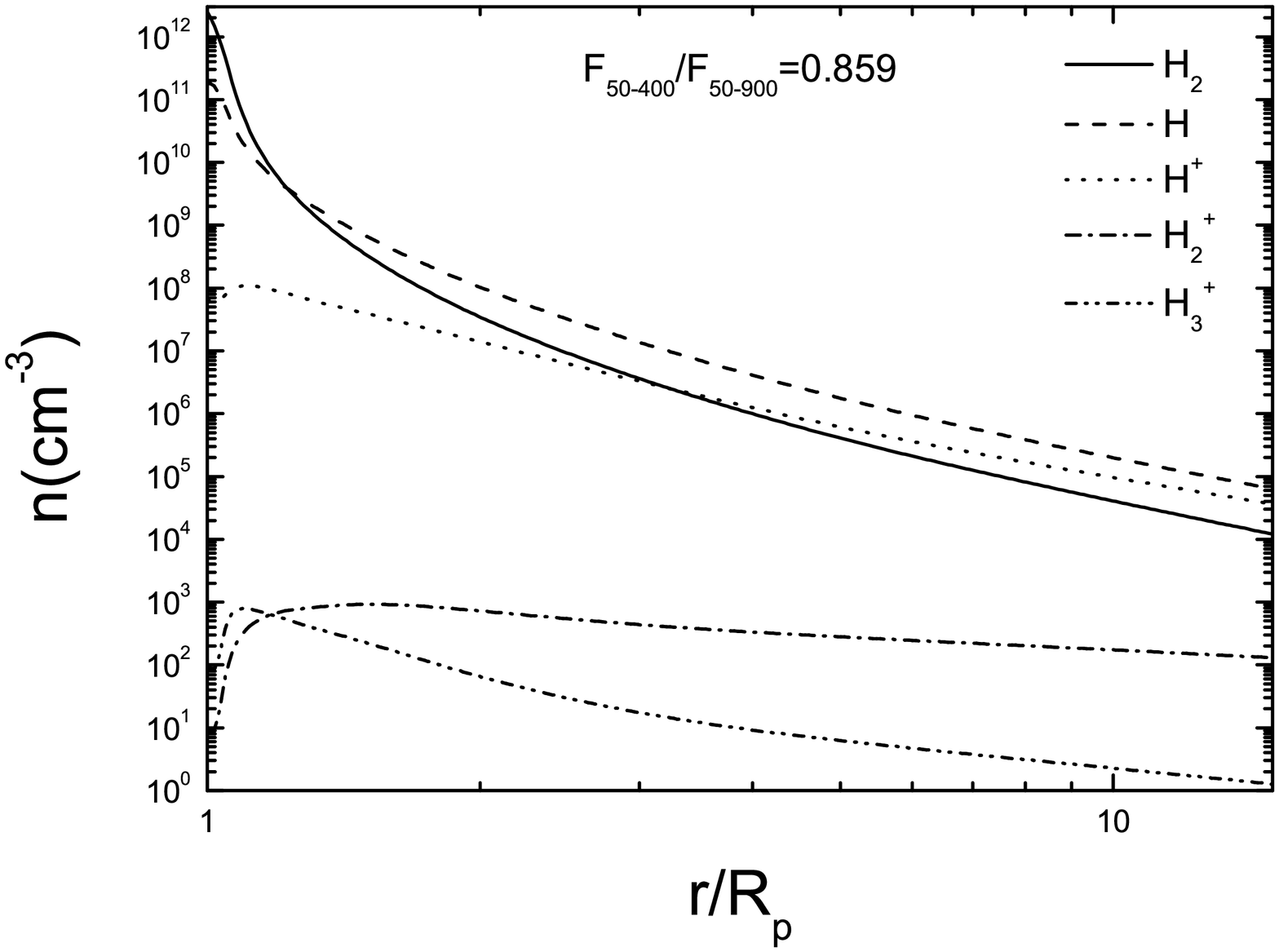}
\end{minipage}
\begin{minipage}[t]{0.5\linewidth}
\centering
\includegraphics[width=3.6in,height=2.6in]{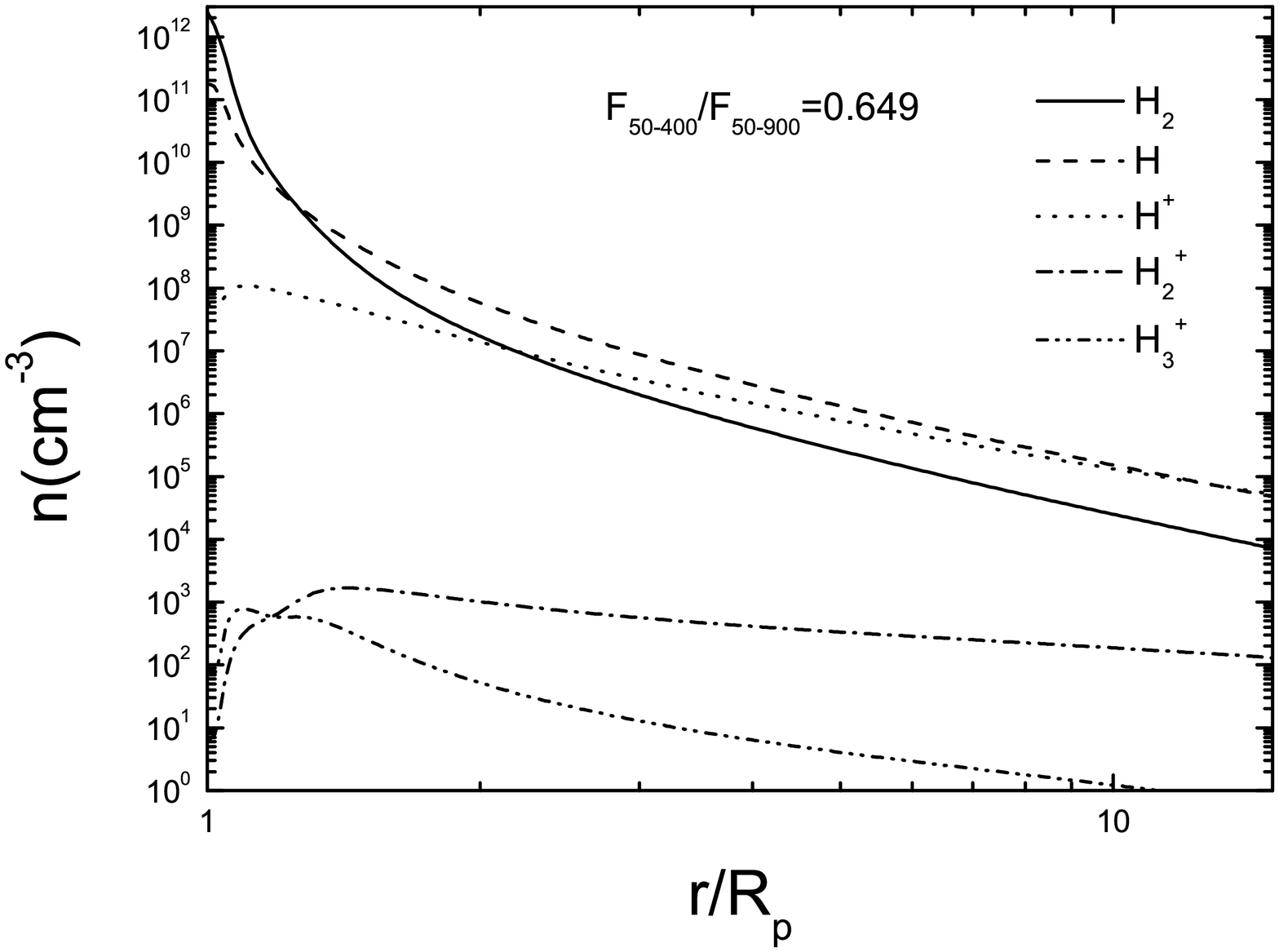}
\end{minipage}
\begin{minipage}[t]{0.5\linewidth}
\centering
\includegraphics[width=3.6in,height=2.6in]{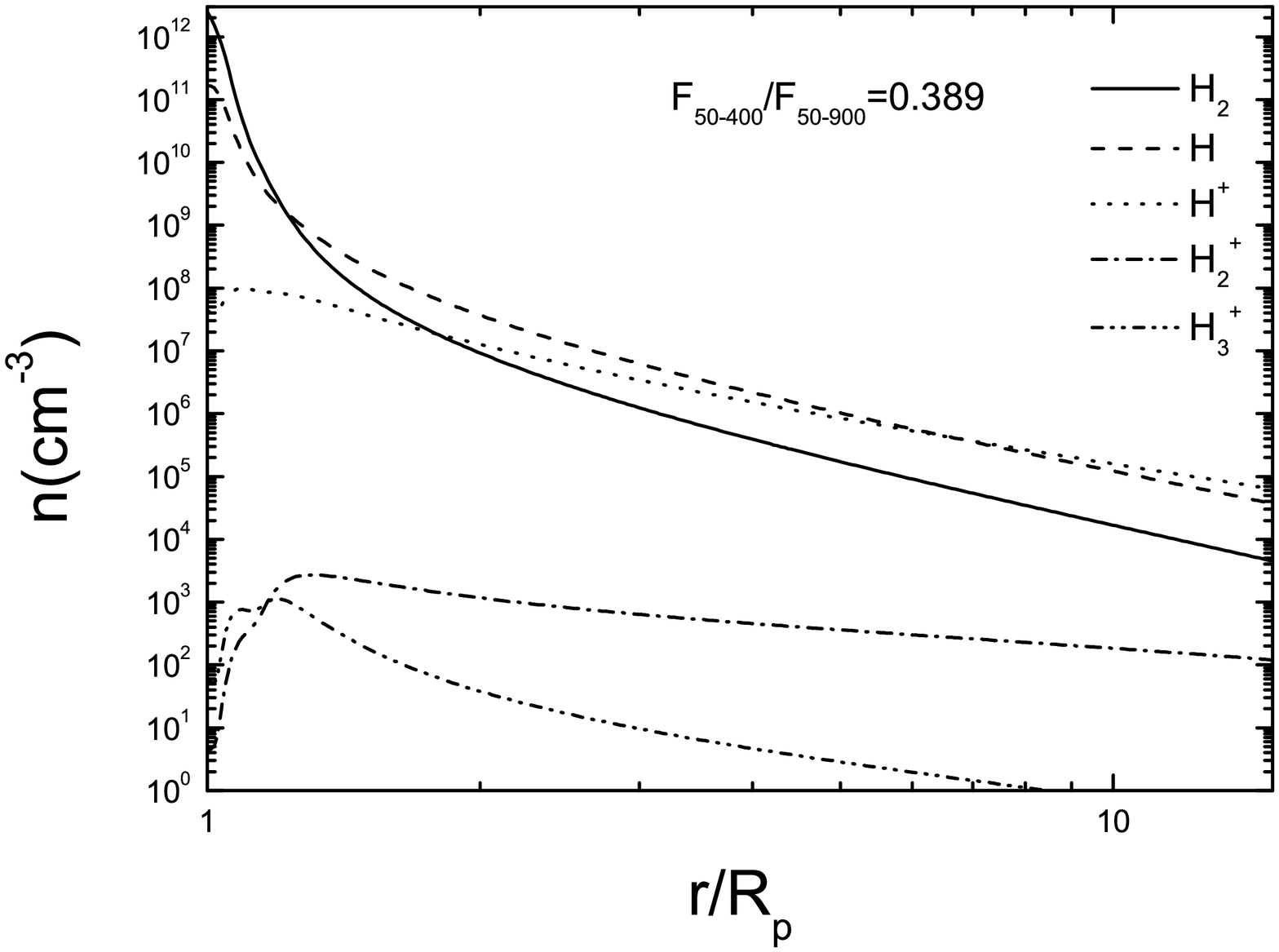}
\end{minipage}
\caption{The composition of Kepler-11b with different spectral index.}
\end{figure}

\begin{figure}
\begin{minipage}[t]{0.5\linewidth}
\centering
\includegraphics[width=3.6in,height=2.6in]{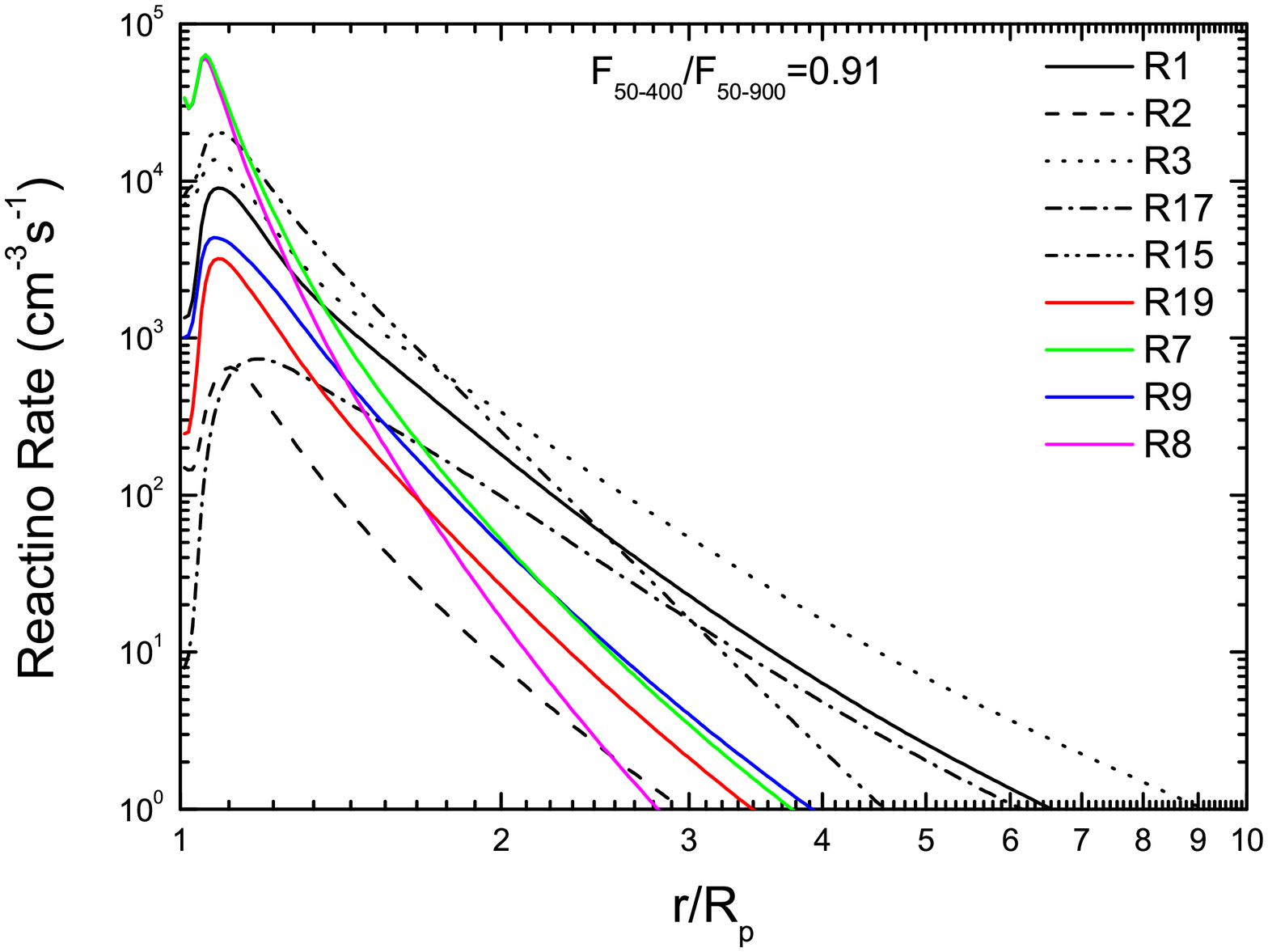}
\end{minipage}
\begin{minipage}[t]{0.5\linewidth}
\centering
\includegraphics[width=3.6in,height=2.6in]{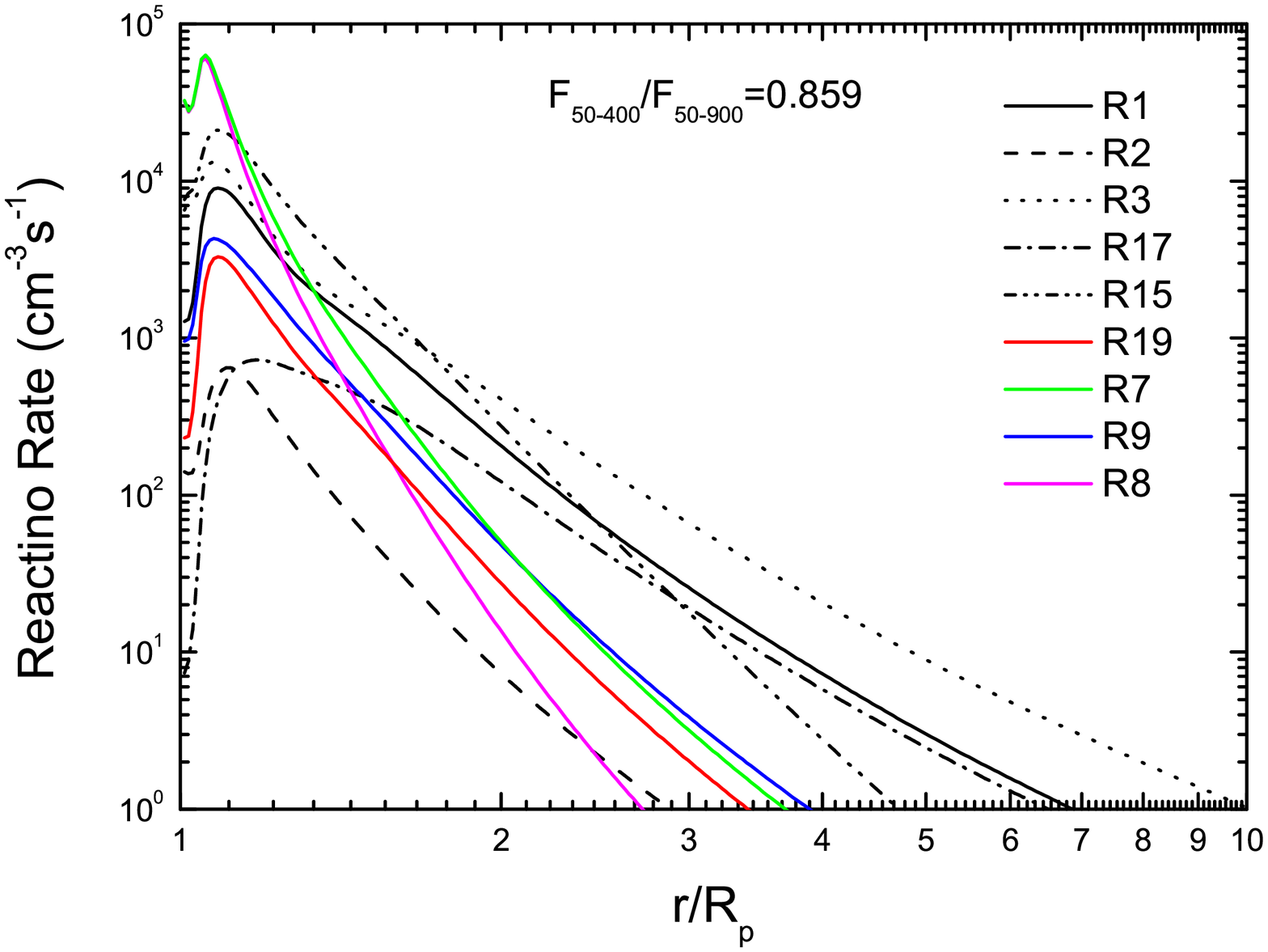}
\end{minipage}
\begin{minipage}[t]{0.5\linewidth}
\centering
\includegraphics[width=3.6in,height=2.6in]{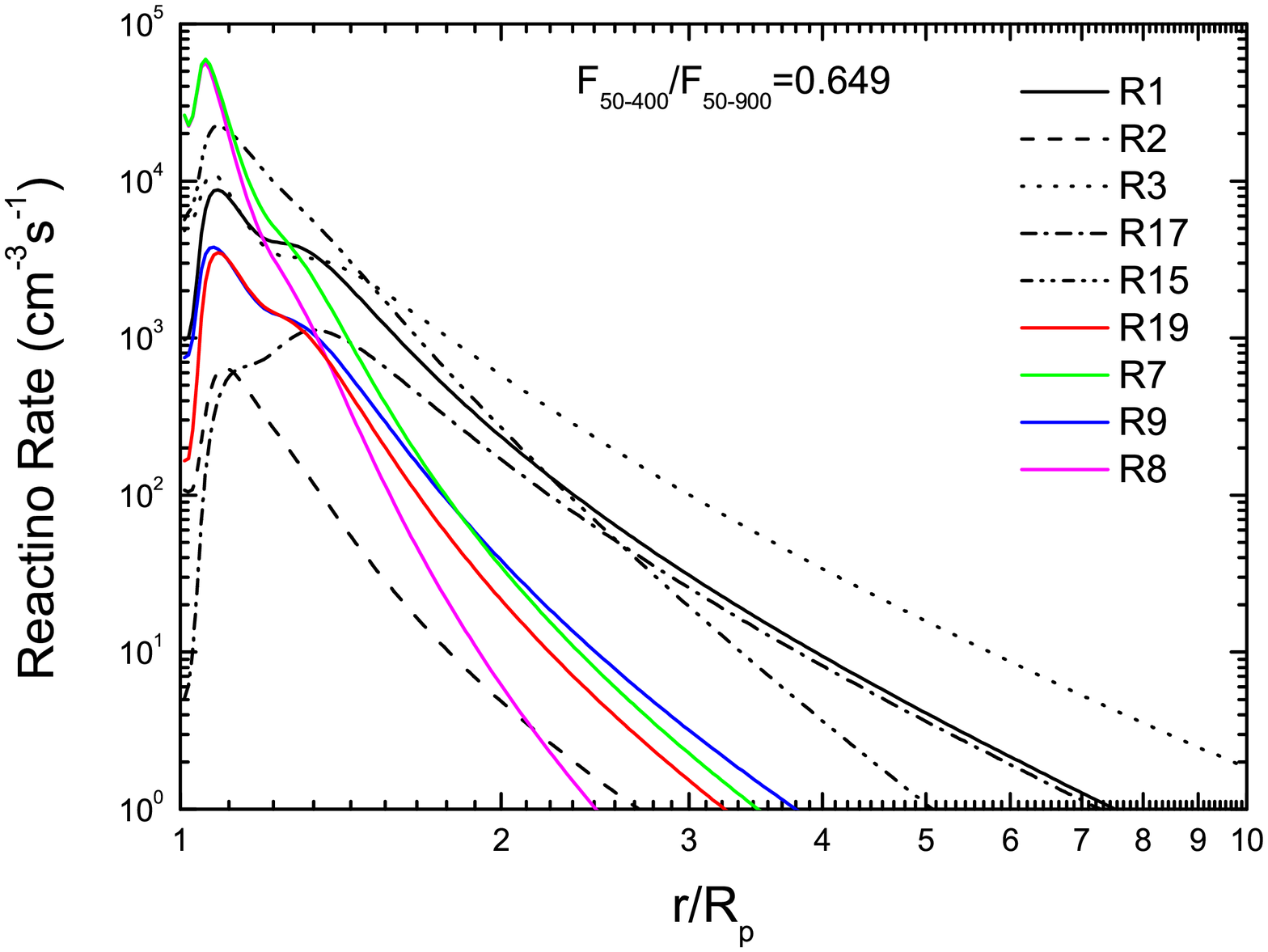}
\end{minipage}
\begin{minipage}[t]{0.5\linewidth}
\centering
\includegraphics[width=3.6in,height=2.6in]{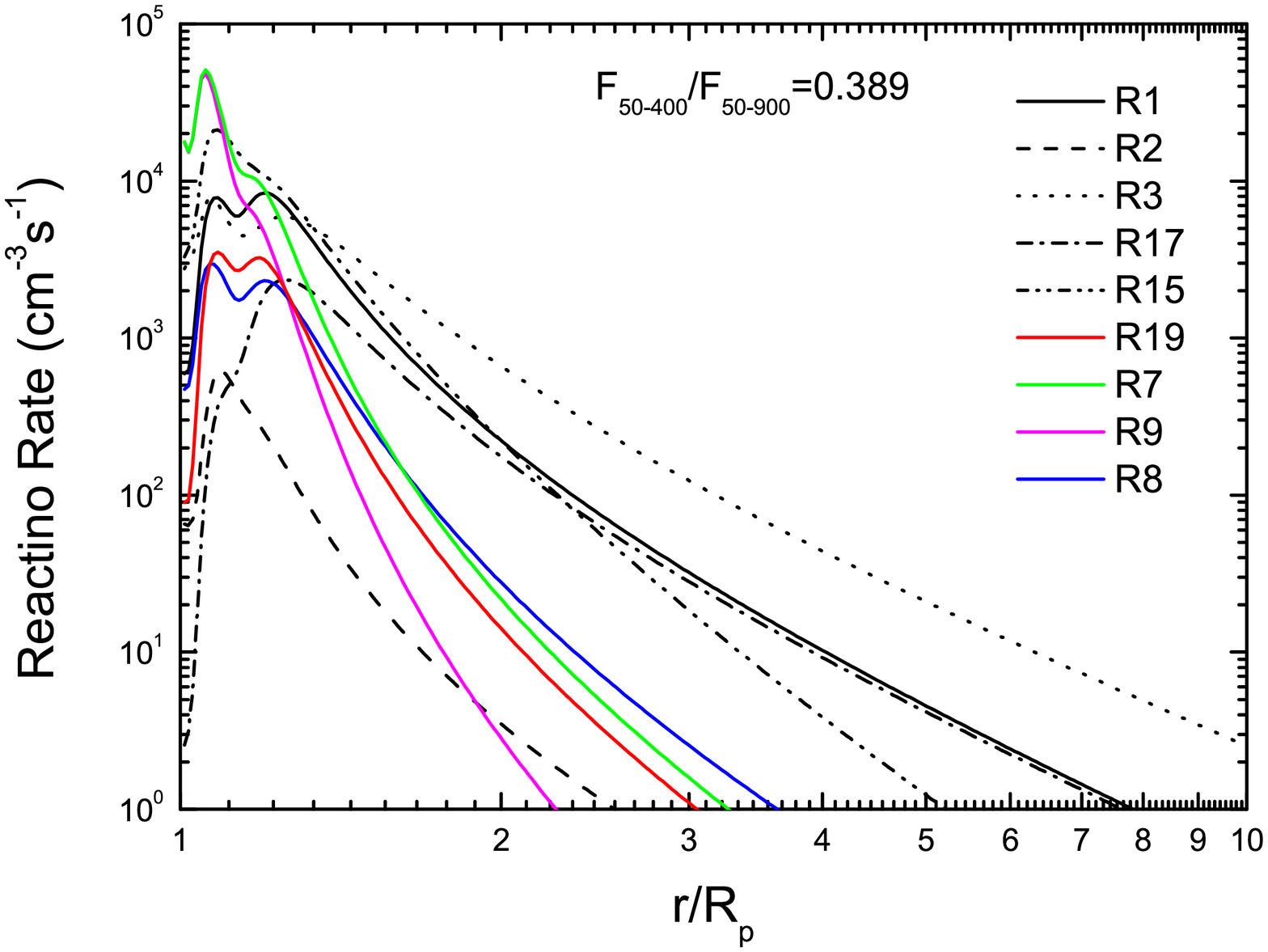}
\end{minipage}
\caption{The primary chemical reaction rates in the atmosphere of Kepler-11b. Note that not all chemical reactions are shown.}
\end{figure}

Compared with HD\, 209458b, and GJ\, 436b, H$_{2}$ contributes a significant portion to the whole atmosphere of Kepler-11b (Figure 7).  The number densities of H$_{2}$ are even higher than the number densities of H$^{+}$ below the altitude of 2-3R$_{p}$. In addition, the number density of hydrogen neutral atoms at all altitudes is higher than that of H$^{+}$ for the case of $\beta$=0.91 and 0.859. When the value of $\beta$ decreases, the transition of H/H$^{+}$ occurs at 13.1 and 6.7R$_{p}$ for the case of $\beta$=0.649 and 0.389, respectively.
For all cases, the number density of H$_{2}^{+}$ is about 10$^{3}$ cm$^{-3}$ at 1.3R$_{p}$ and slowly declines with the altitude. The number density of H$_{3}^{+}$ can attain 10$^{3}$ cm$^{-3}$ at r/R$_{p}$=1.3 and is about 10$^{2}$ -10$^{3}$ cm$^{-3}$ at r$<$2R$_{p}$. This means that the cooling produced by H$_{3}^{+}$ cannot be neglected.

The photochemistry processes obtained for Kepler-11b are similar to those of GJ\, 436b. In the base of the atmosphere, the composition is determined mainly by the chemical reaction R1, R3, R7, R8 and R15 (Figure 8). The photoionization of H becomes more important at higher altitudes and dominates the photochemical processes above 1.4-2R$_{p}$. The photodissociation of H$_{2}$ (R1) and its inverse process (R17) also play important roles in the processes of photochemistry. In contrast to the HD\, 209458b and GJ\, 436b cases, H$_{2}$ and its corresponding photochemical reactions are evidently very important in the atmosphere of Kepler-11b and cannot be neglected.

\subsubsection{HD\, 189733b}

The host star of HD 189733b is a young, active K-star. Its orbital distance is 1.5 times closer to its host star than that of HD 209458b, and the planet receives X-ray radiation from its host star that is 300 times stronger than that received by HD 209458b \citep[]{sanz10}. Thus, one can expect that the main composition of HD189733b should be ions. Figure 9 shows that H and H$^{+}$ fully dominate the composition.
The other species are almost negligible. In the case of $\beta$=0.89, the transitions of H/H$^{+}$ occur at the location of r/R$_{p} \approx$ 2. With the decrease of $\beta$, the transition of H/H$^{+}$ moves to lower altitudes.

The corresponding chemical reactions reflect the fact that the exoplanet's atmosphere is largely controlled by the photoionization of H and recombination of H$^{+}$ (Figure 10). Due to the high number densities of H$^{+}$ and electrons, the recombination of H$^{+}$ becomes comparable with the photoionization of H with the decrease of $\beta$. Those chemical reaction rates involving H$_{2}$ become very low in the regions of r $>$ 1.04R$_{p}$. It is clear from Figures 9 and 10 that the strong EUV irradiation of the host star almost fully ionizes the gas. At the very low altitude, H$_{2}$ rapidly becomes H and H$^{+}$. For the ionized planetary wind, a model of an atomic hydrogen-proton mixture can accurately describe the atmospheric escape of HD 189733b.

\begin{figure}
\begin{minipage}[t]{0.5\linewidth}
\centering
\includegraphics[width=3.6in,height=2.6in]{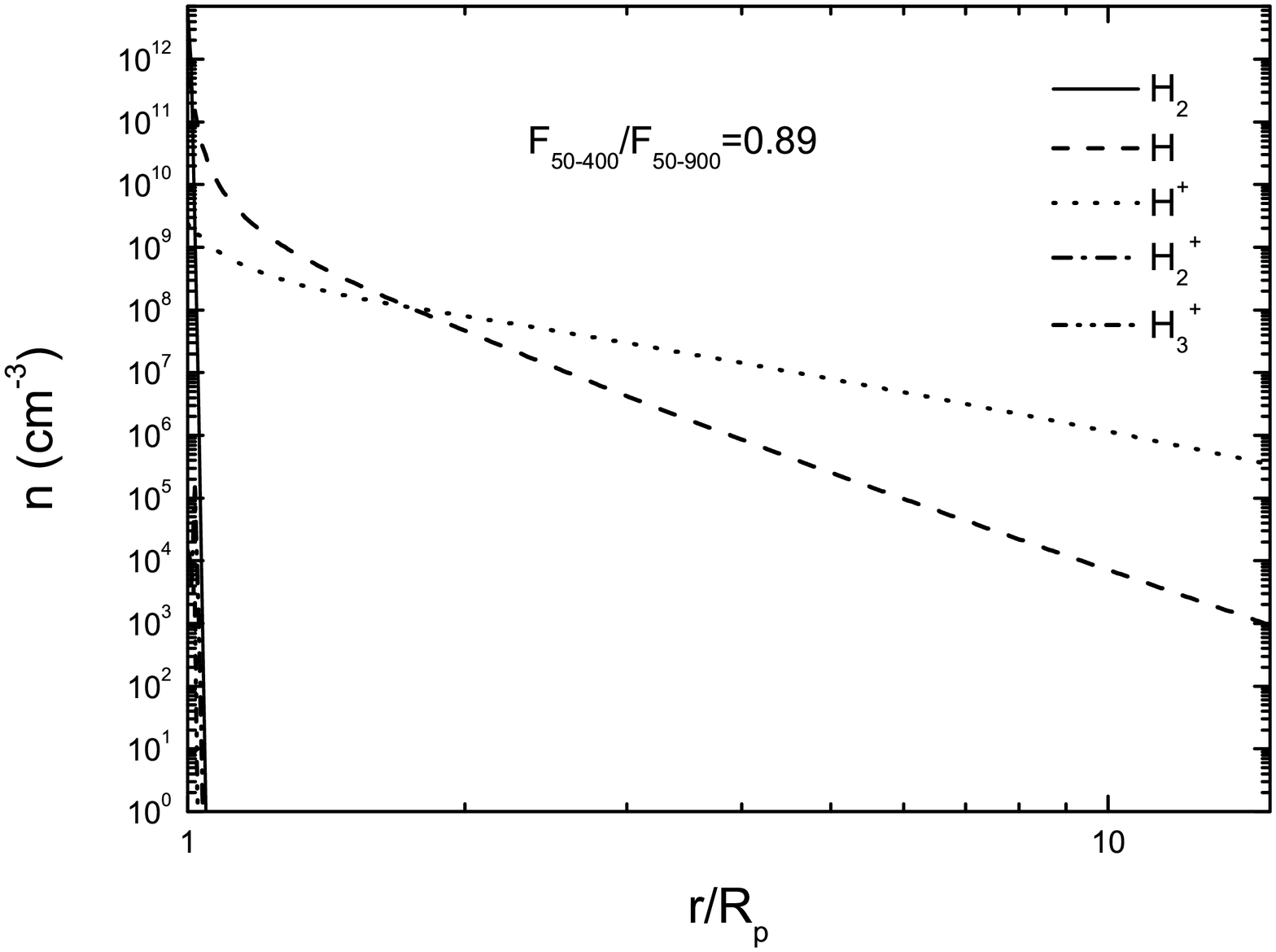}
\end{minipage}
\begin{minipage}[t]{0.5\linewidth}
\centering
\includegraphics[width=3.6in,height=2.6in]{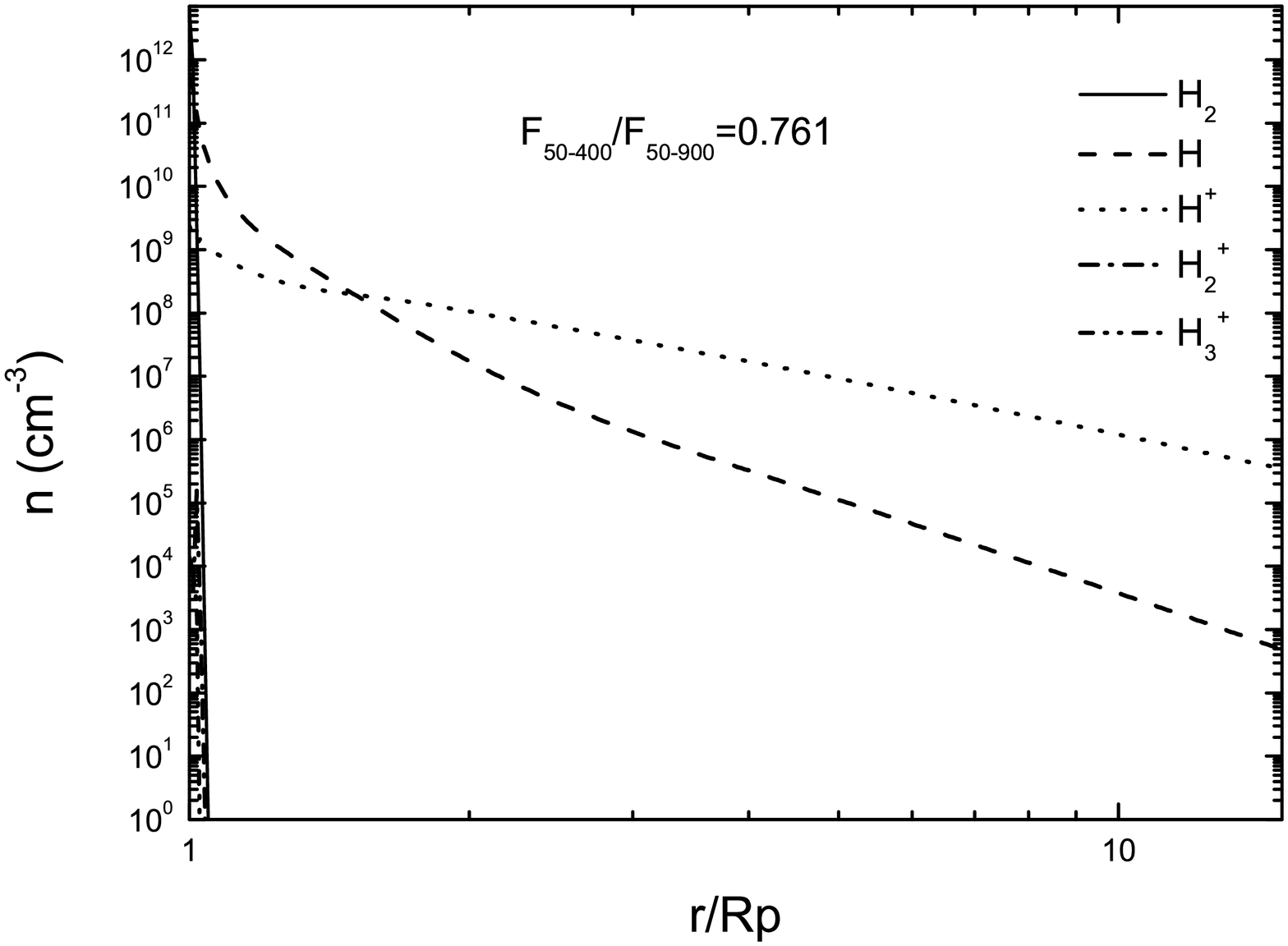}
\end{minipage}
\begin{minipage}[t]{0.5\linewidth}
\centering
\includegraphics[width=3.6in,height=2.6in]{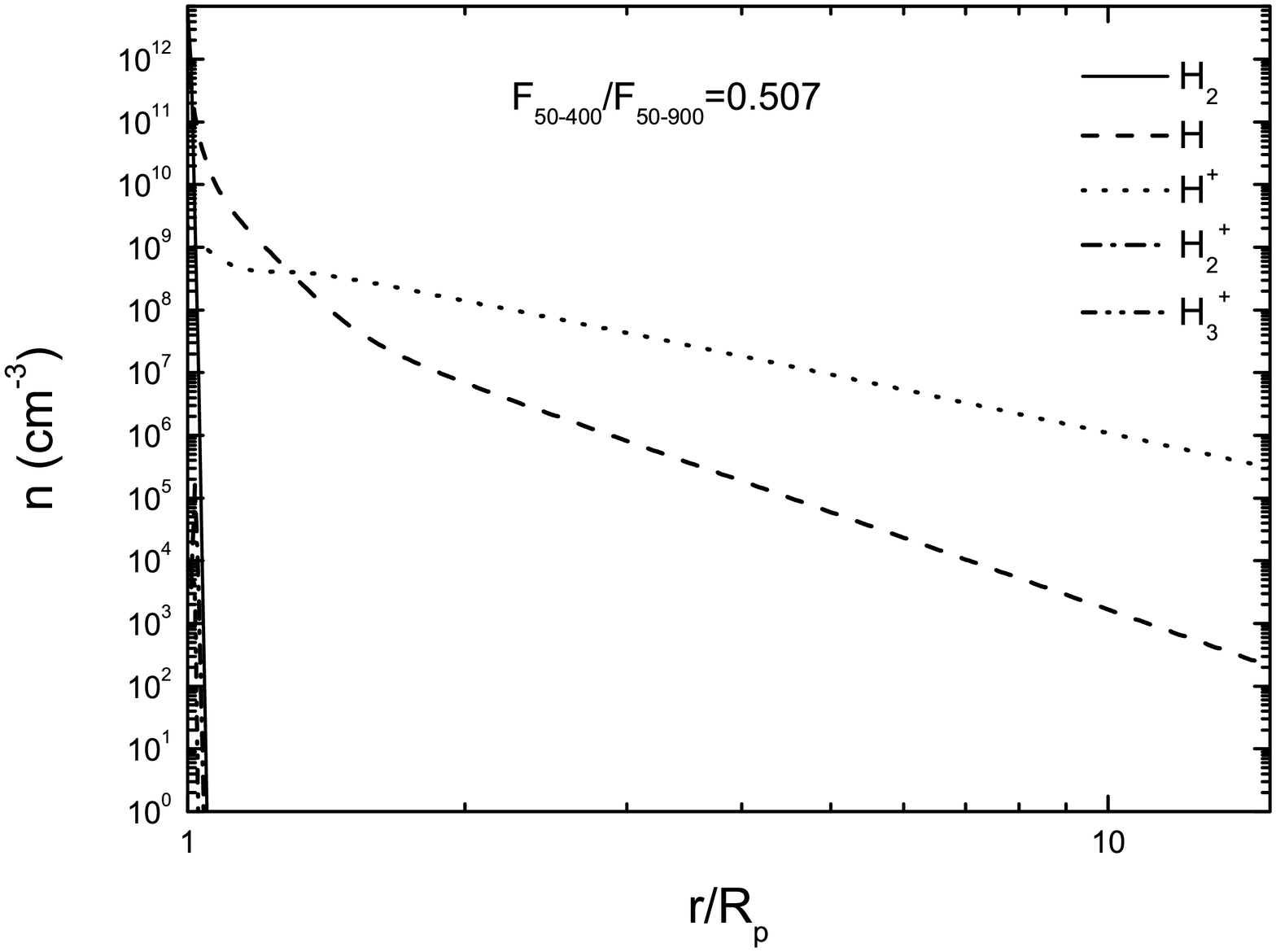}
\end{minipage}
\begin{minipage}[t]{0.5\linewidth}
\centering
\includegraphics[width=3.6in,height=2.6in]{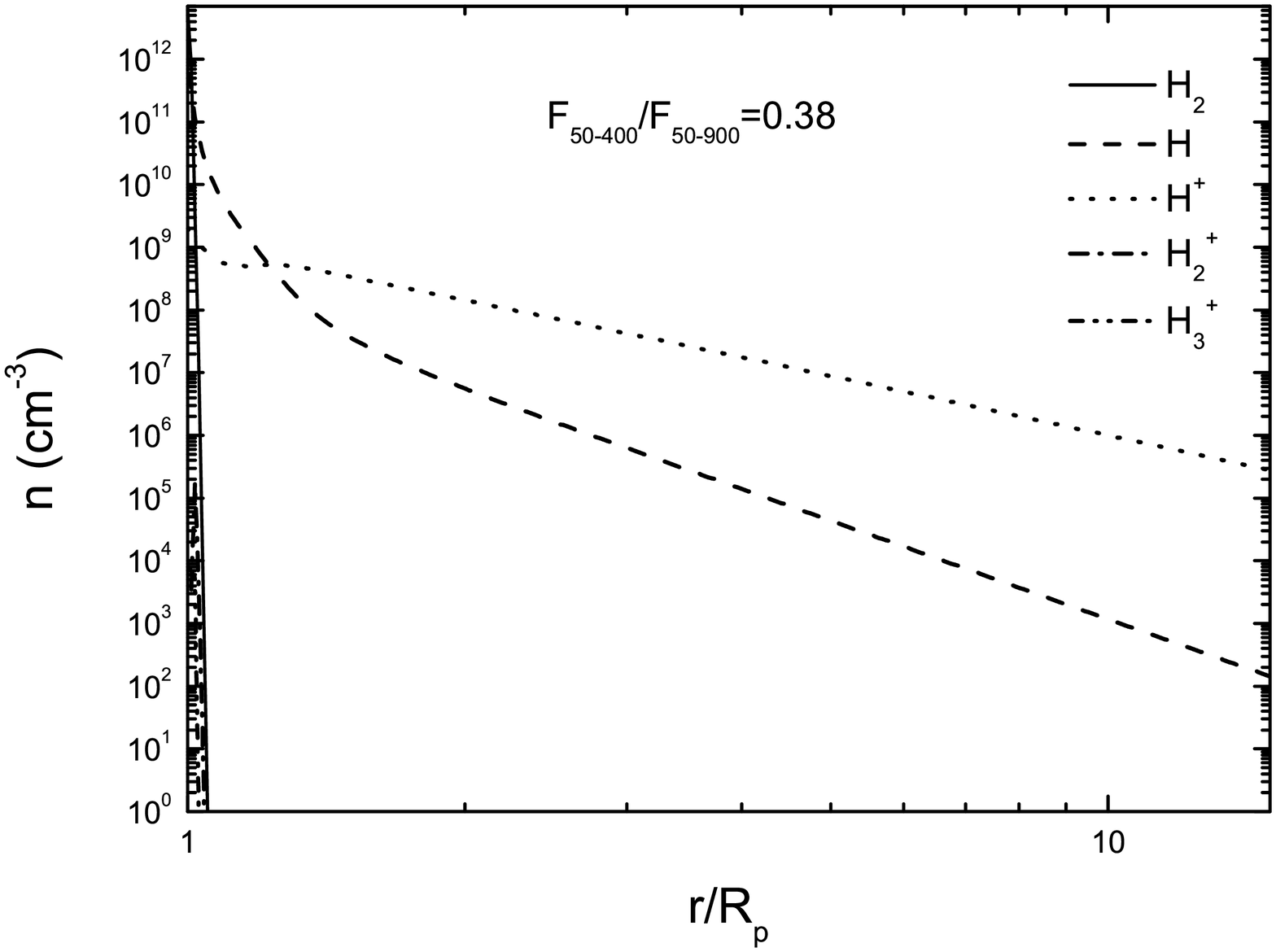}
\end{minipage}
\caption{The composition of HD 189733b with different spectral index.}
\end{figure}

\begin{figure}
\begin{minipage}[t]{0.5\linewidth}
\centering
\includegraphics[width=3.6in,height=2.6in]{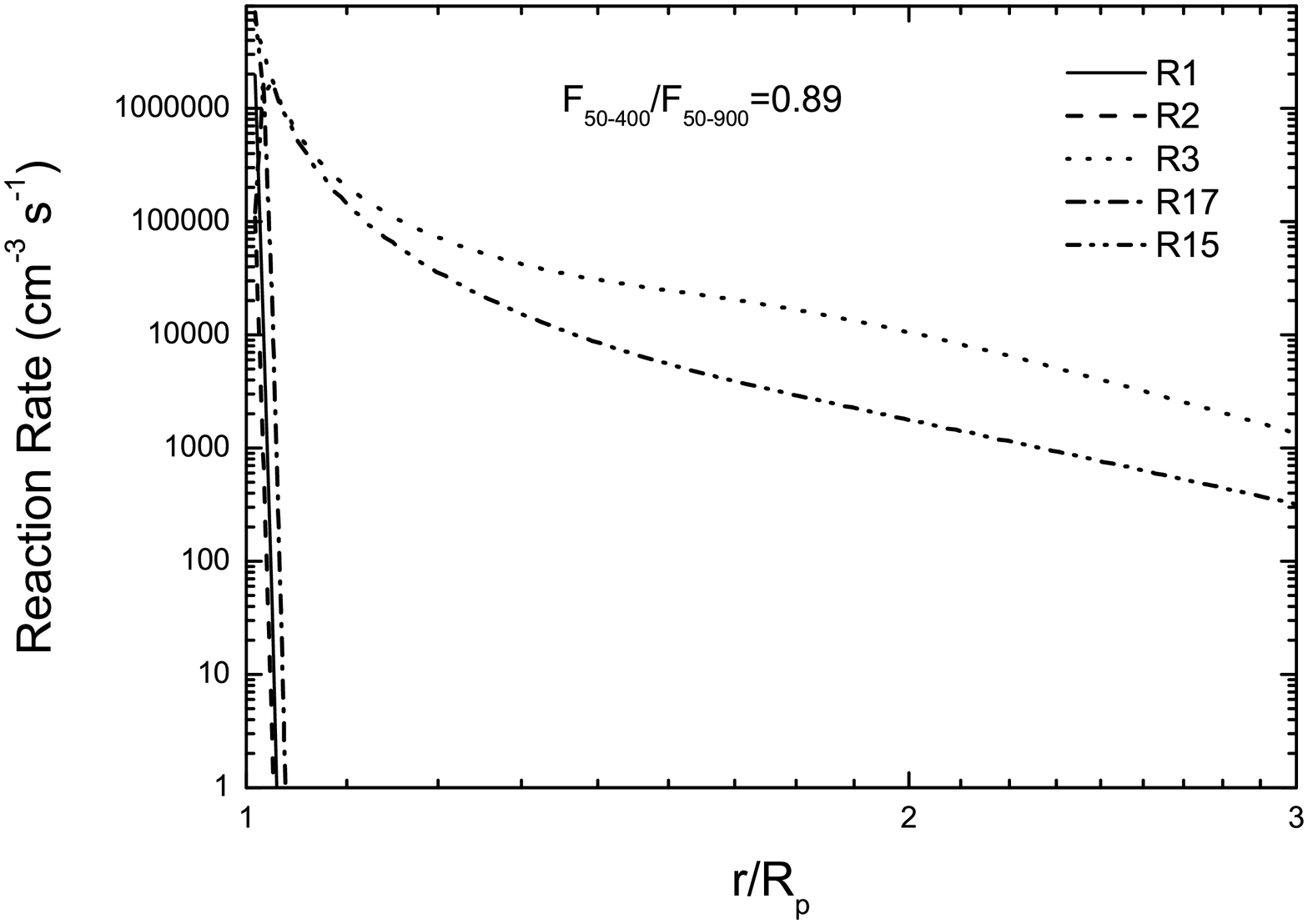}
\end{minipage}
\begin{minipage}[t]{0.5\linewidth}
\centering
\includegraphics[width=3.6in,height=2.6in]{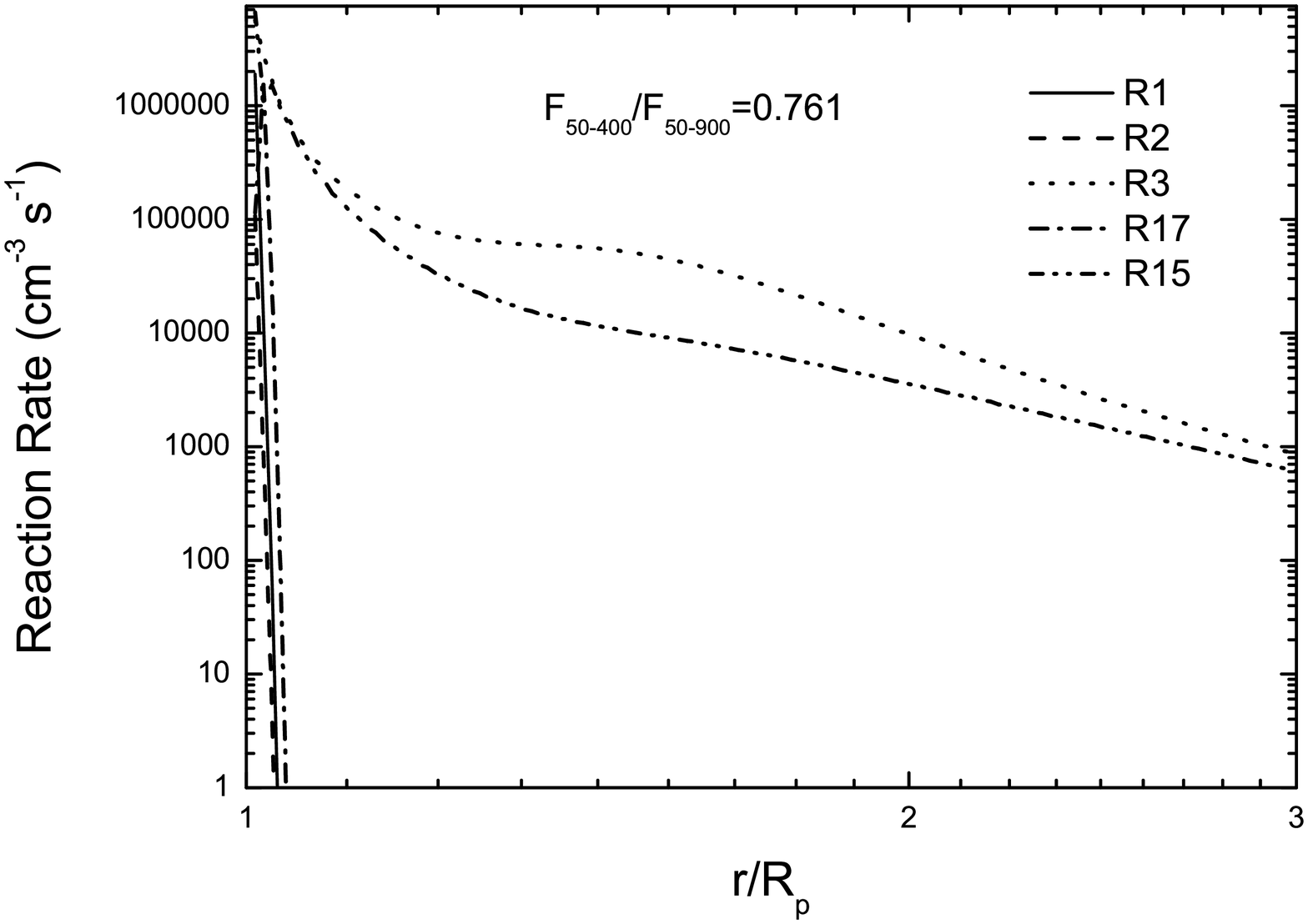}
\end{minipage}
\begin{minipage}[t]{0.5\linewidth}
\centering
\includegraphics[width=3.6in,height=2.6in]{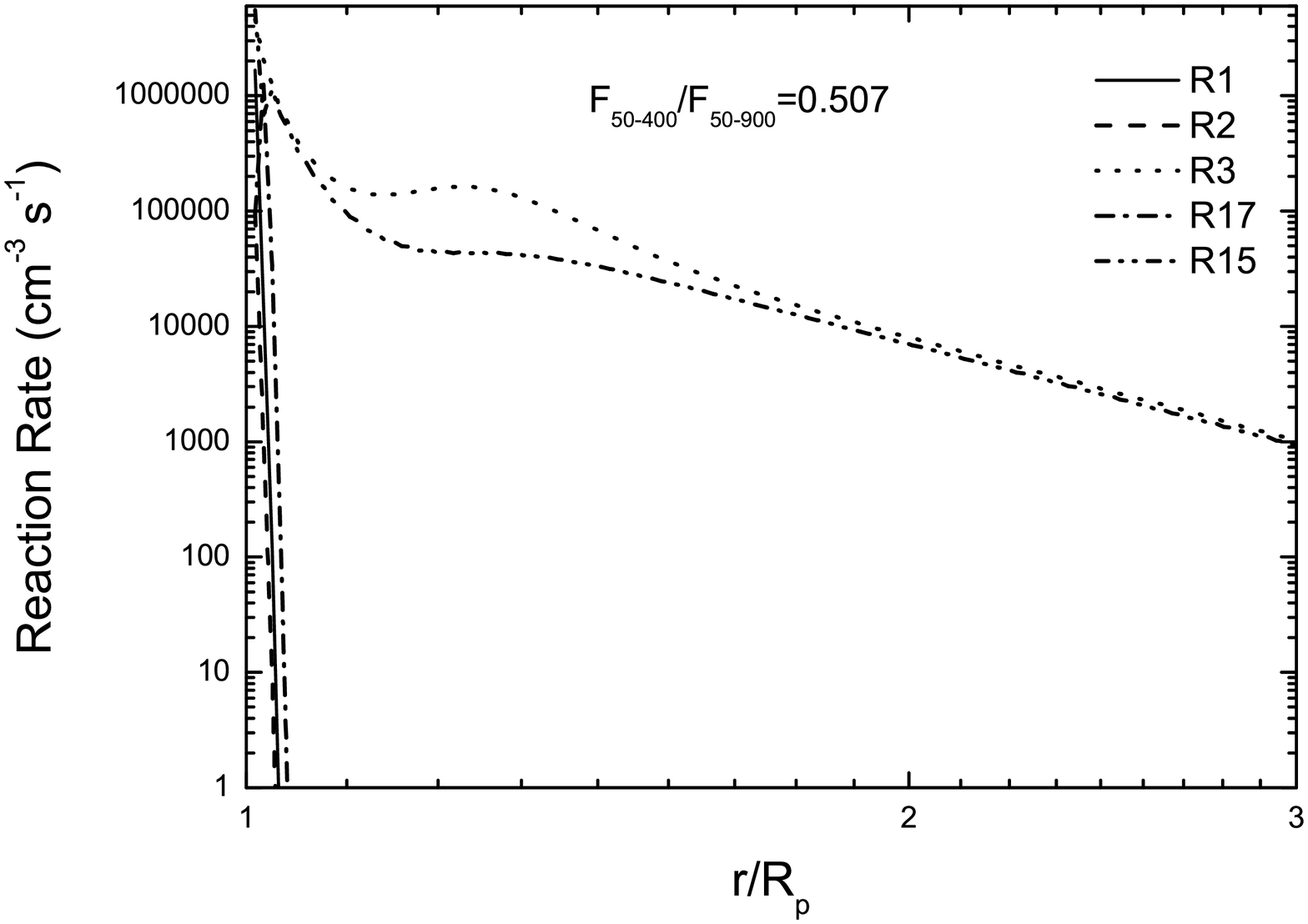}
\end{minipage}
\begin{minipage}[t]{0.5\linewidth}
\centering
\includegraphics[width=3.6in,height=2.6in]{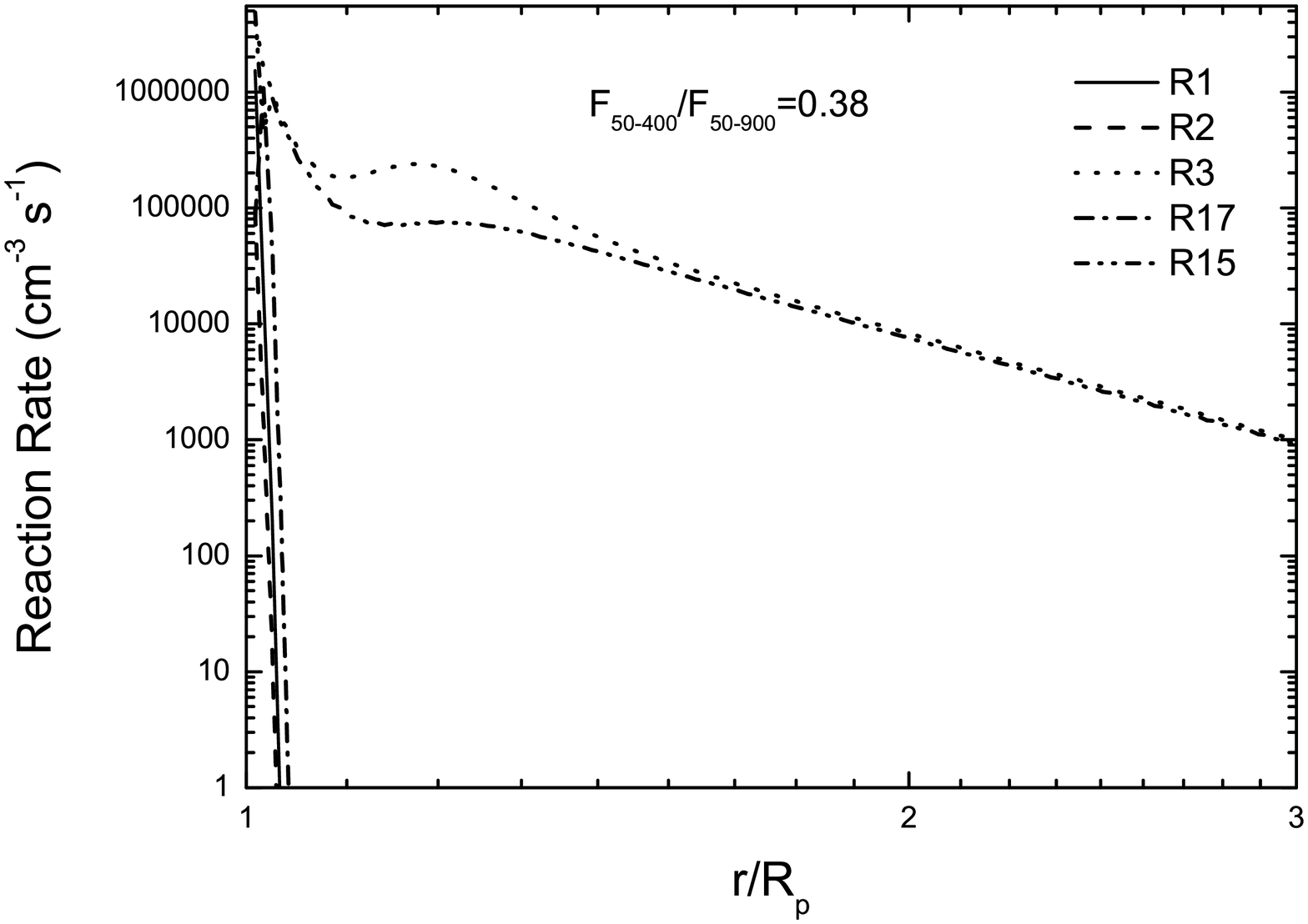}
\end{minipage}
\caption{The primary chemical reaction rates in the atmosphere of HD 189733b. Note that not all chemical reactions are shown.}
\end{figure}

\subsection{The Hydrogen Neutral Atoms Altitude Profile}
Up to now, the atmospheric escape has been evaluated mostly from transit observations of the neutral hydrogen spectral lines. For instance, this was the case for HD 189733b, HD 209458b, and GJ 436b. Thus it is important to check the dependence of the profiles of H on their respective EUV SED. In the following, we first discuss the case of HD 189733b, HD 209458b, and GJ 436b in Section 3.2.1. For Kepler-11b, we check its vertical profiles of H in Section 3.2.2.

\subsubsection{HD 209458b, GJ 436b, and HD 189733b}

It is clear from Figure 11 that the vertical profiles of hydrogen atoms in HD 209458b, GJ 436b, and HD 189733b are greatly affected by the profiles of the EUV SED (see Figures 11a, 11b, and 11c), and the sensitivity of H to the index $\beta$ is altitude-dependent. For HD 209458b, the number densities of H at 1.1 R$_{p}$ exhibit a sensitivity to the different values of $\beta$ due to the strong photoionization. Further, the number densities of H at 2-3R$_{p}$ may decrease an order of magnitude with the decrease of $\beta$. At a higher altitude, the variations of number densities of H are more prominent.
For GJ 436b, the number densities of H at 3R$_{p}$ decrease by a factor of 6 with the decrease of $\beta$. It is clear that the decline in number density of H is steeper in the cases of smaller $\beta$. We also see similar distributions of H in HD 189733b. Figure 11c shows that the number densities of H at r/R$_{p}$=3 decline by a factor of 7 with the decrease of $\beta$: namely, the value is from 4.3$\times$ 10$^{6}$ to 6.3$\times$ 10$^{5}$ cm$^{3}$.

\begin{figure}
\begin{minipage}[t]{0.5\linewidth}
\centering
\includegraphics[width=3.6in,height=2.6in]{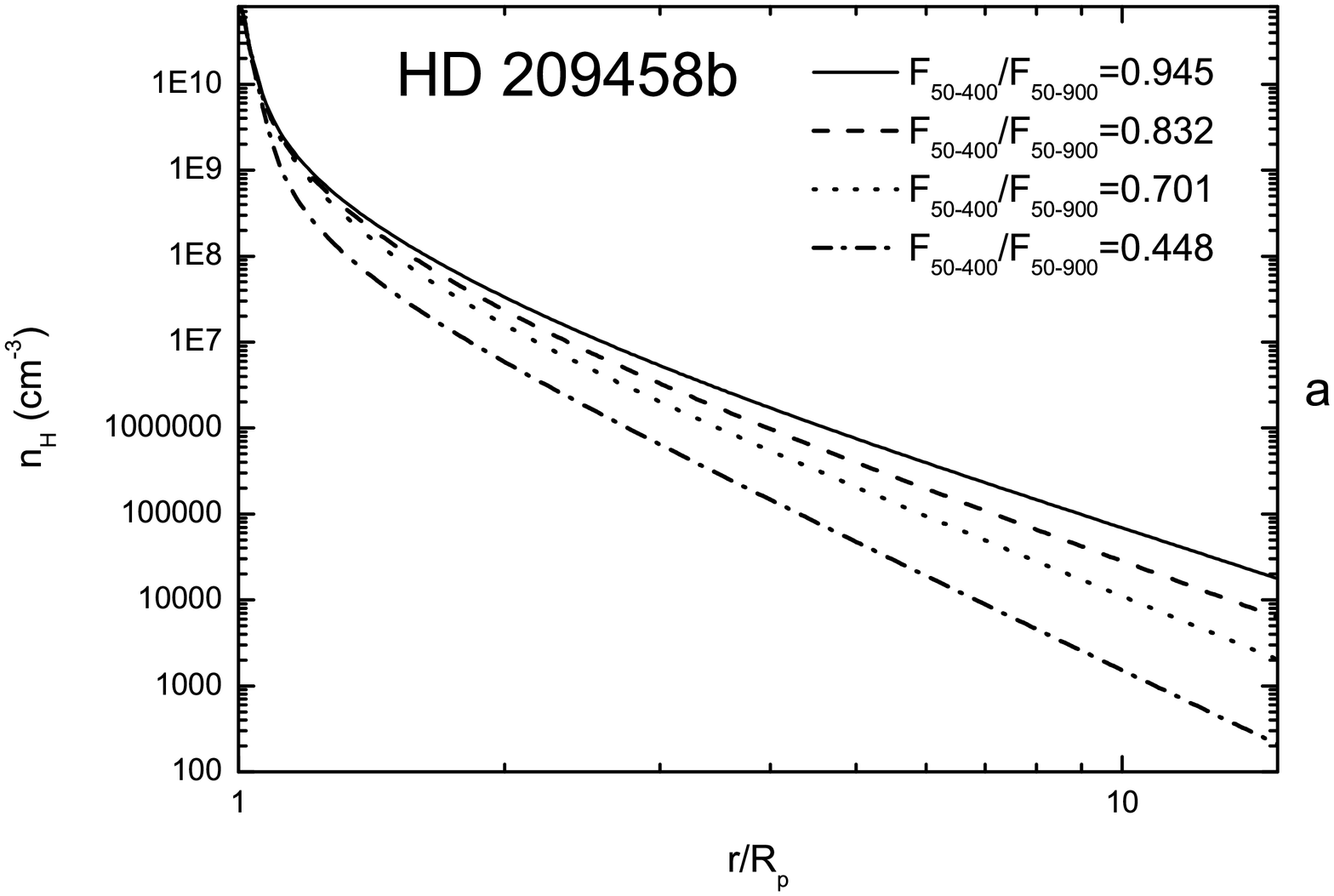}
\end{minipage}
\begin{minipage}[t]{0.5\linewidth}
\centering
\includegraphics[width=3.6in,height=2.6in]{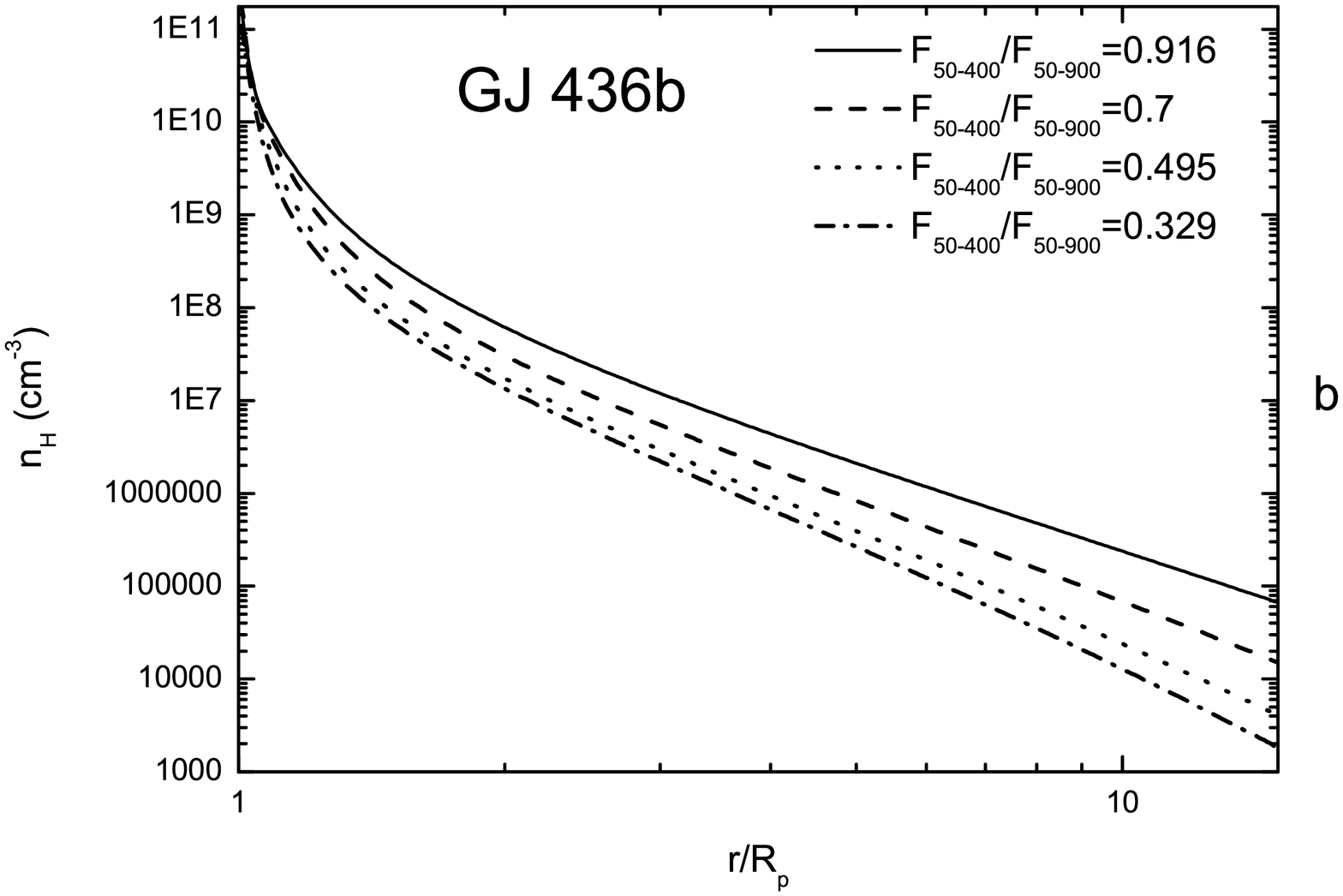}
\end{minipage}
\begin{minipage}[t]{0.5\linewidth}
\centering
\includegraphics[width=3.6in,height=2.6in]{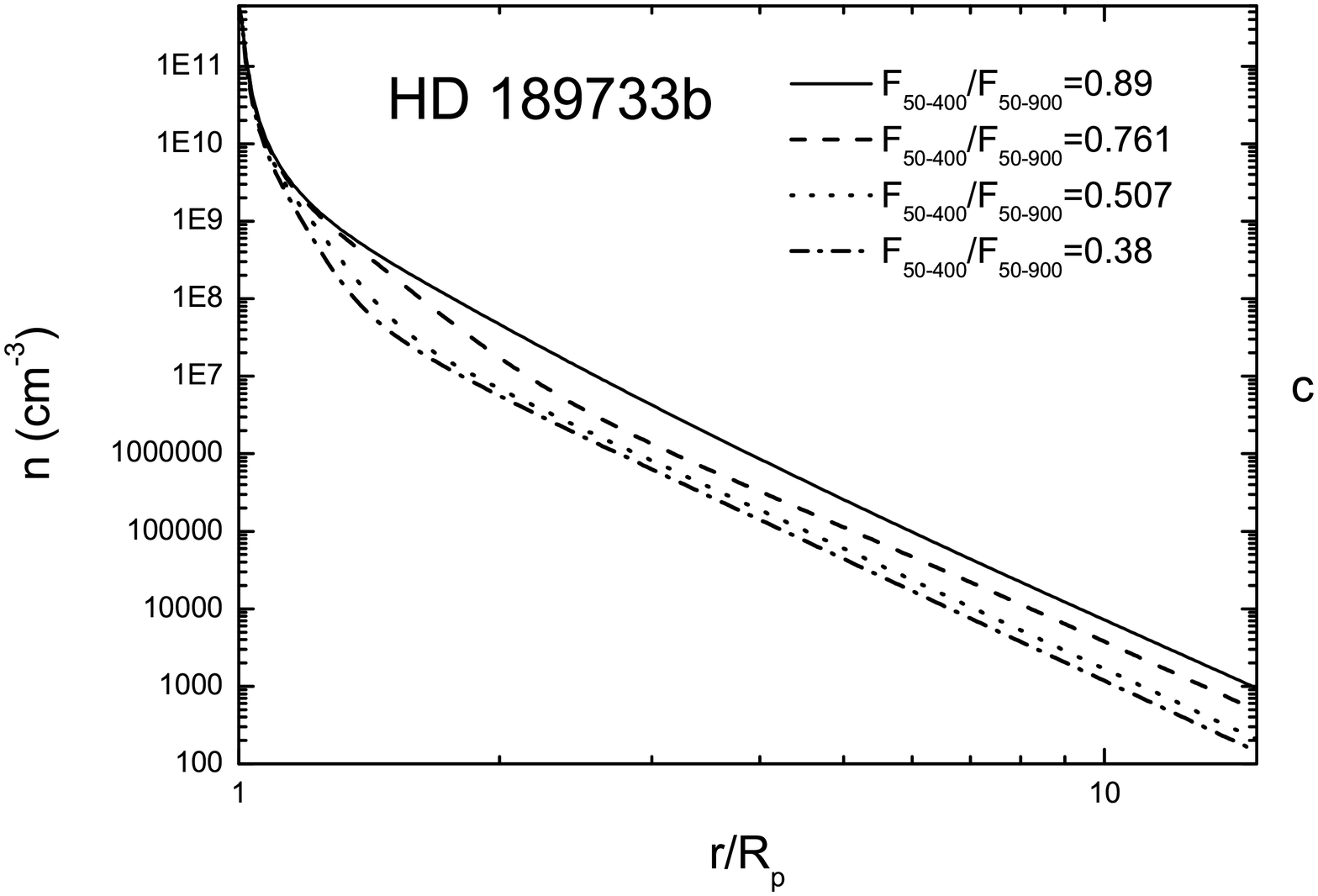}
\end{minipage}
\begin{minipage}[t]{0.5\linewidth}
\centering
\includegraphics[width=3.6in,height=2.6in]{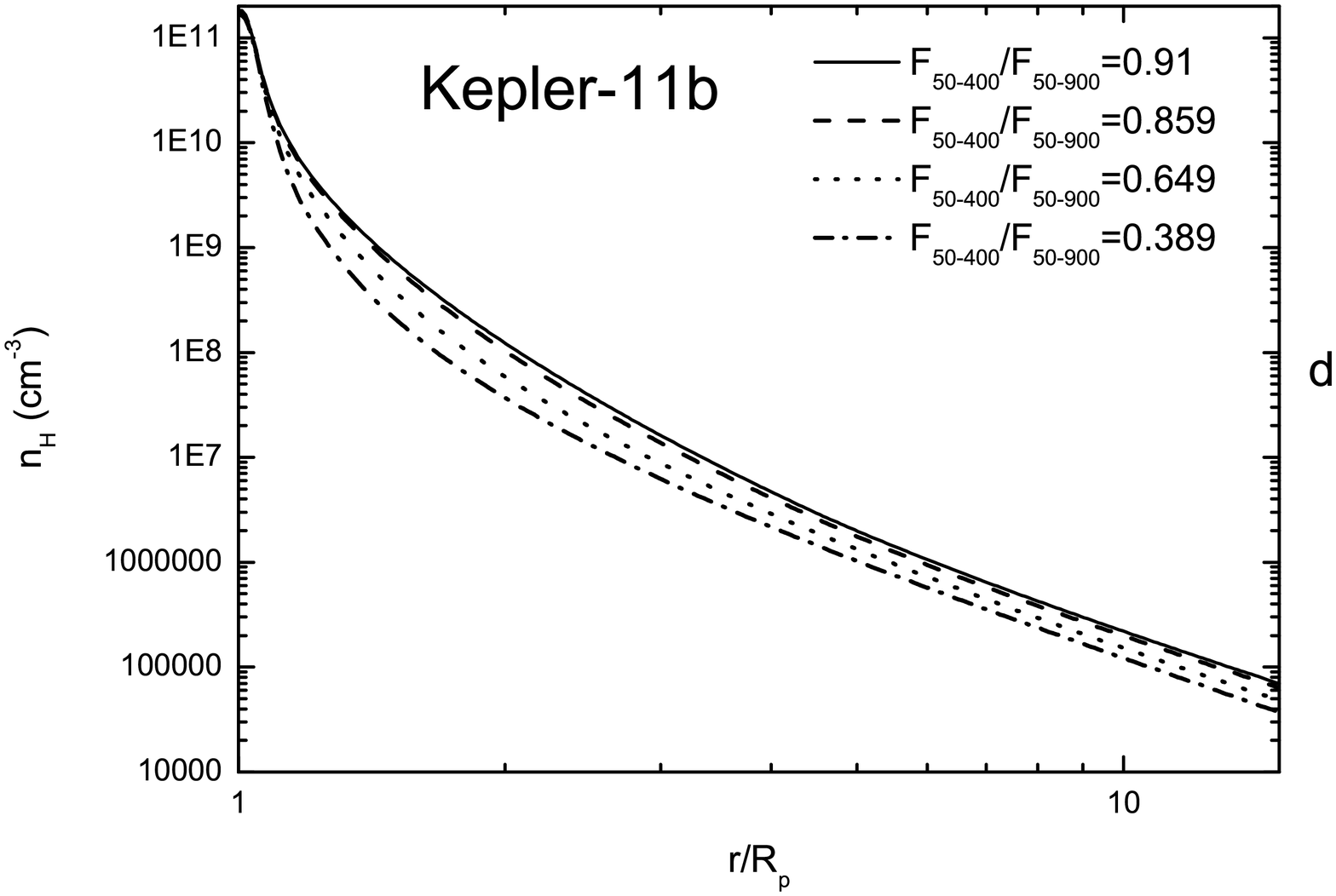}
\end{minipage}
\caption{
The number density profiles of hydrogen neutral atoms.
(a) The density distributions of hydrogen atoms for HD 209458b;
(b) the density distributions of hydrogen atoms for GJ 436b;
(c) the density distributions of hydrogen atoms for HD 189733b; and
(d) the density distributions of hydrogen atoms for Kepler-11b.}
\end{figure}

The dependence of the H altitude distribution on the EUV SEDs results from the simple fact that the H number density is controlled by the processes of the photoionization by the EUV photons. As shown in Figure 4, Figure 6, and Figure 10, the most important photochemical reactions of HD 209458b, GJ 436b, and HD 189733b are the photoionization of H (R3). The photoionization of H becomes dominant at high altitude, and can even attain 10 times the recombination (R15) at 3R$_{p}$. Since the photoionization rate is larger than that of recombination, H becomes H$^{+}$.

\begin{figure}
\begin{minipage}[t]{0.5\linewidth}
\centering
\includegraphics[width=3.6in,height=2.6in]{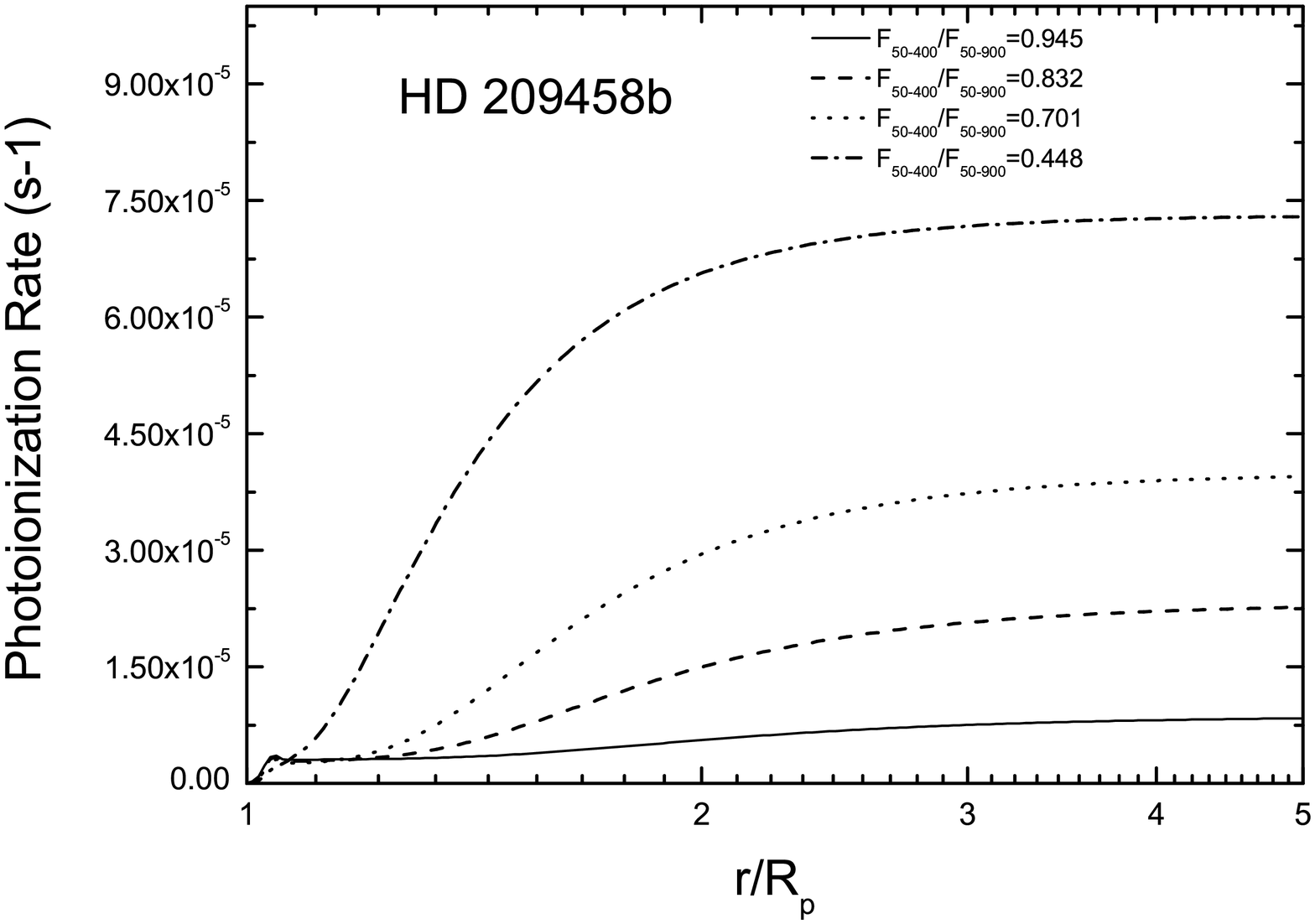}
\end{minipage}
\begin{minipage}[t]{0.5\linewidth}
\centering
\includegraphics[width=3.6in,height=2.6in]{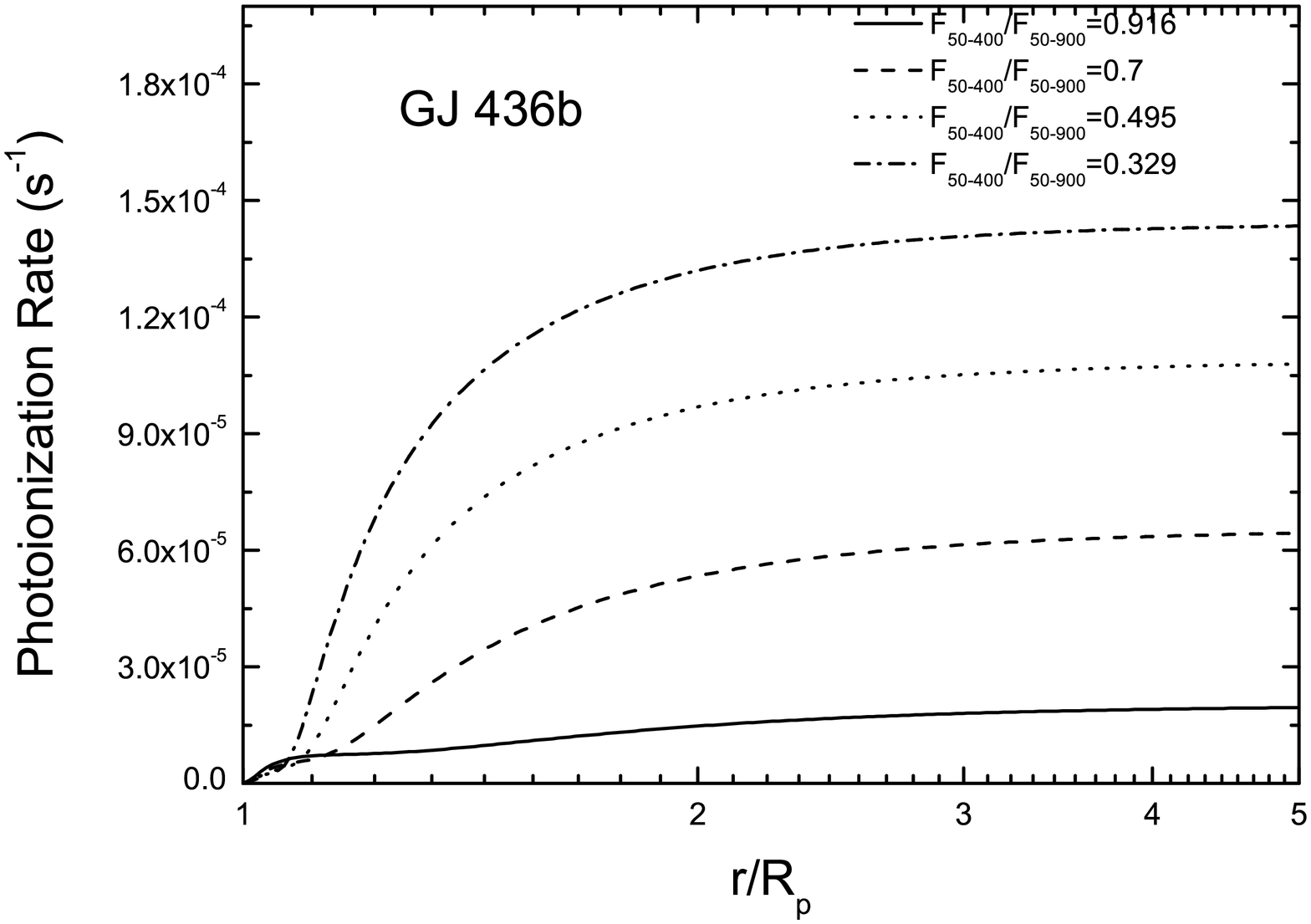}
\end{minipage}
\begin{minipage}[t]{0.5\linewidth}
\centering
\includegraphics[width=3.6in,height=2.6in]{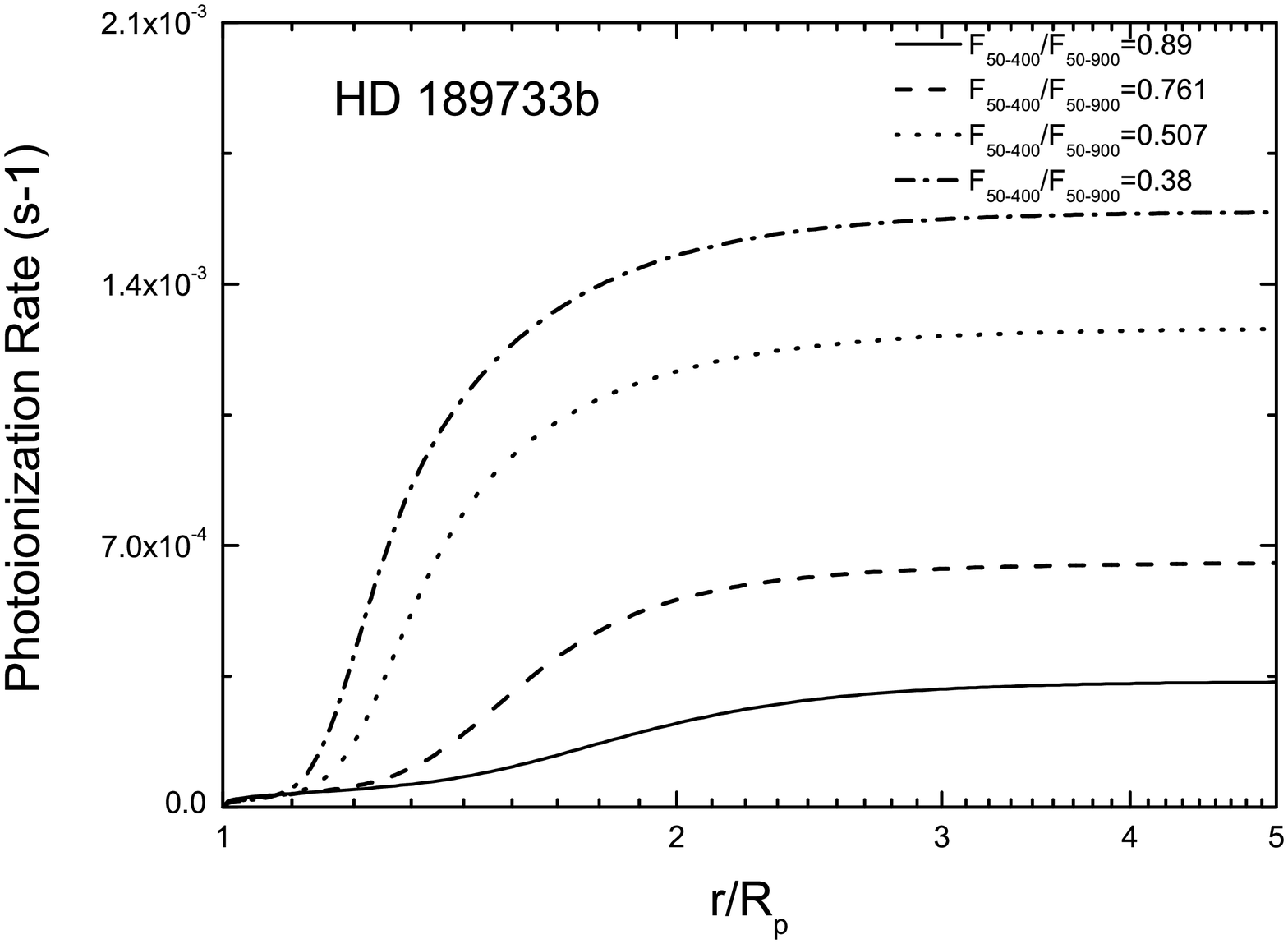}
\end{minipage}
\begin{minipage}[t]{0.5\linewidth}
\centering
\includegraphics[width=3.6in,height=2.6in]{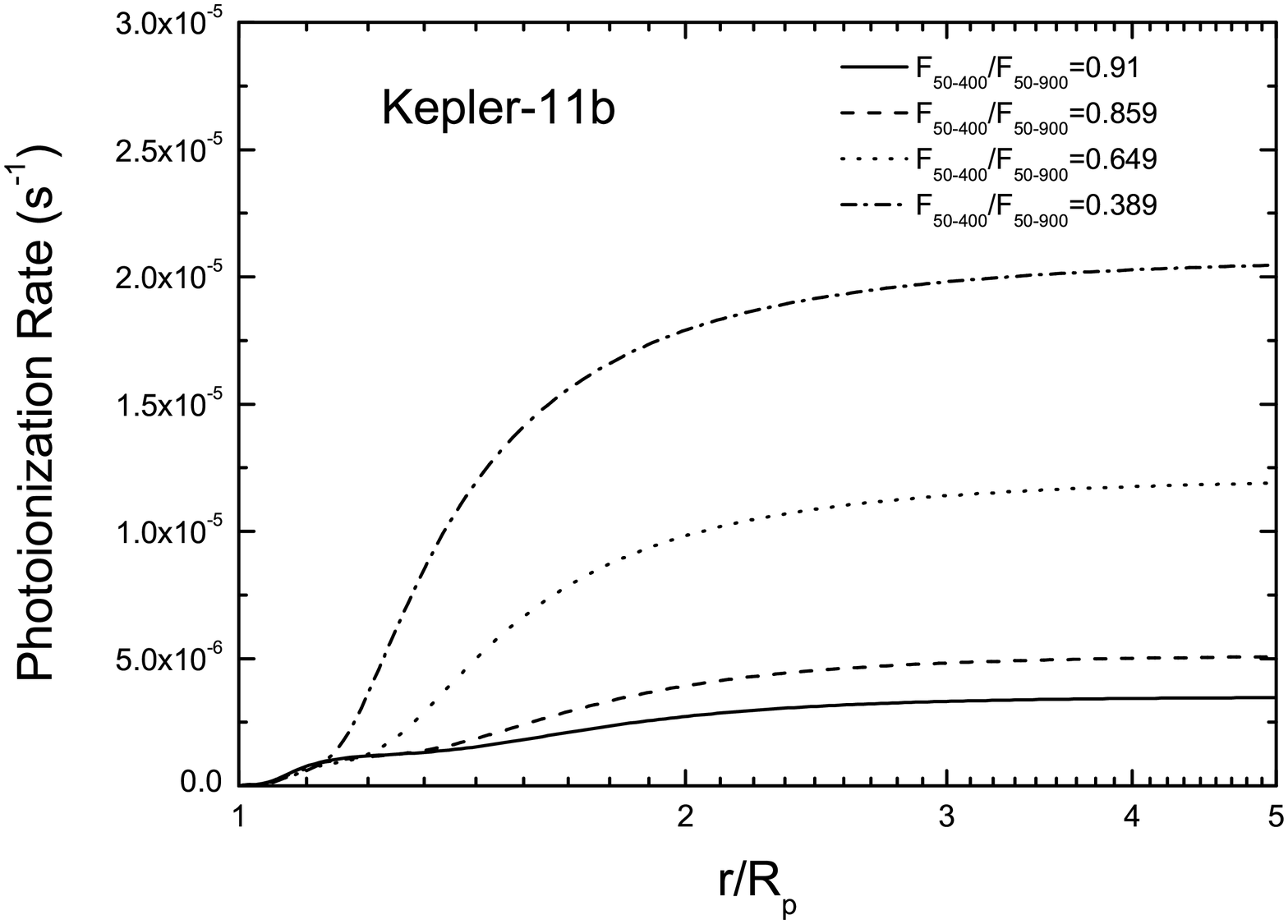}
\end{minipage}
\caption{
The photoionization rate of H for all planet samples.}
\end{figure}

Figure 12 further shows the photoionization rate (s$^{-1}$) of H for HD 209458b, GJ 436b, HD 189733b, and Kepler-11b. The increase of photoionization rate with altitude in the case of low $\beta$ ratio is faster than that of the high $\beta$ ratio. In the case of highest $\beta$, the photoionization rate of H increases by a factor of 2.74, 2.78, 9.4, and 4.83 for HD 209458b, GJ 436b, HD 189733b, and Kepler-11b from 1.1R$_{p}$ to 5R$_{p}$ . In contrast, there is an increase of a factor of 13.75, 7.25, 35, and 37.5 for the same targets for the lowest $\beta$ ratio.

In fact, the photoionization rate of H can be expressed as
\begin{equation}
\sum_{\nu}\frac{F_{\nu}e^{-\tau_{\nu}}}{h\nu}\sigma_{\nu,H}=\sum_{\nu}p_{\nu}e^{-\tau_{\nu}}\sigma_{\nu,H},
\end{equation}
where $p_{\nu}$ is photon flux (cm$^{-2}$s$^{-1}$). Evidently, the photoionization rate of H is related to the photon flux received at every altitude and the absorption cross-sections.
In our model, the integrated fluxes of all stellar targets are fixed as constants. However, the profiles of the EUV SED are very different. This means that the integrated photon fluxes of those targets can be different for different values of $\beta$. For example, the integrated photon flux of HD 209458b is 2.62$\times$10$^{13}$ cm$^{-2}$ s$^{-1}$ in the case of $\beta$=0.448, which is a factor of 2 greater than that of $\beta$=0.945 (1.32$\times$10$^{13}$ cm$^{-2}$ s$^{-1}$). Thus, one can expect that the photoionization rate of H produced by the different EUV SED (or the integrated photon flux) should be roughly proportional to the integrated photon flux regardless of the absorption cross-sections.

However, it is clear from Figure 12 that the ratios of the photoionization rate of the highest $\beta$ to lowest $\beta$ for HD 209458b can attain a factor of 7-10 at r/R$_{p}$=2-3, which is greater than the ratio of the integrated photon flux. A similar ratio ($\sim$ 5-11) can be seen in GJ 436b and HD 189733b. The large difference can be attributed to the variation of absorption cross-sections with the wavelength.
In fact, the absorption cross-sections of photoionization of H increase with the increase of wavelength in the range of 50-900\AA. For the cases of smaller $\beta$, the low-energy photons provide more stellar irradiation and produce a higher ionization degree (or lower H density) due to the larger absorption cross-sections.
For the cases of higher $\beta$, the loss of H produced by photoionization is relatively low because the absorption cross-sections of photons with high energy are a factor of a few (even ten) smaller than those of low-energy photons. Thus, the densities of H for HD 209458b, GJ 436b, and HD 189733b are determined mainly by the absorption cross-sections of photoionization. This explains why the number densities of H decrease a factor of a few (or an order of magnitude) with the decrease of $\beta$.

\subsubsection{Kepler-11b}
We note that the H density profile of Kepler-11b is less affected by the different shapes of the EUV SED (Figure 11d). As seen in the lower left panel of Figure 12, the photoionization rate of $\beta$=0.389 at r/R$_{p}$=3 is a factor of 6 higher than that of $\beta$=0.91. However, the number density ratio of H at r/R$_{p}$=3 is only 2.78 (Figure 11d). Compared to HD 209458b, the photoionization rate of $\beta$=0.448 is 9.46 times of $\beta$=0.945, but the number density ratio of H is 8.87 at r/R$_{p}$=3. One can see that the number density ratio of H for HD 209458b is coincident with the ratio of the photoionization rate, but it is not exactly the same for Kepler-11b. In fact, the H distribution of Kepler-11b is weakly related to photoionization, for which there are two possible explanations.

First, the distance of Kepler-11b from the star (0.091AU) is far greater than those of HD 209458b (0.047AU) and GJ 436b (0.0287AU), such that Kepler-11b only receives one-fourth of the integrated fluxes of HD 209458b and one-sixth of the integrated fluxes of GJ 436b. Evidently, such a low integrated flux results in a decrease of photoionization of H. As shown in Figure 12, the photoionization rate of Kepler-11b is in the order of magnitude of 10$^{-6}$-10$^{-5}$ s$^{-1}$, a factor of a few (or ten) smaller than those of HD 209458b, GJ 436b, and HD 189733b. Thus, one can expect that the H profiles of Kepler-11b are relatively insensitive to the shapes of the EUV SED.

Second, the effect of photoionization of H can be counteracted by other chemical reactions. Figure 8 indeed shows that the effect of the photoionization of H for Kepler-11b is not as prominent as for the other planets. In fact, the photoionization of H is independent of temperature, but the recombination reactions of species (for example, R15, R16, R17, and R19) are dependent on temperature. Our calculations show that the highest temperatures of HD 209458b and GJ 436b can attain 10000K and 4500K, respectively. For Kepler-11b, the highest temperature can only attain $\sim$ 1800K (the right panel of Figure 14, and also see Lammer et al. 2003).
The circumstance of low temperature is more favorable for the recombination reactions.
Thus, recombination counteracts the effect of photoionization to a certain extent (see Figure 8), making the profiles of H moderately sensitive to the profiles of EUV SED.

\subsection{The Mass Loss Rate}
\begin{figure}
  \begin{center}
\begin{minipage}[r]{0.5\linewidth}
\centering
\includegraphics[width=5.2in,height=4.in]{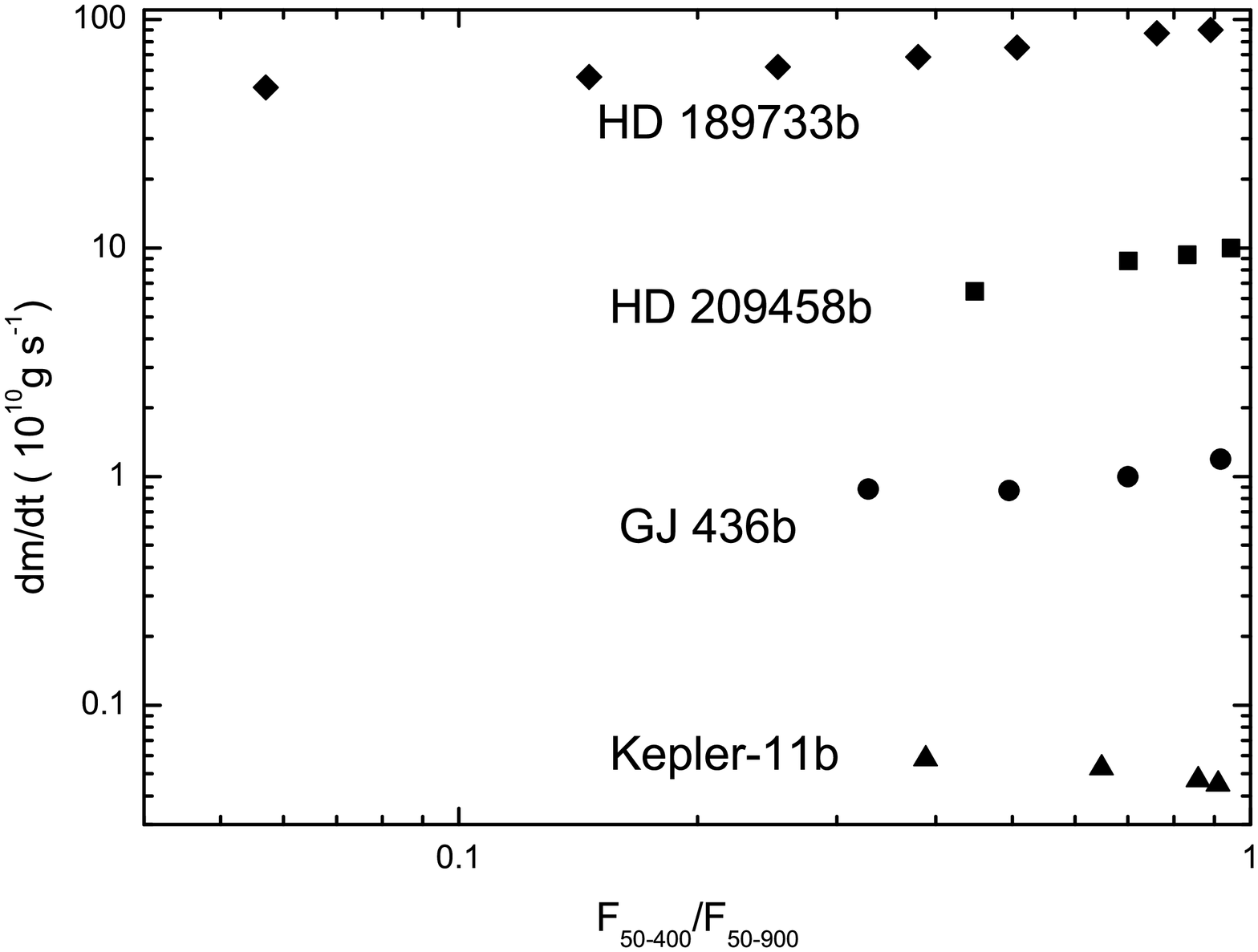}
\end{minipage}
\caption{The variations of the total mass loss rate with the variations of spectral index F$_{50-400}$/F$_{50-900}$.}
  \end{center}
\end{figure}

The total mass loss rates are moderately sensitive to the profiles of the EUV SED (Figure 13). For HD 209458b and HD 189733b, the mass loss rates only decrease 30-40\% when the spectral index $\beta$ decreases almost by a factor of 2. For GJ 436b and Kepler-11b, the variations of their mass loss rates compare well with that of HD 209458b. The mass loss rates of GJ 436b, at the level of 10$^{10}$ g s$^{-1}$, are an order of magnitude higher than those of Kepler-11b.
We calculated several cases for HD 189733b. The mass loss rate is $\sim 9 \times$ 10$^{11}$ g s$^{-1}$ in the case of $\beta$=0.89. The mass loss rate can decrease by a factor of 2 when the spectral index $\beta$ decreases to 0.057. Our calculation results show that the dependence of the mass loss rates on the EUV SED is mild. Thus, using an arbitrary EUV SED profile provides a reasonable estimation for the mass loss rate.

The mass loss rates are determined by the amount of effective absorption of stellar EUV radiation. Thus, the heating is related to the product of number densities of species and the absorption cross-sections (H$_{heat}$ $\sim$ n$\sigma_{\nu}$). High-energy photons can penetrate deeper into the atmosphere due to the smaller cross-sections. In the dense regions of the atmosphere, more species can absorb the stellar irradiation.
On the other hand, the regions where the low-energy photons are absorbed are located at a higher altitude than those where high-energy photons are absorbed.
Due to the decrease in the number density of species with increasing altitudes, the number of particles absorbing the stellar irradiation also decreases. However, the effective heating by high- and low-energy photons is comparable because the heating of the atmosphere is proportional to the number density of particles and the cross-section of photoionization. For atomic hydrogen, the cross-section of photoionization is inversely proportional to the cube of the frequency \citep[]{spitzer78}. This also explains why the total mass loss rates predicted for different EUV spectra are comparable.

\section{Sensitivity of Atmospheric Properties to EUV SEDs: Discussion}
\subsection{The Effect of Hydrodynamic Escape Parameters on the Temperature}
In Table 2, we have shown that the hydrodynamic escape parameter
 $\lambda=\frac{GM_{p}\mu}{R_{p}\kappa T_{0}}$ (where R$_{p}$ and M$_{p}$ are the radius and mass of the planet, and $\mu$ and T$_{0}$ are the mean mass per particle and temperature at the lower boundary of the atmosphere) is smaller for Kepler-11b than for other planets.
The hydrodynamic escape parameter is defined as the ratio of the potential energy to the kinetic energy of the gas. The gas can be bound to the planet if the temperature of the gas is very low or the radius of the planet is small (the mass of the planet is assumed as a constant). In such a case, the value of $\lambda$ is large. For higher temperatures and/or larger R$_{p}$ values, however, the kinetic energy of the gas can be greater than the potential energy. Thus, the atmosphere can escape the bounds of the planet. The latter case corresponds to a smaller $\lambda$.
In an atmosphere heated by XUV radiation of the host star, the temperature of the gas can attain a few thousand Kelvins in the bottom of the atmosphere \citep[]{yelle04,murray09,guo11,guo13}. Thus, the value of $\lambda$ can decrease with the increase of the altitude as a result of high temperatures and a large radius. This means that the gas can escape the planet at some altitudes.
To produce a hydrodynamic wind, the critical value of $\lambda$ is about 10-15 \citep[]{stone09}. Thus, a thermally-driven hydrodynamic wind can be produced easily in the atmosphere with a smaller $\lambda$ because a weak XUV irradiation of the star can heat the gas to a medium (low) temperature, which is enough to decrease the value of $\lambda$ to the critical value.
For HD 209458b, GJ 436b, and HD 189733b, the values of $\lambda$ at the lower boundary can attain 179, 115, and 378. This means that the atmospheres are tightly bound, and very high temperatures are needed in order to overcome the gravitational acceleration.
At the same time, only high temperature can decrease the value of $\lambda$ to the critical level.
For Kepler-11b, a low temperature is enough to drive the escape of atmosphere because the value of $\lambda$ is only 36.

Here we test the dependence of the temperature and H density on the integrated flux and hydrodynamic escape parameter. First, we increase the integrated flux of Kepler-11b by a factor of 3. Figure 14 (left panel) shows the ratio of the H number density of case $\beta$=0.389 to case $\beta$=0.91. It is clear that the number density ratios of H near the bottom are moderately changed by the profiles of the EUV SED if the integrated flux is increased by a factor of 3. With the increase of the altitude, the ratios of the H number density become more significant. The dependence is clear and obvious.
Since $\lambda$ has not changed, a higher temperature is not necessary in driving the escape of the atmosphere. The highest temperature only attains $\sim$ 2000K in that case (right panel, Figure 14). It also hints that chemical reaction rates can slightly decrease, yet not too much. The left panel of Figure 15 shows that the primary photochemical reactions before and after increasing the integrated flux are similar to each other, and it suggests that the photochemical reactions should not be very sensitive to the integrated flux. This explains why the H number density ratios near the base of the atmosphere are not changed significantly by the profiles of the EUV SED.
At high altitudes, the optical depth is so small that the H photoionization produces a stronger effect in ionizing the hydrogen neutral atoms due to the increasing integrated flux. In those regions, the ratio of H altitude profiles for different $\beta$\, are more sensitive to the profiles of EUV SED.

Second, we test a case in which the mass of Kepler-11b is increased by a factor of 3. In that case, the gravitational potential increases three times (the value of $\lambda$ also increases 3 times), and the highest temperature attains $\sim$ 5000K (right panel, Figure 14). As shown in the right panel of Figure 15, some chemical reactions are severely depressed. The composition in the bottom of the atmosphere is controlled by the chemical reaction R7, R8, R9, and R19. The photoionization of H dominates the photochemical processes above 1.37R$_{p}$.
The profiles of chemical reaction rates are similar to those of HD 209458b if the mass of Kepler-11b increases by a factor of 3. The number density ratios of H can attain 0.25 at r$>$1.3R$_{p}$ (left panel, Figure 14). Compared to a case having 3 times the integrated flux, the ratios of H are more sensitive to the profiles of EUV SED.
The results above suggest that the temperature profiles are related to the hydrodynamic escape parameter. Based on those results, we suggest that the atmospheric composition of an exoplanet is moderately affected by the EUV SED if $\lambda$ (or the integrated flux) is small (low). However, the composition of the atmosphere becomes more sensitive to the EUV SED if the hydrodynamic escape parameter (or the integrated flux) increases.
\begin{figure}
\begin{minipage}[t]{0.5\linewidth}
\centering
\includegraphics[width=3.6in,height=2.6in]{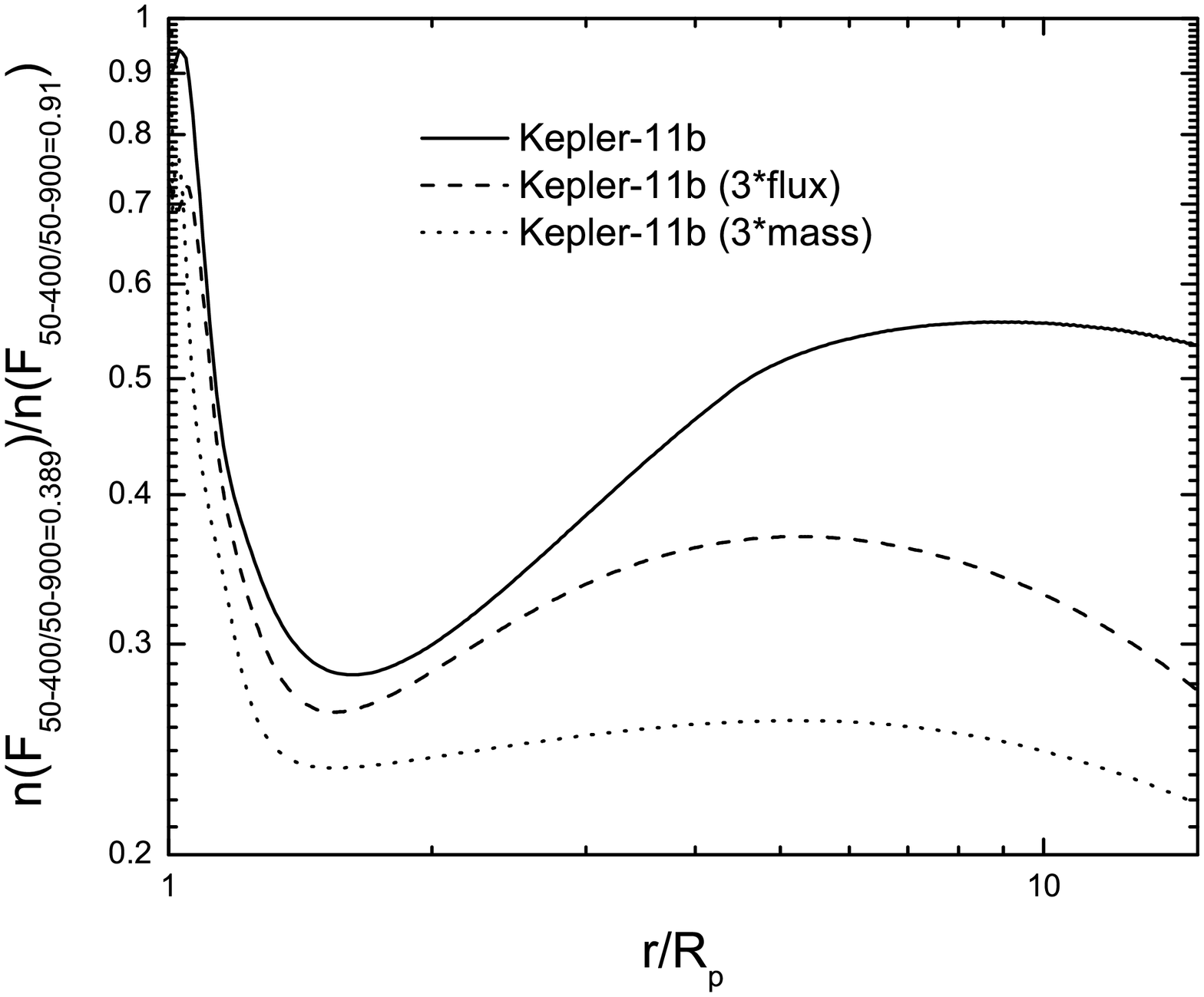}
\end{minipage}
\begin{minipage}[t]{0.5\linewidth}
\centering
\includegraphics[width=3.6in,height=2.6in]{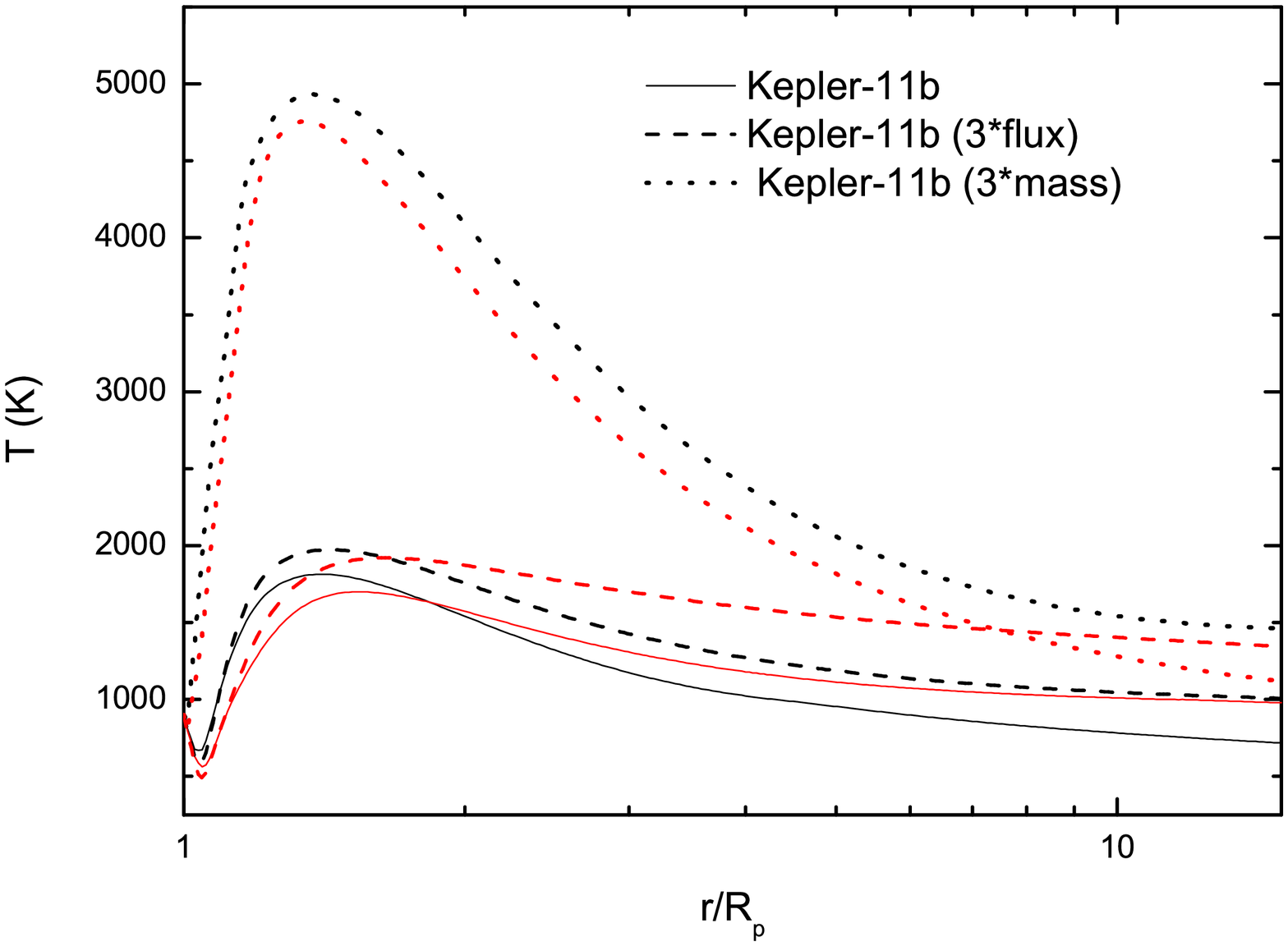}
\end{minipage}
\caption{Left panel: number density ratios of H. Solid line denotes ratio of F$_{50-400}$/F$_{50-900}$=0.389 to
F$_{50-400}$/F$_{50-900}$=0.91 for Kepler-11b. Dashed lines are ratio of F$_{50-400}$/F$_{50-900}$=0.389 to
F$_{50-400}$/F$_{50-900}$=0.91 for the case of 3 times integrated flux of Kepler-11b. Dotted lines are the ratio of F$_{50-400}$/F$_{50-900}$=0.389 to
F$_{50-400}$/F$_{50-900}$=0.91 for the case of 3 times planetary mass of Kepler-11b.
Right panel: temperature distributions of Kepler-11b. Black lines: F$_{50-400}$/F$_{50-900}$=0.91.
Red lines: F$_{50-400}$/F$_{50-900}$=0.389. Solid lines denote the case of Kepler-11b.
Dashed lines are for case of three times integrated flux of Kepler-11b.
Dotted lines are for case of three times planetary mass of Kepler-11b. }
\end{figure}

\begin{figure}
\begin{minipage}[t]{0.5\linewidth}
\centering
\includegraphics[width=3.6in,height=2.6in]{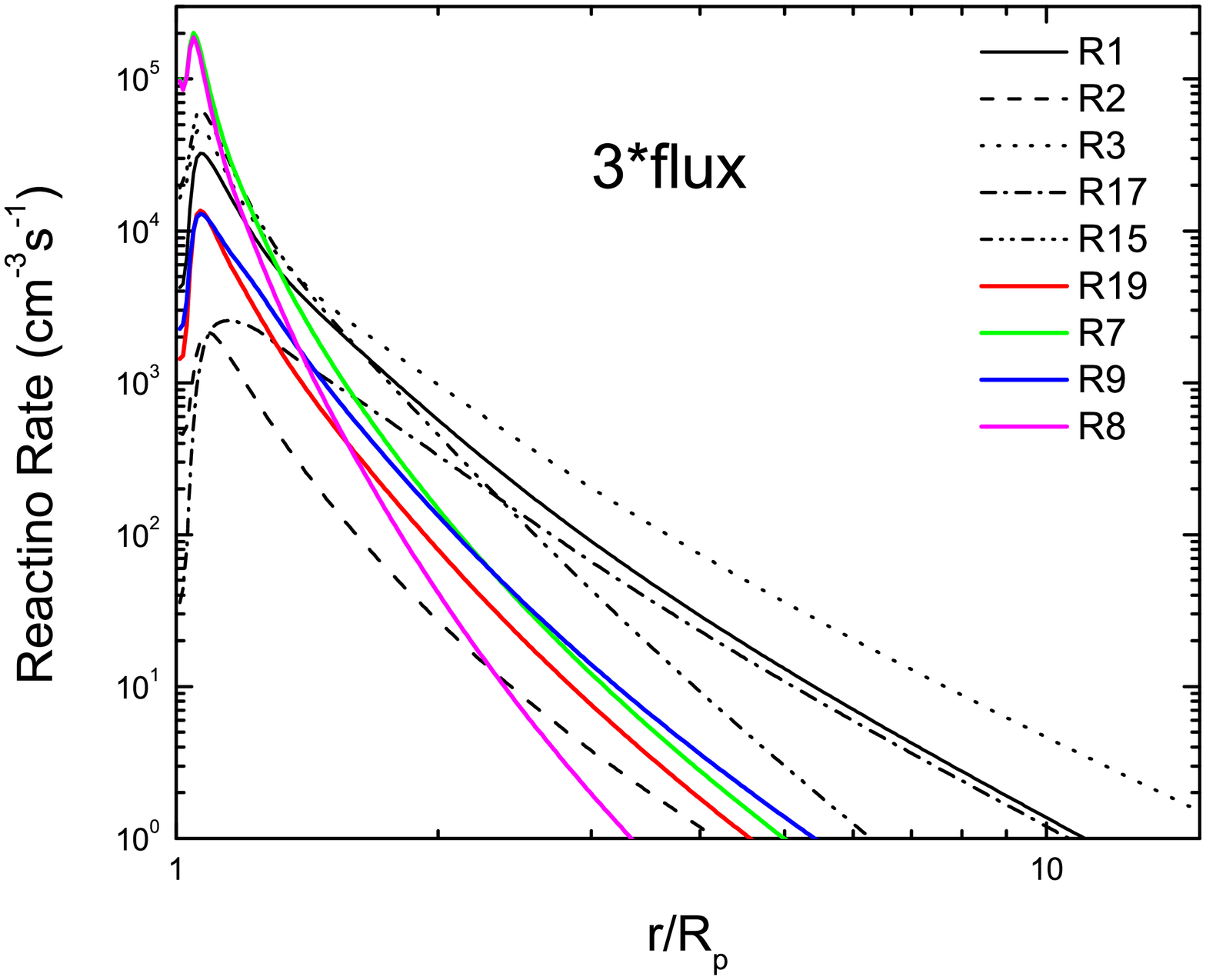}
\end{minipage}
\begin{minipage}[t]{0.5\linewidth}
\centering
\includegraphics[width=3.6in,height=2.6in]{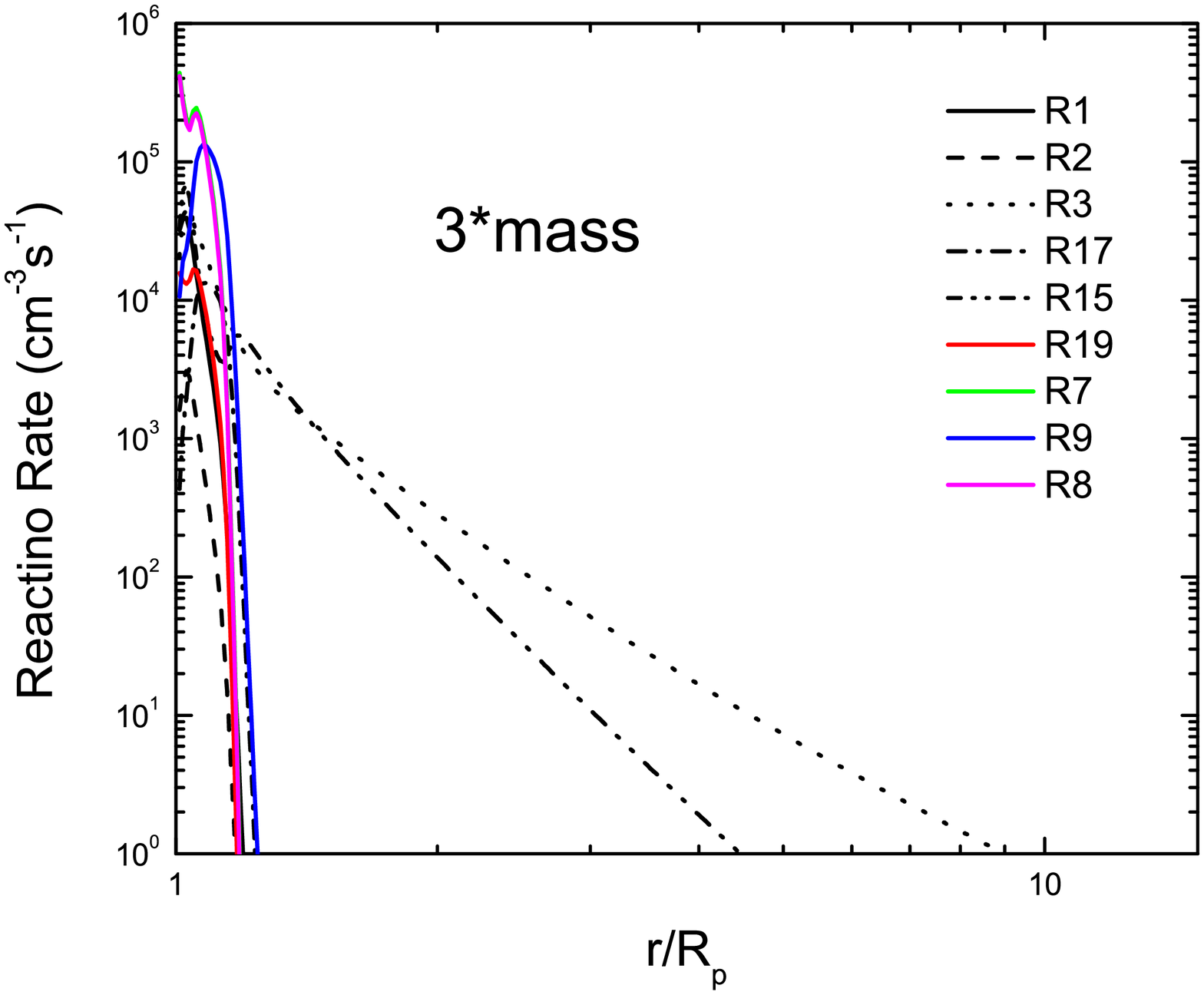}
\end{minipage}
\caption{The photochemical reactions of Kepler-11b for the case of F$_{50-400}$/F$_{50-900}$=0.91.
Left panel: three times integrated flux of Kepler-11b.
Right panel: three times mass of Kepler-11b. Note that not all chemical reactions are shown.}
\end{figure}

\subsection{Effect of Second Ionization}
Generally, only high-energy bands of SED can produce significant second ionization because the energetic photoelectrons produced by photoionization of the bands have enough energy to re-ionize atoms. For the bands of lower energy of SED, the energies of photoelectrons can be lower than the ionization thresholds. Thus, one can expect that the effect of second ionization should be more evident when the SED is dominated by high-energy spectral regions-namely, the case with a higher value of $\beta$.
Using the model results of HD 209458b as the input of number density distribution, we test two cases with different values of $\beta$. The calculation results validate the scenario that the effect of second ionization is indeed important when the value of $\beta$ is higher. The left panel of Figure 16 shows the total ionization rate of H with the case of $\beta$=0.945. It is clear that the second ionization can increase by a factor of 10 for a total ionization rate in the regions of r$<$1.05R$_{p}$ where the gas is almost neutral, confirming the general trend of the importance of second ionization in neutral gases (e.g., Figure 13 of \citet[]{ricotti02}). With the increase of altitude, the ionization rates increase by a factor of 3 near 1.2R$_{p}$. At 1.5R$_{p}$, the ionization rates with second ionization are 50\% more than the case without second ionization. At larger altitudes, the difference of ionization rates is about 10\%.
For the case of $\beta$=0.448, the effect of second ionization is weaker than that of $\beta$=0.945 (right panel, Figure 16). The prominent differences only occur at r$<$1.1R$_{p}$. For larger altitudes, the influence of second ionization is negligible.

\begin{figure}
\begin{minipage}[t]{0.5\linewidth}
\centering
\includegraphics[width=3.6in,height=2.6in]{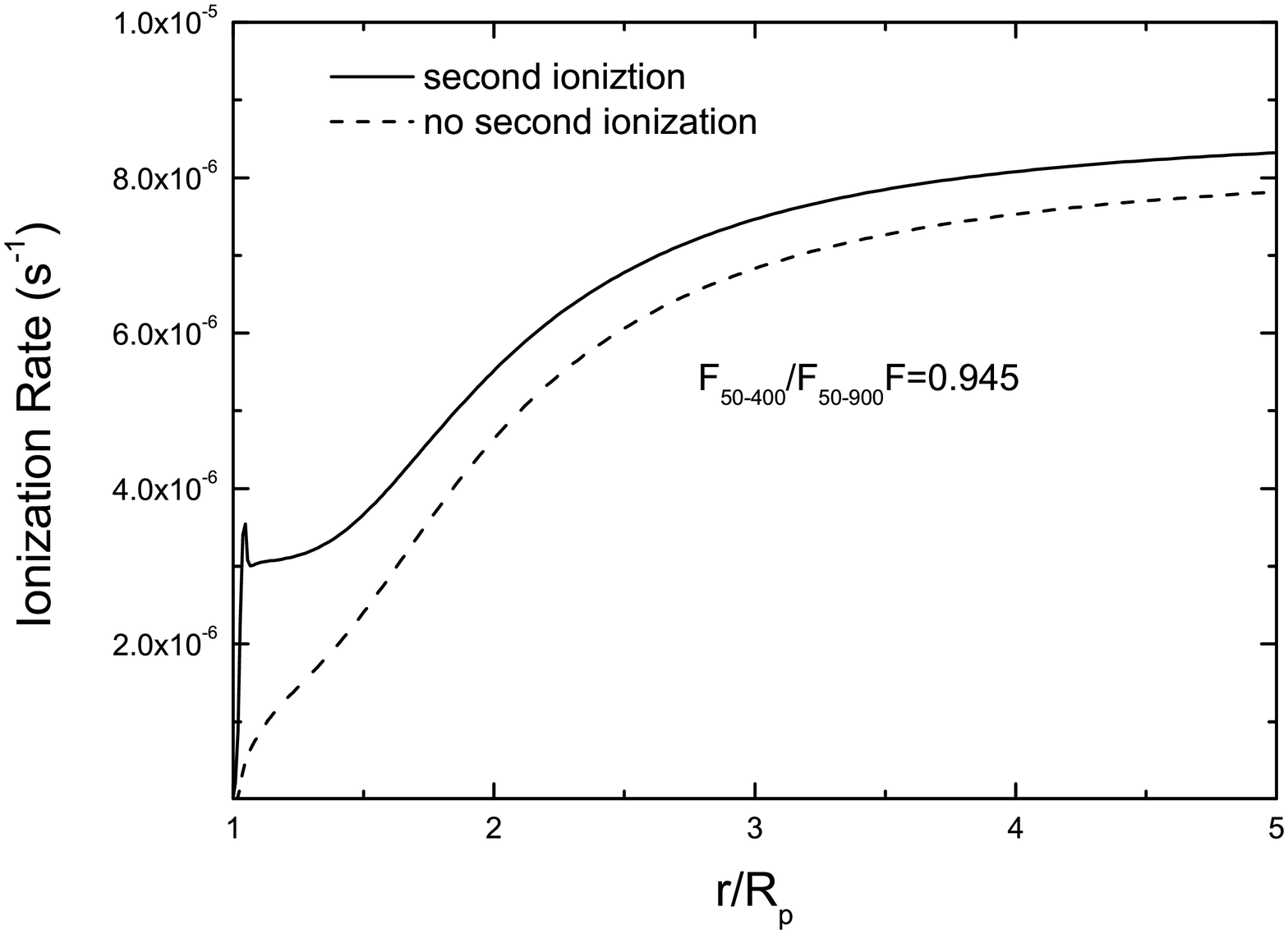}
\end{minipage}
\begin{minipage}[t]{0.5\linewidth}
\centering
\includegraphics[width=3.6in,height=2.6in]{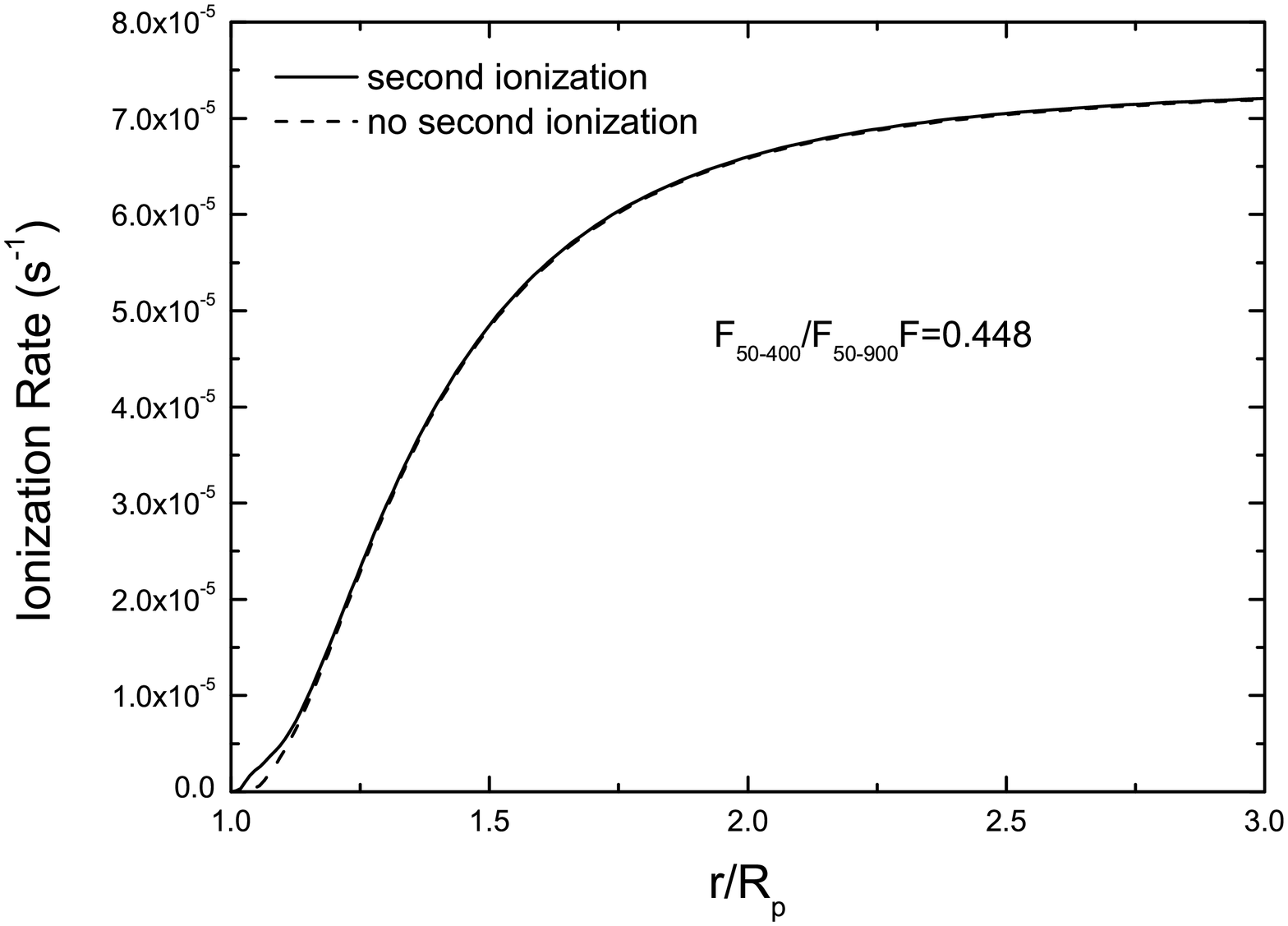}
\end{minipage}
\caption{The ionization rates of H of HD 209458b.
Left panel: case with F$_{50-400}$/F$_{50-900}$=0.945.
Right panel: case with  F$_{50-400}$/F$_{50-900}$=0.448.}
\end{figure}

\subsection{Influence of EUV SEDs on the Lyman $\alpha$ Transit Absorption}
Figures 3, 5, 7, and 9 show that the composition of escaping atmosphere is strongly affected by the profiles of the EUV SED. It is clear from Figure 11 that the number densities of H are also affected by the EUV SED.

The transit absorption of Lyman $\alpha$ is mainly determined by the column density of H along the line of sight as well as the ambient temperature. For a spherical outflow, the optical depth along the line of sight is

\begin{equation}
\tau_{\nu}=\int_{0}^{\infty}n_{h}\sigma_{\nu} dx,
\end{equation}

where x is the path length measured along the line of sight. By the calculations of \citet[]{ben10}, and confirmed by \citet[]{kos13}, the numbers of H below 3 R$_{p}$ produce a significant optical depth for Lyman $\alpha$.
It is clear from Figure 11 that the number densities of atomic hydrogen of HD 209458b and GJ 436b at r/R$_{p}$=2-3 decrease one order of magnitude when the value of $\beta$ decreases by a factor of 2 or 3. At a larger radius, the decrease is more dramatic.
Even the number densities of the H of Kepler-11b at r/R$_{p}$=2-3 also vary a few times when the value of $\beta$ decreases by a factor of 3.
Our results suggest that the optical depth of Lyman $\alpha$ can vary one order of magnitude with the variations of the EUV SEDs even if the absorbing regions are limited in the range of r$<$ 3R$_{p}$.
It is important to stress here that 1D hydrodynamic models provide a first order of magnitude estimation of the transit depth because the atmospheric structures are affected by stellar tides, ionosphere-magnetosphere coupling, multi-dimensional radiative transfer, and stellar winds. However, our present effort represents a reasonable investment in our exploratory investigation of the real conditions that prevail in the upper atmosphere of exoplanets.

The above results mean that it is easier to detect the signals of the atmospheric escape in those stars with higher ratios of $\beta$ because the escaping atmosphere in those systems can eject more hydrogen atoms.
Thus, as a key role in detecting and modeling the atmospheric escape, it is very important to use a credible EUV SED of the star for accurately depicting the atmospheric escape.

It is well known that the direct measurements of the EUV SEDs of the stars are difficult due to the strong extinction of the interstellar medium. \citet[]{linsky14} compared their reconstructed EUV SEDs with those predicted by \citet[]{sanz11}. Using Table 6 of \citet[]{linsky14}, we calculated the spectral indexes $\beta$ of HD 209458 and GJ 436. The results of \citet[]{linsky14} show that the values of $\beta$ of HD 209458 and GJ 436 are 0.7 and 0.71. However, the values of $\beta$ obtained using the method of \citet[]{sanz11} are 0.32 and 0.3 for HD 209458 and GJ 436, respectively.
Evidently, \citet[]{linsky14} predicted higher fluxes in the higher-energy spectral regions, but the fluxes predicted by \citet[]{sanz11} are dominated by the lower-energy spectral regions. The fact that the different methods predicted different EUV profiles indicates the inadequate treatment for the stellar EUV SEDs. Further improvements or confirmation in reconstructing EUV SED are needed.

\subsection{Effect of the Stellar Wind}
To this point, our models do not include the influence of stellar wind.
The impinging stellar wind should drag the ions of planetary wind if exoplanets have weak magnetic fields. At the same time, the charge-exchange between the protons of the stellar wind and the planetary hydrogen atoms leads to the production of cold ions of planetary origin and energetic neutral atoms (ENA). In the process, the majority of ions of planetary origin could be picked up or transported by the stellar wind so that they can escape from the planet. These non-thermal processes can affect the mass loss and the composition.
\citet[]{kis14} estimated that the non-thermal mass loss rates are only a few percent of thermal mass loss rates for Kepler-11b-f. This suggests that the winds of close-in exoplanets are dominated by the thermal mass loss due to the high irradiation.
Thus, the inclusion of the stellar winds should not change the main conclusion of the article.

\section{Application to Study the Time Variation of HD 189733b \lya\ Transit Absorption}
HD\, 189733b represents the unique case of an exoplanet orbiting a young, active K-star and showing a time variation in its transit absorption at \lya\ observed by HST/STIS between 2010 and 2011 \citep{lecavelier12,bou13a}. For reference, we recall that a first report based on low-resolution observations obtained by HST/ACS in 2007-2008 at \lya\ showed a strong transit absorption of $5.0\pm0.8$\% for the integrated line \citep{lecavelier10}.  Interestingly, when looking into the details of the reported detection, one can note that three transits were obtained, two in 2007 around the June period, and a third ten months later in April, 2008. As shown in their Table 3, \citet{lecavelier10} reported a rather strong transit for the 2007 period ($\sim   6.2\pm 1.1\%$ in the average), while for 2008 the \lya\, transit absorption was weak ($\sim 2.7\pm 1.2\%$), yet a flare event may have complicated the diagnostic for that third transit. The variability in the \lya\ transit absorption was later confirmed by HST/STIS medium-resolution observations obtained in 2010 and 2011 \citep{lecavelier12}. For instance, the HD 189733b \lya\ transit showed a variation between 2.9\% in 2010 and 5.1\% in 2011 for the integrated line. In contrast to the 2007/2008 transits, HST/STIS medium observations brought key additional details regarding the spectral distribution of the \lya\ absorptions. In addition, as discussed in detail in \citet{bou13a}, the \lya\ line profile has two peaks that show different transit absorptions during transit and over time.

To assess the HST data handling previously reported, we re-analyze the 2010 and 2011 HST/STIS data and confirm the finding of \citet{lecavelier12} and \citet{bou13a}. However, it is important to stress that the focus on the blue and red wings of the line is misleading because of the absorption by the interstellar medium and the earth's geocoronal emission contamination that varied between 2010 and 2011. As a matter of fact, the geocoronal contamination was stronger and spectrally more extended for the 2010 observation than for 2011. Similarly, the reference to a transit absorption of the "whole line" is also misleading because a large fraction of the \lya\ is useless due to the quoted contaminations, particularly the time-variable Earth geocoronal emission. This leads us to omit showing the spectral region contaminated by the Earth's geocorona in order to highlight clearly the true signal available for analysis.

\citet{bou13a} carefully analyzed the transit absorption for specific spectral ranges of the \lya\ line, showing distinct results for the spectral range -230 to -140 km/s (on the blue wing of the line) compared to the range 60 to 140 km/s (on the red wing of the line). For instance, strong absorptions are found for the red wing of the \lya\, line in both 2010 ($\sim 5.8\pm2.6 \% $) and 2011 ($\sim 7.7\pm2.7 \%$) (e.g., Table 4 of \citet{bou13a}). However, for the blue wing, only 2011 showed a strong absorption in the indicated spectral range ($\sim 14.4\pm3.6\%$) compared to $\sim 0.5\pm3.8\%$ in 2010. In that frame, it is misleading to conclude that no absorption was found during the 2010 April observation when an indication of a strong absorption appears on the red wing of the line ($\sim 5.8\pm2.6 \% $) \citep{lecavelier10,lecavelier12,bou13a}. Despite the large uncertainty attached, the 2010 red wing absorption is consistent both with the 2011 detection ($\sim 7.7\pm2.7 \%$) and the fact that the geocoronal emission contamination was stronger and spectrally more extended in 2010 than in 2011. This also means that the time-variation observed by HST/ACS during the 2007-2008 period may also correspond to transit absorptions due to the same process spectrally revealed by the 2010/2011 HST/STIS medium-resolution observations.  Here, it is important to stress that the spectral ranges that are available (due to ISM and geocorona contaminations) to study the \lya\ transit absorption are not symmetric with respect to the center of the line, and they do not overlap. This means that the two wavelength ranges do not probe the same opacity regimes in the absorbing planetary atmosphere \citep{ben10}, an important fact to remember in the data interpretation.

\begin{figure}
\begin{minipage}[t]{0.5\linewidth}
\centering
\includegraphics[width=3.5in,height=2.6in]{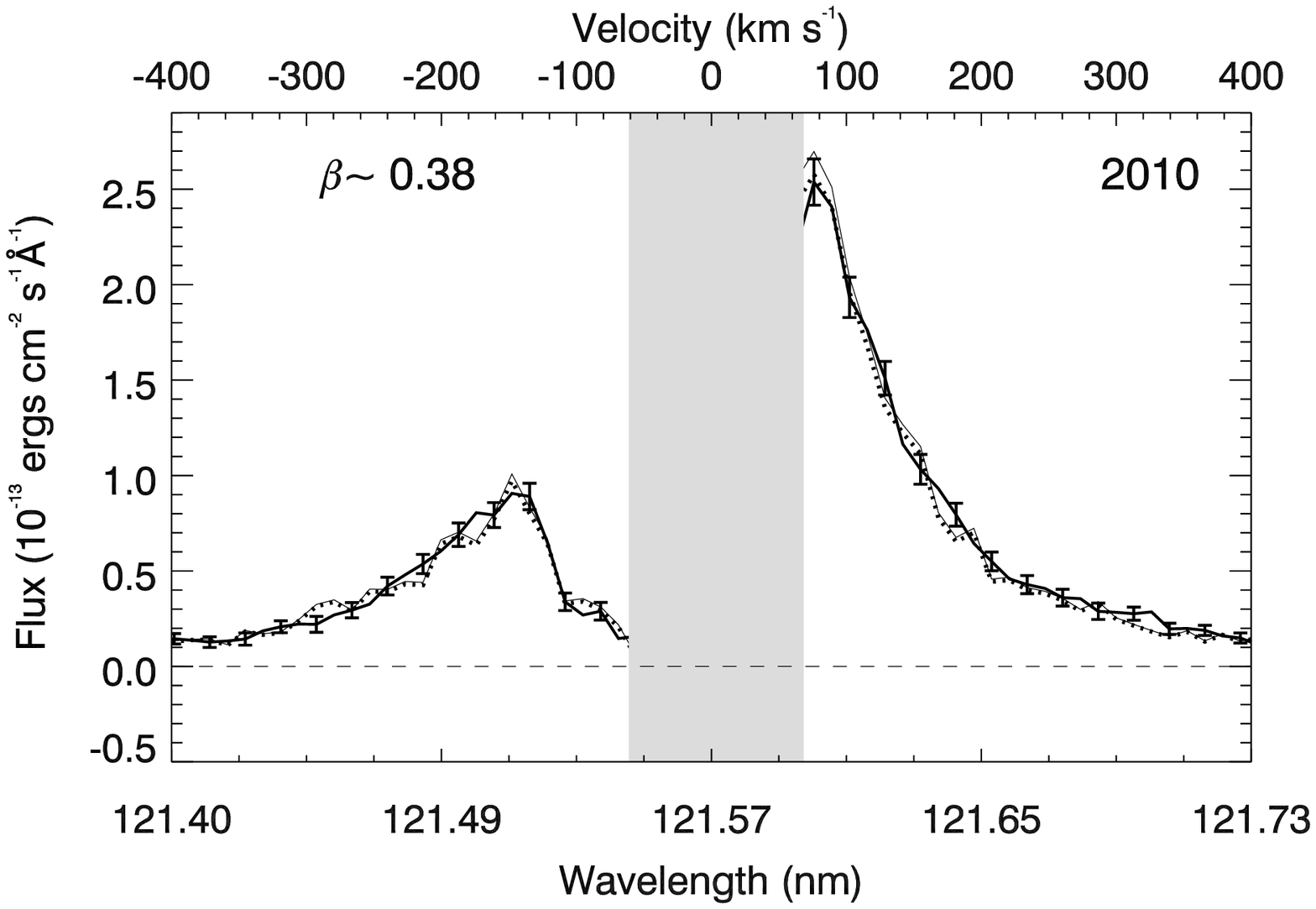}
\end{minipage}
\begin{minipage}[t]{0.5\linewidth}
\centering
\includegraphics[width=3.5in,height=2.6in]{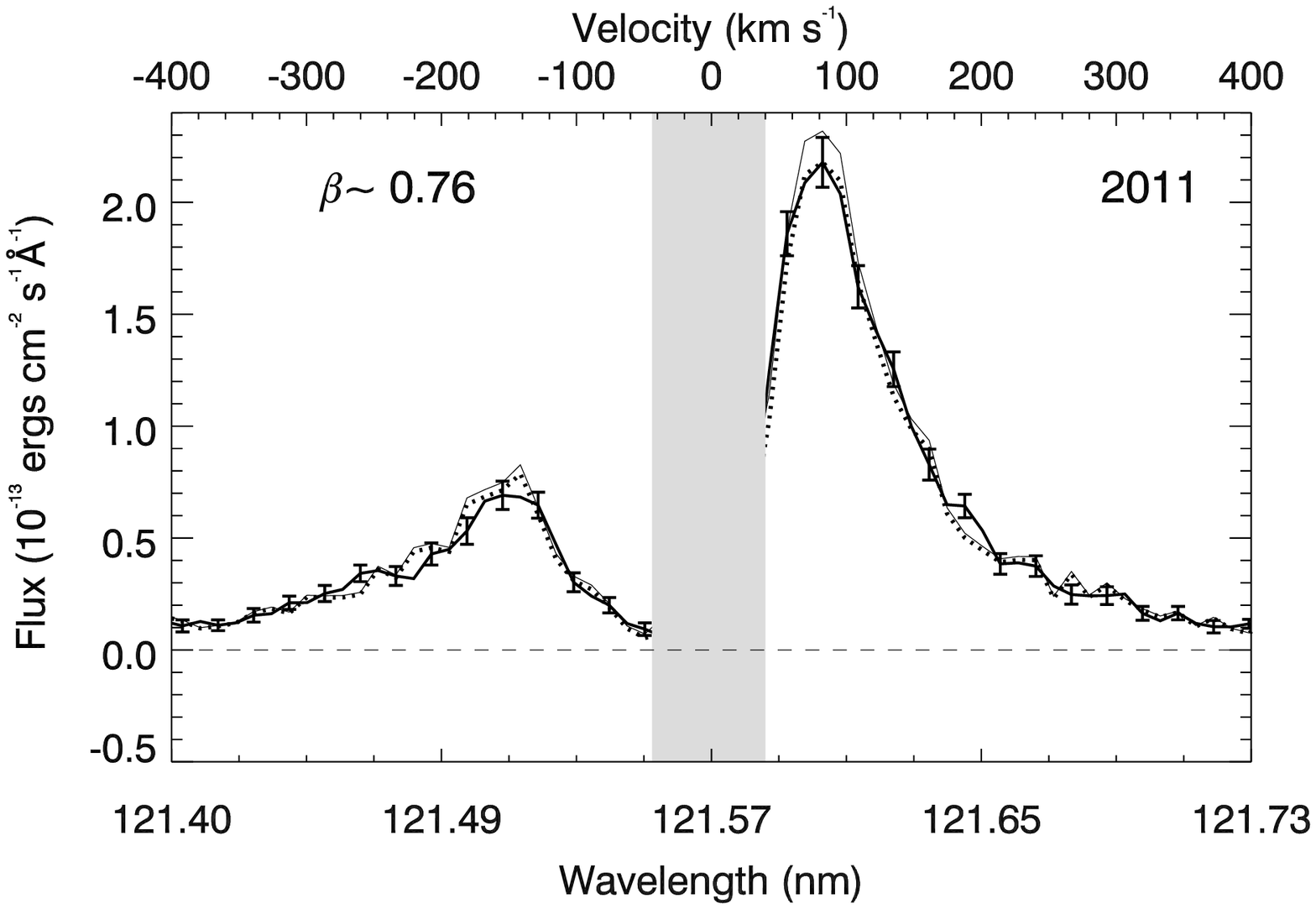}
\end{minipage}
\caption{HST/STIS medium-resolution observation of HD 189733b \lya\ transit. Comparison of model in-transit (dots) to observed in-transit (thick line) and to out-transit (thin line) line profiles. Error bars are shown for reference. The shaded area corresponds to the Earth's geocorona \lya\, emission contamination. (Left) 2010 model with EUV SED index $\beta\sim0.38$. (Right) 2011 model with EUV SED index $\beta\sim0.76$. }
\end{figure}

In the following we vary the index beta of the HD 189733 EUV SED to obtain different HI radial distributions and compute the corresponding \lya\, transit absorption following the concept described in \citet{ben08}. In contrast to previous interpretations in terms of radiation pressure that produce a cometary tail and hot atoms produced from the stellar wind \citep{bou13b}, we postulate that the observed variation of the \lya\, transit absorption could result from a variation of the EUV SED of the star that modified the upper atmosphere of the planet differently. As the star is a young, active K-star, one may assume that the beta index may strongly vary with time. As shown in Figure 17, we can obtain a good fit to the \lya\, line profiles observed during transit in 2010 and 2011, respectively, using $\beta \sim 0.38$ and $\sim 0.76$.

For the obtained solutions, the HI density vertical column varied from $[HI]\sim {\rm 7.3\times 10^{19}cm^{-2}}$ in 2010 to $\sim  {\rm 9.7\times 10^{19}cm^{-2}}$ in 2011, and its radial distribution expanded almost $\sim1.5$R$_p$  outward between 2010 and 2011. This means that the derived increase in the EUV SED index enhanced and expanded the HI population enough to produce the transit absorption depicted in Figure 17.
Indeed, if the $\beta$ increases by a factor of 2 from 2010 to 2011, the photoionization rate of H decreases (Figure. 12). Consequently, the number density of H  increases by a factor of few at different altitudes (e.g., Figure 11), which explains the enhanced and expanded absorbing HI layer for 2011 compared to 2010.

In addition, our solution naturally explains the absence of absorption on the blue wing of the \lya\, line in 2010 because the total atmosphere opacity is smaller, which results in a much fainter absorption for the far wavelength range -230 to -140 km/s, compared to much closer wavelength range (with respect to the line center) on the red side between 60 and 140 km/s. For reference, the blue wing wavelength band is located $\sim 15$ Doppler units (assuming a reference temperature T$\sim 10^4$\,K) from the line center to compare to  $\sim 8$ Doppler units from the center for the red wing band. In terms of the classical atomic absorption Voigt profile, the absorption probability is three to four times stronger for the 60-140 km/s band than for the -230 to -140 km/s band. As shown in Figure 17, this simple explanation properly describes the change observed during transit between 2010 and 2011, particularly for the farther blue wing.

Our findings seems to indicate that much like for HD 209458b \citep{ben08,ben10,kos13} ,the thermal HI population is an appropriate explanation of the HD 189733b \lya\, transit without the need for any additional process, independently of the time variation thus far observed that is likely related to the EUV SED variation. In that frame, we provide the EUV spectra of HD 189733 that would result from that interpretation for the 2010 and 2011 time periods (Figure 18). It is interesting to see that the new EUV spectra are not that different from the reference spectrum of $\epsilon$ Eri, yet the 2010 spectrum could be used as a reference for future comparisons between models and observations. If confirmed, this would, for the first time, offer the possibility of directly monitoring the EUV flux activity of a K-star by observing the \lya\, transit of the its planetary companion, a further confirmation of the importance of the \lya\, stellar diagnostic proposed by \citet{linsky14}.

\begin{figure}
\centering
\includegraphics[width=14cm,height=10cm]{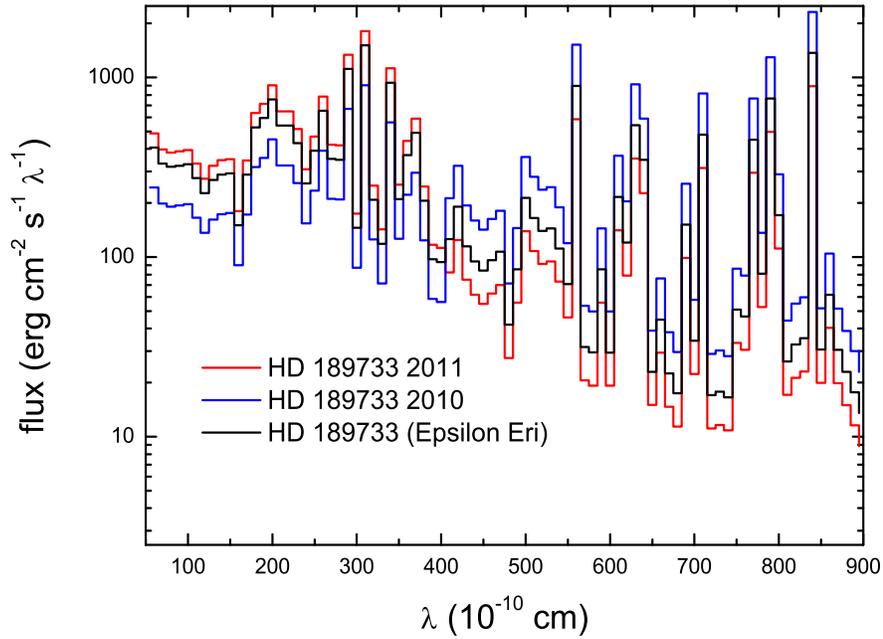}
\caption{EUV SED of HD 189733 derived from the \lya\, transit absorption of the planetary companion at the exoplanet orbital position. The 2010 (blue) EUV spectrum was obtained for index $\beta\sim0.38$, and the 2011(red) EUV spectrum for index $\beta\sim0.76$. The change observed between 2010 and 2011 is reasonable considering the early age of the K-star. For reference, the nominal $\epsilon$\, Eri spectrum (black) is also shown.}
\end{figure}

\section{CONCLUSION}
We calculated the influences of the stellar EUV SEDs on the respective compositions of the atmosphere for HD 209458b, GJ 436b, HD 189733b, and Kepler-11b. This sample includes Hot-Jupiter, Hot-Neptune and super-Earth planets. According to our model calculations, the total mass loss rates of the sample are moderately sensitive to the profiles of the EUV SED. However, the density distributions of atomic hydrogen show two different cases. For HD 209458b, GJ 436b, and HD 189733b, the density distributions of atomic hydrogen are tightly related to the profiles of the EUV SED because the photoionization is dominant in the photochemical processes. For Kepler-11b, due to the small hydrodynamic escape parameter, the atmospheric temperature is low. For that reason, its composition is controlled by many photochemical reactions rather than by the photoionization solely. For that reason, the composition of Kepler-11b is moderately sensitive to the EUV SEDs. Our results also show that a credible EUV SED of the star can be important in detecting the signals of atmospheric escape because the amount of hydrogen atoms in the atmosphere is strongly influenced by the shape of the EUV SED.

To illustrate how the measure of the \lya\, transit absorption of an exoplanet may help constrain the EUV SED of its host star, we tried to derive the stellar flux index (the flux ratio of high to total energy spectral ranges) that allows a fit to the observed in-transit \lya\, line profile. Our finding is that a factor-two jump in the stellar EUV $\beta$ index is necessary to explain the time-variation observed between 2010 and 2011. In addition, for each of the individual transits, we could fit the transit absorption invoking only the thermal HI population with a vertical distribution that expanded $\sim 1.5$\,R$_p$ outward between 2010 and 2011, which led to the stronger absorption observed in 2011. Finally, the EUV SED derived here for each date shows reasonable shape compared to the nominal EUV spectrum initially assumed for the K-star (e.g., Figure 18). Our proposed technique provides a straightforward and easy-to-follow proxy to connect the EUV SED of the star with the planetary companion \lya\ transit absorption, the monitoring of which may provide a direct measure of the stellar EUV flux.

\section*{Acknowledgments}
For J. Guo, this work was supported by National Natural Science Foundation of
China (Nos.11273054; Nos 10973035). L. Ben-Jaffel knowledges support from CNES and from the French Embassy in China (Programme D\'ecouverte Chine).
The authors would like to thank the referee for several suggestions that helped improve
the manuscript.

\clearpage

\end{document}